\documentclass[twocolumn]{aastex63}

\usepackage{subfigure}
\usepackage{amsmath}
\usepackage{amsfonts}
\usepackage{gensymb} 

\newcounter{affilct}
\makeatletter
\newcommand{\affilref}[1]{%
  \@ifundefined{c@#1}%
    {\newcounter{#1}%
     \setcounter{#1}{\theaffilct}%
     \refstepcounter{affilct}%
     \label{#1}%
     }{}%
  \ref{#1}%
 }
\makeatother

\makeatletter
\newcommand*\affilreftxt[2]{%
  \@ifundefined{c@#1txt}
    {\newcounter{#1txt}%
     \setcounter{#1txt}{1}
     \altaffiltext{\ref{#1}}{#2}
     }{
     }
  }
\makeatother


\usepackage{lineno}

\def\macs1149{MACS\,J1149.5+2223}

\def\rasx{11$^{\rm h}$49$^{\rm m}$36.022$^{\rm s}$}
\def\decsx{22$^{\circ}$23$'$48.10$''$}

\def\chisq1radioactive{27202.565}
\def\params1radioactive{13}
\def\aic1radioactive{27228.565}

\def\delaysxsonethreetwodolphot{374.956267}

\def\redchisqthreetwodolphot{1.70}
\def\chisqthreetwodolphot{1009.7}
\def\paramsthreetwodolphot{22}
\def\aicthreetwodolphot{1053.7}

\def\delaysxsonetwoonedolphot{372.254953}

\def\redchisqtwoonedolphot{1.66}
\def\chisqtwoonedolphot{993.1}
\def\paramstwoonedolphot{18}
\def\aictwoonedolphot{1029.1}

\def\delaysxsonetwotwodolphot{372.251261}

\def\redchisqtwotwodolphot{1.66}
\def\chisqtwotwodolphot{993.1}
\def\paramstwotwodolphot{20}
\def\aictwotwodolphot{1033.1}

\def\delaysxsonethreeonedolphot{374.454781}

\def\redchisqthreeonedolphot{1.65}
\def\chisqthreeonedolphot{984.4}
\def\paramsthreeonedolphot{20}
\def\aicthreeonedolphot{1024.4}

\def\delaysxsonefouronedolphot{379.730981}

\def\redchisqfouronedolphot{1.61}
\def\chisqfouronedolphot{958.0}
\def\paramsfouronedolphot{22}
\def\aicfouronedolphot{1002.0}

\def\delaysxsonefourtwodolphot{381.557306}

\def\redchisqfourtwodolphot{1.65}
\def\chisqfourtwodolphot{980.9}
\def\paramsfourtwodolphot{24}
\def\aicfourtwodolphot{1028.9}

\def\delaysxsonefiveonedolphot{375.789274}

\def\redchisqfiveonedolphot{1.61}
\def\chisqfiveonedolphot{952.9}
\def\paramsfiveonedolphot{24}
\def\aicfiveonedolphot{1000.9}

\def\delaysxsonefivetwodolphot{367.115460}

\def\redchisqfivetwodolphot{1.75}
\def\chisqfivetwodolphot{1036.8}
\def\paramsfivetwodolphot{26}
\def\aicfivetwodolphot{1088.8}

\def\delaysxsonesixonedolphot{365.728932}

\def\redchisqsixonedolphot{1.76}
\def\chisqsixonedolphot{1037.9}
\def\paramssixonedolphot{26}
\def\aicsixonedolphot{1089.9}

\def\delaysxsonesixtwodolphot{380.966961}

\def\redchisqsixtwodolphot{1.69}
\def\chisqsixtwodolphot{993.1}
\def\paramssixtwodolphot{28}
\def\aicsixtwodolphot{1049.1}

\def\rmsstwosonedelayRPthreerdmedianchabrier{7.1}
\def\outlierfractionstwosonedelayRPthreerdmedianchabrier{0.029}

\def\rmsstwosonedelayThorpmedianchabrier{5.2}
\def\outlierfractionstwosonedelayThorpmedianchabrier{0.031}

\def\rmsstwosonedelayKellymedianchabrier{4.6}
\def\outlierfractionstwosonedelayKellymedianchabrier{0.038}

\def\rmsstwosonedelayMillonsplmedianchabrier{4.9}
\def\outlierfractionstwosonedelayMillonsplmedianchabrier{0.048}

\def\rmssthreesonedelayRPthreerdmedianchabrier{6.6}
\def\outlierfractionsthreesonedelayRPthreerdmedianchabrier{0.008}

\def\rmssthreesonedelayThorpmedianchabrier{5.0}
\def\outlierfractionsthreesonedelayThorpmedianchabrier{0.015}

\def\rmssthreesonedelayKellymedianchabrier{4.3}
\def\outlierfractionsthreesonedelayKellymedianchabrier{0.019}

\def\rmssthreesonedelayMillonsplmedianchabrier{4.6}
\def\outlierfractionsthreesonedelayMillonsplmedianchabrier{0.019}

\def\rmssfoursonedelayRPthreerdmedianchabrier{11.1}
\def\outlierfractionsfoursonedelayRPthreerdmedianchabrier{0.019}

\def\rmssfoursonedelayThorpmedianchabrier{10.3}
\def\outlierfractionsfoursonedelayThorpmedianchabrier{0.026}

\def\rmssfoursonedelayKellymedianchabrier{5.8}
\def\outlierfractionsfoursonedelayKellymedianchabrier{0.019}

\def\rmssfoursonedelayMillonsplmedianchabrier{8.1}
\def\outlierfractionsfoursonedelayMillonsplmedianchabrier{0.030}

\def\rmssxsonedelayRPthreerdmedianchabrier{8.6}
\def\outlierfractionsxsonedelayRPthreerdmedianchabrier{0.009}

\def\rmssxsonedelayThorpmedianchabrier{7.7}
\def\outlierfractionsxsonedelayThorpmedianchabrier{0.006}

\def\rmssxsonedelayKellymedianchabrier{5.5}
\def\outlierfractionsxsonedelayKellymedianchabrier{0.009}

\def\rmssxsonedelayMillonsplmedianchabrier{7.6}
\def\outlierfractionsxsonedelayMillonsplmedianchabrier{0.040}
\def\weightstwosonedelayRPthreerdmedianchabrier{0.005}
\def\weightstwosonedelayThorpmedianchabrier{0.217}
\def\weightstwosonedelayKellymedianchabrier{0.540}
\def\weightstwosonedelayMillonsplmedianchabrier{0.238}

\def\weightsthreesonedelayRPthreerdmedianchabrier{0.027}
\def\weightsthreesonedelayThorpmedianchabrier{0.173}
\def\weightsthreesonedelayKellymedianchabrier{0.598}
\def\weightsthreesonedelayMillonsplmedianchabrier{0.203}

\def\weightsfoursonedelayRPthreerdmedianchabrier{0.023}
\def\weightsfoursonedelayThorpmedianchabrier{0.015}
\def\weightsfoursonedelayKellymedianchabrier{0.840}
\def\weightsfoursonedelayMillonsplmedianchabrier{0.122}

\def\weightsxsonedelayRPthreerdmedianchabrier{0.053}
\def\weightsxsonedelayThorpmedianchabrier{0.038}
\def\weightsxsonedelayKellymedianchabrier{0.885}
\def\weightsxsonedelayMillonsplmedianchabrier{0.024}

\def\rmsstwosonemuRPthreerdmedianchabrier{0.30}
\def\outlierfractionstwosonemuRPthreerdmedianchabrier{0.022}

\def\rmsstwosonemuThorpmedianchabrier{0.24}
\def\outlierfractionstwosonemuThorpmedianchabrier{0.019}

\def\rmsstwosonemuKellymedianchabrier{0.29}
\def\outlierfractionstwosonemuKellymedianchabrier{0.028}

\def\rmsstwosonemuMillonsplmedianchabrier{0.27}
\def\outlierfractionstwosonemuMillonsplmedianchabrier{0.023}

\def\rmssthreesonemuRPthreerdmedianchabrier{0.35}
\def\outlierfractionsthreesonemuRPthreerdmedianchabrier{0.034}

\def\rmssthreesonemuThorpmedianchabrier{0.32}
\def\outlierfractionsthreesonemuThorpmedianchabrier{0.035}

\def\rmssthreesonemuKellymedianchabrier{0.33}
\def\outlierfractionsthreesonemuKellymedianchabrier{0.034}

\def\rmssthreesonemuMillonsplmedianchabrier{0.33}
\def\outlierfractionsthreesonemuMillonsplmedianchabrier{0.035}

\def\rmssfoursonemuRPthreerdmedianchabrier{0.08}
\def\outlierfractionsfoursonemuRPthreerdmedianchabrier{0.019}

\def\rmssfoursonemuThorpmedianchabrier{0.11}
\def\outlierfractionsfoursonemuThorpmedianchabrier{0.005}

\def\rmssfoursonemuKellymedianchabrier{0.08}
\def\outlierfractionsfoursonemuKellymedianchabrier{0.019}

\def\rmssfoursonemuMillonsplmedianchabrier{0.08}
\def\outlierfractionsfoursonemuMillonsplmedianchabrier{0.010}

\def\rmssxsonemuRPthreerdmedianchabrier{0.08}
\def\outlierfractionsxsonemuRPthreerdmedianchabrier{0.016}

\def\rmssxsonemuThorpmedianchabrier{0.08}
\def\outlierfractionsxsonemuThorpmedianchabrier{0.014}

\def\rmssxsonemuKellymedianchabrier{0.08}
\def\outlierfractionsxsonemuKellymedianchabrier{0.016}

\def\rmssxsonemuMillonsplmedianchabrier{0.08}
\def\outlierfractionsxsonemuMillonsplmedianchabrier{0.016}
\def\weightstwosonemuRPthreerdmedianchabrier{0.000}
\def\weightstwosonemuThorpmedianchabrier{0.872}
\def\weightstwosonemuKellymedianchabrier{0.128}
\def\weightstwosonemuMillonsplmedianchabrier{0.000}

\def\weightsthreesonemuRPthreerdmedianchabrier{0.092}
\def\weightsthreesonemuThorpmedianchabrier{0.535}
\def\weightsthreesonemuKellymedianchabrier{0.373}
\def\weightsthreesonemuMillonsplmedianchabrier{0.000}

\def\weightsfoursonemuRPthreerdmedianchabrier{0.308}
\def\weightsfoursonemuThorpmedianchabrier{0.054}
\def\weightsfoursonemuKellymedianchabrier{0.583}
\def\weightsfoursonemuMillonsplmedianchabrier{0.055}

\def\weightsxsonemuRPthreerdmedianchabrier{0.000}
\def\weightsxsonemuThorpmedianchabrier{0.014}
\def\weightsxsonemuKellymedianchabrier{0.252}
\def\weightsxsonemuMillonsplmedianchabrier{0.734}

\def\corrtwosigdelaystwosoneRPthreerdchabriermedian{4.28 (-13.97, -3.10, 3.23, 8.74, 17.61)}

\def\corrtwosigdelaystwosoneThorpchabriermedian{11.09 (-0.80, 7.95, 11.61, 15.83, 23.48)}

\def\corrtwosigdelaystwosoneKellychabriermedian{8.49 (-6.67, 5.49, 8.57, 11.80, 17.75)}

\def\corrtwosigdelaystwosoneMillonsplchabriermedian{12.89 (-9.96, 9.60, 12.80, 15.93, 20.95)}

\def\corrtwosigdelaysthreesoneRPthreerdchabriermedian{12.93 (-0.85, 7.23, 13.52, 19.58, 28.07)}

\def\corrtwosigdelaysthreesoneThorpchabriermedian{11.13 (2.24, 8.04, 11.74, 16.25, 23.06)}

\def\corrtwosigdelaysthreesoneKellychabriermedian{7.52 (-0.07, 4.12, 7.46, 10.72, 17.02)}

\def\corrtwosigdelaysthreesoneMillonsplchabriermedian{11.73 (2.99, 8.15, 11.65, 14.77, 19.83)}

\def\corrtwosigdelaysfoursoneRPthreerdchabriermedian{16.84 (-15.09, 5.24, 16.27, 27.06, 36.89)}

\def\corrtwosigdelaysfoursoneThorpchabriermedian{55.48 (38.12, 49.51, 58.52, 70.77, 85.43)}

\def\corrtwosigdelaysfoursoneKellychabriermedian{19.24 (4.92, 14.19, 19.29, 24.36, 31.75)}

\def\corrtwosigdelaysfoursoneMillonsplchabriermedian{31.25 (0.65, 24.37, 31.22, 37.83, 44.79)}

\def\corrtwosigdelaysxsoneRPthreerdchabriermedian{380.83 (360.53, 370.68, 379.58, 386.68, 394.58)}

\def\corrtwosigdelaysxsoneThorpchabriermedian{383.83 (367.04, 375.96, 383.62, 390.23, 400.15)}

\def\corrtwosigdelaysxsoneKellychabriermedian{376.02 (365.31, 370.63, 375.74, 380.99, 387.37)}

\def\corrtwosigdelaysxsoneMillonsplchabriermedian{382.43 (367.43, 376.45, 382.53, 389.14, 450.11)}

\def\corrtwosigmustwosoneRPthreerdchabriermedian{1.021 (0.451, 0.910, 1.056, 1.308, 1.949)}

\def\corrtwosigmustwosoneThorpchabriermedian{1.057 (0.767, 0.944, 1.081, 1.296, 1.794)}

\def\corrtwosigmustwosoneKellychabriermedian{1.021 (0.618, 0.915, 1.048, 1.287, 1.921)}

\def\corrtwosigmustwosoneMillonsplchabriermedian{1.051 (0.667, 0.952, 1.079, 1.315, 1.885)}

\def\corrtwosigmusthreesoneRPthreerdchabriermedian{0.889 (0.281, 0.721, 0.933, 1.206, 2.008)}

\def\corrtwosigmusthreesoneThorpchabriermedian{0.919 (0.506, 0.803, 0.977, 1.285, 2.025)}

\def\corrtwosigmusthreesoneKellychabriermedian{0.907 (0.357, 0.799, 0.969, 1.250, 2.021)}

\def\corrtwosigmusthreesoneMillonsplchabriermedian{0.931 (0.396, 0.827, 0.996, 1.279, 2.044)}

\def\corrtwosigmusfoursoneRPthreerdchabriermedian{0.278 (0.176, 0.247, 0.285, 0.345, 0.495)}

\def\corrtwosigmusfoursoneThorpchabriermedian{0.357 (0.239, 0.298, 0.356, 0.420, 0.525)}

\def\corrtwosigmusfoursoneKellychabriermedian{0.296 (0.199, 0.268, 0.303, 0.363, 0.529)}

\def\corrtwosigmusfoursoneMillonsplchabriermedian{0.333 (0.248, 0.310, 0.344, 0.397, 0.538)}

\def\corrtwosigmusxsoneRPthreerdchabriermedian{0.275 (0.198, 0.253, 0.289, 0.343, 0.447)}

\def\corrtwosigmusxsoneThorpchabriermedian{0.311 (0.203, 0.264, 0.311, 0.362, 0.455)}

\def\corrtwosigmusxsoneKellychabriermedian{0.287 (0.211, 0.264, 0.297, 0.350, 0.459)}

\def\corrtwosigmusxsoneMillonsplchabriermedian{0.305 (0.194, 0.274, 0.309, 0.355, 0.437)}

\def\corrpmdelaysxsoneCombinedchabriermedian{$376.0_{-5.5}^{+5.6}$ days}

\def\corrpmmusxsoneCombinedchabriermedian{$0.30_{-0.03}^{+0.05}$}

\def\rmsstwosonedelayRPthreerdmediansalpeter{9.0}
\def\outlierfractionstwosonedelayRPthreerdmediansalpeter{0.055}

\def\rmsstwosonedelayThorpmediansalpeter{6.7}
\def\outlierfractionstwosonedelayThorpmediansalpeter{0.045}

\def\rmsstwosonedelayKellymediansalpeter{7.3}
\def\outlierfractionstwosonedelayKellymediansalpeter{0.068}

\def\rmsstwosonedelayMillonsplmediansalpeter{7.6}
\def\outlierfractionstwosonedelayMillonsplmediansalpeter{0.087}

\def\rmssthreesonedelayRPthreerdmediansalpeter{7.7}
\def\outlierfractionsthreesonedelayRPthreerdmediansalpeter{0.011}

\def\rmssthreesonedelayThorpmediansalpeter{6.0}
\def\outlierfractionsthreesonedelayThorpmediansalpeter{0.015}

\def\rmssthreesonedelayKellymediansalpeter{5.9}
\def\outlierfractionsthreesonedelayKellymediansalpeter{0.027}

\def\rmssthreesonedelayMillonsplmediansalpeter{6.1}
\def\outlierfractionsthreesonedelayMillonsplmediansalpeter{0.030}

\def\rmssfoursonedelayRPthreerdmediansalpeter{12.3}
\def\outlierfractionsfoursonedelayRPthreerdmediansalpeter{0.024}

\def\rmssfoursonedelayThorpmediansalpeter{11.5}
\def\outlierfractionsfoursonedelayThorpmediansalpeter{0.048}

\def\rmssfoursonedelayKellymediansalpeter{7.8}
\def\outlierfractionsfoursonedelayKellymediansalpeter{0.042}

\def\rmssfoursonedelayMillonsplmediansalpeter{9.6}
\def\outlierfractionsfoursonedelayMillonsplmediansalpeter{0.060}

\def\rmssxsonedelayRPthreerdmediansalpeter{9.2}
\def\outlierfractionsxsonedelayRPthreerdmediansalpeter{0.016}

\def\rmssxsonedelayThorpmediansalpeter{8.5}
\def\outlierfractionsxsonedelayThorpmediansalpeter{0.011}

\def\rmssxsonedelayKellymediansalpeter{6.0}
\def\outlierfractionsxsonedelayKellymediansalpeter{0.014}

\def\rmssxsonedelayMillonsplmediansalpeter{8.7}
\def\outlierfractionsxsonedelayMillonsplmediansalpeter{0.078}

\def\rmsstwosonemuRPthreerdmediansalpeter{0.40}
\def\outlierfractionstwosonemuRPthreerdmediansalpeter{0.014}

\def\rmsstwosonemuThorpmediansalpeter{0.32}
\def\outlierfractionstwosonemuThorpmediansalpeter{0.010}

\def\rmsstwosonemuKellymediansalpeter{0.38}
\def\outlierfractionstwosonemuKellymediansalpeter{0.012}

\def\rmsstwosonemuMillonsplmediansalpeter{0.35}
\def\outlierfractionstwosonemuMillonsplmediansalpeter{0.012}

\def\rmssthreesonemuRPthreerdmediansalpeter{0.37}
\def\outlierfractionsthreesonemuRPthreerdmediansalpeter{0.026}

\def\rmssthreesonemuThorpmediansalpeter{0.34}
\def\outlierfractionsthreesonemuThorpmediansalpeter{0.027}

\def\rmssthreesonemuKellymediansalpeter{0.36}
\def\outlierfractionsthreesonemuKellymediansalpeter{0.032}

\def\rmssthreesonemuMillonsplmediansalpeter{0.36}
\def\outlierfractionsthreesonemuMillonsplmediansalpeter{0.031}

\def\rmssfoursonemuRPthreerdmediansalpeter{0.11}
\def\outlierfractionsfoursonemuRPthreerdmediansalpeter{0.009}

\def\rmssfoursonemuThorpmediansalpeter{0.11}
\def\outlierfractionsfoursonemuThorpmediansalpeter{0.010}

\def\rmssfoursonemuKellymediansalpeter{0.11}
\def\outlierfractionsfoursonemuKellymediansalpeter{0.011}

\def\rmssfoursonemuMillonsplmediansalpeter{0.09}
\def\outlierfractionsfoursonemuMillonsplmediansalpeter{0.011}

\def\rmssxsonemuRPthreerdmediansalpeter{0.08}
\def\outlierfractionsxsonemuRPthreerdmediansalpeter{0.018}

\def\rmssxsonemuThorpmediansalpeter{0.10}
\def\outlierfractionsxsonemuThorpmediansalpeter{0.015}

\def\rmssxsonemuKellymediansalpeter{0.09}
\def\outlierfractionsxsonemuKellymediansalpeter{0.014}

\def\rmssxsonemuMillonsplmediansalpeter{0.09}
\def\outlierfractionsxsonemuMillonsplmediansalpeter{0.016}

\def\corrtwosigdelaystwosoneRPthreerdchabriermedian{4.28 (-13.97, -3.10, 3.23, 8.74, 17.61)}
\def\corrtwosigdelaystwosoneThorpchabriermedian{11.09 (-0.80, 7.95, 11.61, 15.83, 23.48)}
\def\corrtwosigdelaystwosoneKellychabriermedian{8.49 (-6.67, 5.49, 8.57, 11.80, 17.75)}
\def\corrtwosigdelaystwosoneMillonsplchabriermedian{12.89 (-9.96, 9.60, 12.80, 15.93, 20.95)}
\def\corrtwosigdelaysthreesoneRPthreerdchabriermedian{12.93 (-0.85, 7.23, 13.52, 19.58, 28.07)}
\def\corrtwosigdelaysthreesoneThorpchabriermedian{11.13 (2.24, 8.04, 11.74, 16.25, 23.06)}
\def\corrtwosigdelaysthreesoneKellychabriermedian{7.52 (-0.07, 4.12, 7.46, 10.72, 17.02)}
\def\corrtwosigdelaysthreesoneMillonsplchabriermedian{11.73 (2.99, 8.15, 11.65, 14.77, 19.83)}
\def\corrtwosigdelaysfoursoneRPthreerdchabriermedian{16.84 (-15.09, 5.24, 16.27, 27.06, 36.89)}
\def\corrtwosigdelaysfoursoneThorpchabriermedian{55.48 (38.12, 49.51, 58.52, 70.77, 85.43)}
\def\corrtwosigdelaysfoursoneKellychabriermedian{19.24 (4.92, 14.19, 19.29, 24.36, 31.75)}
\def\corrtwosigdelaysfoursoneMillonsplchabriermedian{31.25 (0.65, 24.37, 31.22, 37.83, 44.79)}
\def\corrtwosigdelaysxsoneRPthreerdchabriermedian{380.83 (360.53, 370.68, 379.58, 386.68, 394.58)}
\def\corrtwosigdelaysxsoneThorpchabriermedian{383.83 (367.04, 375.96, 383.62, 390.23, 400.15)}
\def\corrtwosigdelaysxsoneKellychabriermedian{376.02 (365.31, 370.63, 375.74, 380.99, 387.37)}
\def\corrtwosigdelaysxsoneMillonsplchabriermedian{382.43 (367.43, 376.45, 382.53, 389.14, 450.11)}
\def\corrtwosigmustwosoneRPthreerdchabriermedian{1.021 (0.451, 0.910, 1.056, 1.308, 1.949)}
\def\corrtwosigmustwosoneThorpchabriermedian{1.057 (0.767, 0.944, 1.081, 1.296, 1.794)}
\def\corrtwosigmustwosoneKellychabriermedian{1.021 (0.618, 0.915, 1.048, 1.287, 1.921)}
\def\corrtwosigmustwosoneMillonsplchabriermedian{1.051 (0.667, 0.952, 1.079, 1.315, 1.885)}
\def\corrtwosigmusthreesoneRPthreerdchabriermedian{0.889 (0.281, 0.721, 0.933, 1.206, 2.008)}
\def\corrtwosigmusthreesoneThorpchabriermedian{0.919 (0.506, 0.803, 0.977, 1.285, 2.025)}
\def\corrtwosigmusthreesoneKellychabriermedian{0.907 (0.357, 0.799, 0.969, 1.250, 2.021)}
\def\corrtwosigmusthreesoneMillonsplchabriermedian{0.931 (0.396, 0.827, 0.996, 1.279, 2.044)}
\def\corrtwosigmusfoursoneRPthreerdchabriermedian{0.278 (0.176, 0.247, 0.285, 0.345, 0.495)}
\def\corrtwosigmusfoursoneThorpchabriermedian{0.357 (0.239, 0.298, 0.356, 0.420, 0.525)}
\def\corrtwosigmusfoursoneKellychabriermedian{0.296 (0.199, 0.268, 0.303, 0.363, 0.529)}
\def\corrtwosigmusfoursoneMillonsplchabriermedian{0.333 (0.248, 0.310, 0.344, 0.397, 0.538)}
\def\corrtwosigmusxsoneRPthreerdchabriermedian{0.275 (0.198, 0.253, 0.289, 0.343, 0.447)}
\def\corrtwosigmusxsoneThorpchabriermedian{0.311 (0.203, 0.264, 0.311, 0.362, 0.455)}
\def\corrtwosigmusxsoneKellychabriermedian{0.287 (0.211, 0.264, 0.297, 0.350, 0.459)}
\def\corrtwosigmusxsoneMillonsplchabriermedian{0.305 (0.194, 0.274, 0.309, 0.355, 0.437)}
\def\corrtwosigdelaystwosoneCombinedchabriermedian{9.69 (-8.97, 5.96, 9.92, 14.06, 20.56)}
\def\corrtwosigdelaysthreesoneCombinedchabriermedian{7.92 (0.47, 5.16, 8.95, 13.56, 20.19)}
\def\corrtwosigdelaysfoursoneCombinedchabriermedian{19.44 (5.94, 14.56, 20.28, 27.45, 43.15)}
\def\corrtwosigdelaysxsoneCombinedchabriermedian{376.02 (365.12, 370.50, 376.03, 381.65, 390.23)}
\def\corrtwosigmustwosoneCombinedchabriermedian{1.06 (0.63, 0.94, 1.08, 1.30, 1.78)}
\def\corrtwosigmusthreesoneCombinedchabriermedian{0.91 (0.50, 0.80, 0.97, 1.30, 2.07)}
\def\corrtwosigmusfoursoneCombinedchabriermedian{0.28 (0.18, 0.26, 0.30, 0.37, 0.52)}
\def\corrtwosigmusxsoneCombinedchabriermedian{0.30 (0.22, 0.27, 0.31, 0.35, 0.45)}

\def\corrtwosigrefereereportgrillodelaysxsoneCombinedchabriermedian{376.62 (365.18, 370.72, 376.14, 381.67, 389.91)}

\def\corrtwosigrefereereportsharondelaysxsoneCombinedchabriermedian{376.42 (364.71, 370.97, 376.55, 383.12, 393.22)}

\def\corrtwosigrefereereportogurioheightdelaysxsoneCombinedchabriermedian{375.82 (363.57, 370.39, 376.19, 382.29, 392.17)}

\def\corrtwosigrefereereportoguridoublekappasdelaysxsoneCombinedchabriermedian{376.22 (363.17, 370.17, 376.22, 382.03, 391.19)}

\newcommand{\UCLA}{Department of Physics and Astronomy, University of California, Los Angeles, CA 90095}
\newcommand{\USC}{Department of Physics and Astronomy, University of South Carolina, 712 Main St., Columbia, SC 29208, USA}

\newcommand{\TokyoPhys}{Department of Physics, University of Tokyo, 7-3-1 Hongo, Bunkyo-ku, Tokyo 113-0033, Japan}
\newcommand{\IPMU}{Kavli Institute for the Physics and Mathematics of the Universe (Kavli IPMU, WPI), University of Tokyo, 5-1-5 Kashiwanoha, Kashiwa, Chiba 277-8583, Japan}

\newcommand{\EarlyTokyo}{Research Center for the Early Universe, University of Tokyo, Tokyo 113-0033, Japan}
\newcommand{\DARK}{Dark Cosmology Centre, Niels Bohr Institute, University of Copenhagen, Juliane Maries Vej 30, DK-2100 Copenhagen, Denmark}

\newcommand{\Berkeley}{Department of Astronomy, University of California, Berkeley, CA 94720-3411, USA}
\newcommand{\STScI}{Space Telescope Science Institute, 3700 San Martin Dr., Baltimore, MD 21218, USA}

\newcommand{\UCSB}{Department of Physics, University of California, Santa Barbara, CA 93106-9530, USA}

\newcommand{\KICPStanford}{Kavli Institute for Particle Astrophysics and Cosmology, Stanford University, 452 Lomita Mall, Stanford, CA 94305, USA}

\newcommand{\AMNH}{Department of Astrophysics, American Museum of Natural History, Central Park West and 79th Street, New York, NY 10024, USA}

\newcommand{\Rutgers}{Department of Physics and Astronomy, Rutgers, The State University of New Jersey, Piscataway, NJ 08854, USA}

\newcommand{\LCO}{Las Cumbres Observatory, 6740 Cortona Dr., Suite 102, Goleta, CA 93117, USA}

\newcommand{\Cantabria}{IFCA, Instituto de F\'isica de Cantabria (UC-CSIC), Av. de Los Castros s/n, 39005 Santander, Spain}

\newcommand{\IKERBASQUE}{Ikerbasque Foundation, University of the Basque Country, DIPC Donostia,Spain}

\newcommand{\UCSC}{Department of Astronomy and Astrophysics, UCO/Lick Observatory, University of California, 1156 High Street, Santa Cruz, CA 95064, USA}

\newcommand{\DurhamExtra}{Centre for Extragalactic Astronomy, Department of Physics, Durham University, Durham DH1 3LE, U.K.}
\newcommand{\DurhamCosmo}{Institute for Computational Cosmology, Durham University, South Road, Durham DH1 3LE, U.K.}
\newcommand{\Durban}{Astrophysics and Cosmology Research Unit, School of Mathematical Sciences, University of KwaZulu-Natal, Durban 4041, South Africa}
\newcommand{\EPFL}{Laboratoire d'Astrophysique, Ecole Polytechnique Federale de Lausanne (EPFL), Observatoire de Sauverny, CH-1290 Versoix, Switzerland}

\newcommand{\AIP}{Leibniz-Institut fur Astrophysik Potsdam (AIP), An der Sternwarte 16, 14482 Potsdam, Germany}
\newcommand{\UMich}{University of Michigan, Department of Astronomy, 1085 South University Avenue, Ann Arbor, MI 48109-1107, USA}
\newcommand{\UMN}{Minnesota Institute for Astrophysics, University of Minnesota, 116 Church Street SE, Minneapolis, MN 55455, USA}

\newcommand{\BenGurion}{Physics Department, Ben-Gurion University of the Negev, P.O. Box 653, Beer-Sheva 8410501, Israel}

\newcommand{\Miller}{Miller Institute for Basic Research in Science, University of California, Berkeley, CA 94720, USA}

\newcommand{\Stromlo}{The Research School of Astronomy and Astrophysics, Mount Stromlo Observatory, Australian National University, Canberra, ACT 2611, Australia}
\newcommand{\Excellence}{National Centre for the Public Awareness of Science, Australian National University, Canberra, ACT 2611, Australia}
\newcommand{\ASTROThreeD}{The ARC Centre of Excellence for All-Sky Astrophysics in 3 Dimension (ASTRO 3D), Australia}
\newcommand{\Toronto}{Department of Astronomy and Astrophysics, University of Toronto, 50 St. George Street, Toronto, ON, M5S 3H4, Canada}
\newcommand{\Cambridge}{Institute of Astronomy and Kavli Institute for Cosmology, Madingley Road, Cambridge, CB3 0HA, UK}
\newcommand{\Portsmouth}{Institute of Cosmology and Gravitation, University of Portsmouth, Portsmouth, PO1 3FX, UK}
\newcommand{\StatLabCambridge}{Statistical Laboratory, DPMMS, University of Cambridge, Wilberforce Road, Cambridge, CB3 0WB, UK}
\newcommand{\Sorbonne}{Institut d'Astrophysique de Paris, CNRS-Sorbonne Universit\'e, 98 bis boulevard Arago, F-75014 Paris, France}
\newcommand{\ARI}{Astronomisches Rechen-Institut (ARI), Zentrum für Astronomie der Universität Heidelberg (ZAH),  Mönchhofstr. 12-14, 69120 Heidelberg, Germany}
\newcommand{\ISSI}{International Space Science Institute (ISSI), Hallerstrasse 6, 3012 Bern, Switzerland}
\newcommand{\StonyBrook}{Department of Physics and Astronomy, Stony Brook University, Stony Brook, NY 11794, USA}
\newcommand{\Carnegie}{The Observatories of the Carnegie Institution for Science, 813 SantaBarbara Street, Pasadena, CA 91101, USA}

\begin{document}

\shorttitle{The Most Delayed Image of Supernova Refsdal}

\title{The Magnificent Five Images of Supernova Refsdal: Time Delay and Magnification Measurements}

\author{Patrick L. Kelly}
\affiliation{\UMN}
\author{Steven Rodney}
\affiliation{\USC}
\author{Tommaso Treu}
\affiliation{\UCLA}
\author{Simon Birrer}
\affiliation{\KICPStanford}
\author{Vivien Bonvin}
\affiliation{\EPFL}
\author{Luc Dessart}
\affiliation{\Sorbonne}
\author{Ryan J. Foley}
\affiliation{\UCSC}
\author{Alexei V. Filippenko}
\affiliation{\Berkeley}
\affiliation{\Miller}
\author{Daniel Gilman}
\affiliation{\Toronto}
\affiliation{\UCLA}
\author{Saurabh Jha}
\affiliation{\Rutgers}
\author{Jens Hjorth}
\affiliation{\DARK}
\author{Kaisey Mandel}
\affiliation{\Cambridge}
\affiliation{\StatLabCambridge}
\author{Martin Millon}
\affiliation{\EPFL}
\author{Justin Pierel}
\affiliation{\USC}
\author{Stephen Thorp}
\affiliation{\Cambridge}
\author{Adi Zitrin}
\affiliation{\BenGurion}
\author{Tom Broadhurst}
\affiliation{\IKERBASQUE}
\author{Wenlei Chen}
\affiliation{\UMN}
\author{Jose M. Diego}
\affiliation{\Cantabria}
\author{Alan Dressler}
\affiliation{\Carnegie}
\author{Or Graur}
\affiliation{\Portsmouth}
\affiliation{\AMNH}
\author{Mathilde Jauzac}
\affiliation{\DurhamExtra}
\affiliation{\DurhamCosmo}
\affiliation{\Durban}
\author{Matthew A. Malkan}
\affiliation{\UCLA}
\author{Curtis McCully}
\affiliation{\LCO}
\affiliation{\UCSB}
\author{Masamune Oguri}
\affiliation{\EarlyTokyo}
\affiliation{\TokyoPhys}
\affiliation{\IPMU}
\author{Marc Postman}
\affiliation{\STScI}
\author{Kasper Borello Schmidt}
\affiliation{\AIP}
\author{Keren Sharon}
\affiliation{\UMich}
\author{Brad E. Tucker}
\affiliation{\Stromlo}
\affiliation{\Excellence}
\affiliation{\ASTROThreeD}
\author{Anja von der Linden}
\affiliation{\StonyBrook}
\author{Joachim Wambsganss}
\affiliation{\ARI}
\affiliation{\ISSI}

\begin{abstract} 
In late 2014, four images of Supernova (SN) ``Refsdal,'' the first known example of a strongly lensed SN with multiple resolved images, were detected in the MACS\,J1149 galaxy-cluster field. Following the images' discovery, the SN was predicted to reappear within hundreds of days at a new position $\sim$8$''$ away in the field. The observed reappearance in late 2015 makes it possible to carry out Refsdal's (\citeyear{refsdal64}) original proposal to use a multiply imaged SN to measure the Hubble constant H$_0$, since the time delay between appearances should vary inversely with H$_0$. Moreover, the position, brightness, and timing of the reappearance enable a novel test of the blind predictions of galaxy-cluster models, which are typically constrained only by the positions of multiply-imaged galaxies. We have developed a new photometry pipeline that uses {\tt DOLPHOT} to measure the fluxes of the five images of SN Refsdal from difference images. We apply four separate techniques to perform a blind measurement of the relative time delays and magnification ratios 
between the last image SX and the earlier images S1--S4.
We measure the relative time delay of SX--S1 to be \corrpmdelaysxsoneCombinedchabriermedian\ and the relative magnification to be \corrpmmusxsoneCombinedchabriermedian. This corresponds to a 1.5\% precision on the time delay and 17\% precision for the magnification ratios, and includes uncertainties due to millilensing and microlensing.  
In an  accompanying paper, we place initial and blind constraints on the value of the Hubble constant. 
\vspace{3mm}
\end{abstract}

\section{Introduction}
By deflecting the trajectories of photons and increasing their travel time, a sufficiently massive object can produce multiple, magnified images of a well-aligned background object.
\citet{einstein36} first described the possibility of a strong gravitational lens system, envisioning a hypothetical foreground star that creates multiple images of a second, more distant star.
\citet{zwicky37a,zwicky37b} subsequently conjectured that
external galaxies (``extragalactic nebulae'') and galaxy clusters 
should magnify and produce multiple images of background galaxies far beyond the Milky Way.  

Even before the first multiply-imaged source had been discovered, 
\citet{refsdal64} explored the idea that a supernova (SN) 
could explode in a strongly lensed galaxy. 
The light curves of the SN observed in each of its multiple images
could be used to measure the relative time delays among the images.
Sjur Refsdal's analysis demonstrated that these relative time delays could be used
to infer the Hubble constant (H$_0$) given an accurate model for the mass distribution of the lens.

\begin{figure*}
\fig{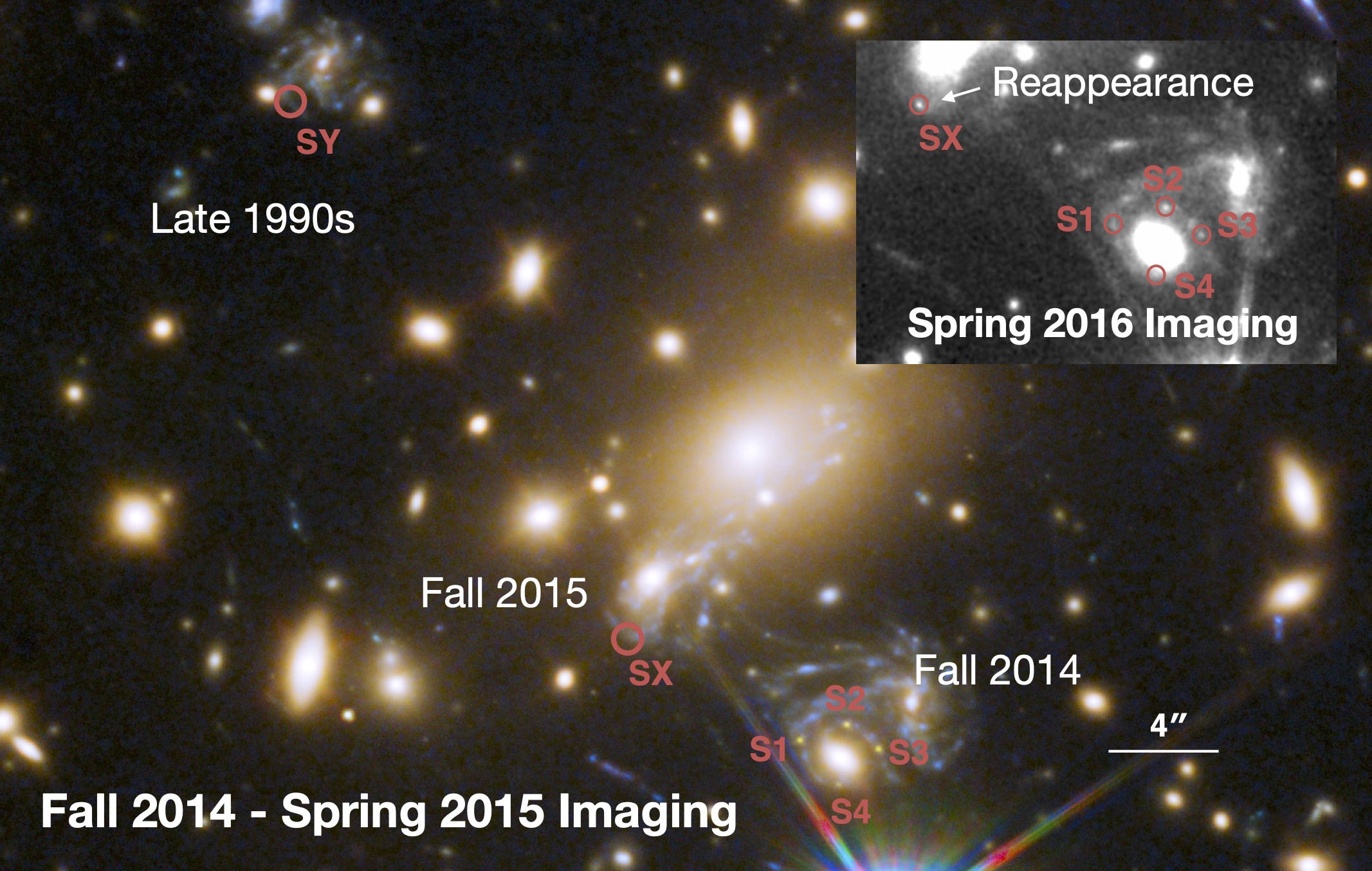}{\textwidth}{}
\caption{Image of the MACS\,J1149 galaxy-cluster field and the locations and timing of the appearances of SN Refsdal.
A first appearance occurred according to model predictions in the host-galaxy image at the upper left in the late 1990s,
but was missed.  We detected images S1--S4 in an Einstein-cross configuration in November 2014 \citep{kellyrodneytreu15}.
Following this appearance, lens modelers predicted the SN's reappearance which was detected in late 2015 \citep{kellyrodneytreu16}.
Inset at upper right shows a coadded WFC3 IR F125W image of the field following the reappearance of the SN in image SX. }
\label{fig:mosaic}
\end{figure*}

Despite the scientific potential of such a strongly lensed SN with multiple resolved images, none was identified in the fifty years after Refsdal's paper was published.
In {\it Hubble Space Telescope (HST)} images taken on 10 November 2014 (UT dates are used throughout this paper),
\citet{kellyrodneytreu15} identified the first example of such a strongly lensed SN with multiple resolved images. This event, dubbed {\it SN Refsdal}, appeared behind the  \macs1149\ cluster \citep{ebelingedgehenry01,ebelingbarrettdonovan07} in imaging collected for the 
Grism Lens-Amplified Survey from Space (GLASS; GO-13459; PI Treu; 
\citealt{schmidttreubrammer14}; \citealt{treuschmidtbrammer15}).
When SN Refsdal was detected in 2014, it appeared as four images (named S1--S4) in an Einstein-cross configuration 
around an early-type galaxy in the foreground, as shown in Figure~\ref{fig:mosaic}.
Lens models predicted that it should subsequently reappear on a timescale of hundreds of days 
in a trailing image (named SX) of its host galaxy that forms close to the cluster's center \citep{kellyrodneytreu15,oguri15,sharonjohnson15, diegobroadhurstchen16,grillokarmansuyu16,jauzacrichardlimousin16, kawamataoguriishigaki16,treubrammerdiego16}. 
As demonstrated by \citet{rodneystrolgerkelly16} in their analysis of the light curves of the Einstein-cross images, SN Refsdal is the first example of a strongly lensed SN where relative time delays 
can be measured. When the SN reappeared in {\it HST} imaging taken on 11 December 2015 (\citealt{kellyrodneytreu16}; GO-14199; PI Kelly),
SN Refsdal became the first (and only, to date) SN whose appearance on the sky was predicted in advance.
The MACS\,J1149 galaxy cluster produces a third, complete image of SN Refsdal's host galaxy,
and lens models indicate that an image of the SN appeared within that galaxy image in the late 1990s.
   
The first multiply-imaged SN turned out to be even more scientifically useful than originally envisioned by Refsdal (see \citealt{oguri19} for a review of applications of strongly lensed transients). From the very first observations of the SN's reappearance, \citet{kellyrodneydiego16} found approximate and preliminary constraints on image SX's relative time delay and magnification. Using a Bayesian framework and available model predictions, \citet{vegaferrerodiegomiranda18} obtained a $\sim$30\% constraint on the value of H$_0$ from the multiply imaged SN across a 68\% confidence interval.
In addition to its potential to measure the Hubble constant, the SN can be used to test the accuracy of lens models on the scale of a galaxy cluster \citep{treubrammerdiego16}. 
Furthermore, the monitoring of SN Refsdal has led to the characterization of a rare core-collapse explosion of a blue supergiant star at redshift $z=1.49$ \citep{kellybrammerselsing16} and to the serendipitous discovery of an individual star in SN Refsdal's host galaxy extremely magnified by microlensing \citep{kellydiegorodney18}. 

In this paper, we present a blinded analysis for measurement of the time delay and relative magnifications between SX and S1--S4. In an accompanying paper \citep{papertwo}, we then use these high-precision measurements to obtain an estimate of H$_0$ from SN Refsdal using pre-reappearance lens models updated using only the reappearance's position.  In  the accompanying paper, we also use the measurements presented here to evaluate MACS J1149 cluster lens models.

This paper is organized as follows.
Section~\ref{sec:observations} summarizes the observations and 
previous analyses of SN Refsdal. 
The methods used in data processing and
photometry are presented in Section~\ref{sec:methods}.
In Section~\ref{sec:micromillilensing}, we include a description of our simulations of microlensing and millilensing that affect images S1--SX.
Section~\ref{sec:fitting} describes the process of fitting the SN Refsdal light curves
to determine time delays and relative magnifications. 
Our measurements of the time delays and relative magnifications are presented in Section~\ref{sec:td_mr}.
We summarize the results and discuss future work in Section~\ref{sec:Summary}.
Throughout this paper, we use AB magnitudes \citep{okegunn83}.

\section{Follow-up Observations and Early Analysis}
\label{sec:observations}

Following the discovery of the four images S1--S4 of SN Refsdal in GLASS imaging \citep{kellyrodneytreu15},
the {\it Hubble} Frontier Fields (HFF) program \citep[DD-13504; PI Lotz;][]{lotzkoekemoercoe17}
fortuitously began its already scheduled imaging campaign of the MACS\,J1149 galaxy-cluster field.  
To complement the early HFF imaging, we acquired additional imaging as
part of the FrontierSN program (GO-13386; PI Rodney) to characterize 
better the spectral energy distribution (SED) of the SN from photometry of images S1--S4.
These early measurements of the SN color and light curve left open the 
possibility that the SN could be a Type Ia SN whose luminosity could be calibrated within $\sim$10\%, 
and we obtained thirty orbits of WFC3-IR G141 grism spectroscopy
to classify spectroscopically the SN before it 
reached maximum brightness as part of a 
Director's Discretionary Time program (DD-14041; PI Kelly; \citealt{kellybrammerselsing16}).
Independent analysis of the SN's explosion site using MUSE spectroscopy showed that it erupted in a highly ionized region with
low metallicity \citep{yuankobayashikewley15, karmangrillobalestra16}.

Three of the images of SN Refsdal that form the Einstein cross (S1--S3) are magnified by factors of up to $\sim15$--22,
while the fourth (S4) was predicted to have a magnification of $\sim1$--4 \citep{kellyrodneytreu15,oguri15,sharonjohnson15, diegobroadhurstchen16,grillokarmansuyu16,jauzacrichardlimousin16, kawamataoguriishigaki16,treubrammerdiego16}.
By fitting photometry of imaging acquired from November 2014 through December 2015, 
\citet{rodneystrolgerkelly16} measured the time delays and magnification ratios 
of the four images S1--S4 that form the Einstein cross. Those measurements used both polynomials and templates
of SN 1987A-like supernovae (SNe), and compared the resulting measurements to lens model predictions.

To detect the reappearance of SN Refsdal in image SX, as shown in Figure~\ref{fig:mosaic}, and 
measure its time delay and magnification relative to S1--S4, we executed a multicycle 
``Refsdal Redux'' monitoring campaign  (GO-14199; PI Kelly).  
This program continued to 
monitor the MACS\,J1149 field at a cadence of approximately every two weeks whenever the field 
was visible to {\it HST} (from approximately late October 
through July).  In May 2016 these observations 
captured a microlensing peak of a
highly magnified blue supergiant star (nicknamed {\it Icarus}) in SN Refsdal's host galaxy 
\citep{kellydiegorodney18}. This prompted additional {\it HST} follow-up observations 
of the MACS\,J1149 field (DD-14528; DD-14872; PI Kelly).

\begin{deluxetable*}{llll}
\label{tab:irdata}
  \tablecolumns{2}
  \tablecaption{\sc HST WFC3-IR Imaging Data}
  \tablehead{  \multicolumn{1}{c}{Dates}  &
    \multicolumn{1}{c}{Short Program Name} &
    \multicolumn{1}{c}{HST Program ID$+$PI} &
    \multicolumn{1}{c}{Reference} }
\startdata
2010-12-05..2011-03-10 & CLASH & GO-12068; M. Postman & \citealt{postmancoebenitez12}\\
2013-11-02..2015-01-06 &  HFF & DD/GO-13504; J. Lotz & \citealt{lotzkoekemoercoe17}\\
2014-11-03..2014-11-11 &  GLASS & GO-13459; T. Treu & \citealt{treuschmidtbrammer15}\\
2014-11-30..2015-07-21 &  FrontierSN & GO-13790; S. Rodney & \citealt{rodneypatelscolnic15}\\
2014-12-23..2015-01-05 & Refsdal-DD & DD-14041; P. Kelly & \citealt{kellyrodneytreu15}\\
2015-10-30..2016-10-30 & Refsdal Redux & GO-14199; P. Kelly & \citealt{kellyrodneytreu16}\\
2016-07-11..2016-07-23 & Icarus-DD1 & DD-14528; P. Kelly & \citealt{kellydiegorodney18}\\
2016-11-28..2017-01-04 & Icarus-DD2 & DD-14872; P. Kelly  & \citealt{kellydiegorodney18}\\
\enddata
\end{deluxetable*}

Table~\ref{tab:irdata} provides a summary of all these observation programs.  
As many of these programs were designed to help measure the relative time delays and magnifications of the images of SN Refsdal, 
a large fraction of the observing time 
used a consistent instrumental setup. The {\it HST} Wide Field Camera 3 (WFC3) infrared (IR) detector was employed, using filters F125W ({\it J}) and F160W ({\it H}), which correspond to approximately the rest-frame {\it V} and {\it R} bands. This filter choice facilitates comparison with
observations of nearby SNe.  

\section{Image Processing and Photometry}
\label{sec:methods}

In this Section we describe the processing of the WFC3-IR imaging, 
an approach to subtract stellar diffraction spikes, and photometry of the five SN Refsdal images. 

\subsection{Astrometry and Alignment}
We processed all {\it HST} images using the {\tt DrizzlePac}
software suite.\footnote{\url{http://drizzlepac.stsci.edu}}
All images were first aligned to a common world coordinate system (WCS) using 
{\tt TweakReg}. We adopted the WCS used for the CLASH, GLASS, and HFF images and catalogs,\footnote{\url{http://www.stsci.edu/hst/campaigns/frontier-fields/}} which is also the WCS generally used by published lens models for this cluster.  We then coadded and resampled each image set to a 
scale of 0.03\arcsec~pix$^{-1}$ using {\tt AstroDrizzle} \citep{fruchter10}.

\subsection{Subtraction of Stellar Diffraction Spikes}
In the WFC3-IR imaging of the MACS\,J1149 field, a bright foreground star ($J =14.0$\,mag Vega) offset by $\sim6''$ from 
images S1--S4 produces four bright diffraction spikes, two of which can be seen near the bottom of Figure~\ref{fig:mosaic}.  
We requested telescope roll angles for which the stellar diffraction spikes fell
away from the SN images. However, for a significant fraction of observations no such angle was available.

The diffraction spikes require additional effort to obtain a measurement of the SN flux when they overlap the SN image, but we have found that it is possible to remove the diffraction spikes almost entirely.  We performed all photometry on ``difference images" which were constructed by subtracting a pre-SN ``template" image from the coadded images of each visit containing the SNe.  To create a template image without diffraction spikes, 
we constructed masks that enclose pixels contaminated by the diffraction spikes. 
We next constructed template images by combining all WFC3-IR images acquired before 30 October 2014 
(preceding the first detection of the SN), and applying the pixel masks. 
The images that we use to construct the template include data acquired at 
multiple telescope roll angles, so the coadded template images have no regions that are completely masked.

As shown by \citet{rodneystrolgerkelly16}, the similarity among the star's four diffraction spikes
and rotation of the diffraction spikes makes it possible to remove effectively if not entirely 
diffracted light at SN positions. 
We first coadd the single exposures
taken during each visit and then subtract the template image from the resulting coadded image. 
As described above, the template images we created lack diffraction spikes. By subtracting the template from each each single-epoch image, we removed the light from all galaxies and the sky background. 
Consequently, the difference images only show light collected during the follow-up epoch, and only from two sources: the multiply-imaged SN and the diffraction spikes.

We next rotate the difference images by 90\degree,
after setting all regions outside of those contaminated by the diffraction spikes to equal
zero. Given the similarity among the four diffraction spikes, each rotated spike  provides an excellent model
for the adjacent spikes.  
After subtracting this masked and rotated image of the spikes, we are left with a difference image that has an increase in noise in a narrow region along the spike, but without substantial contamination.

\subsection{Positions of the SN Images}
In Table~\ref{tab:positions}, we list the coordinates of images S1--SX measured in the CLASH astrometric system from coadditions of WFC3-IR F125W imaging acquired in 2015 and 2016. 

\begin{deluxetable}{ccccc}
  \tablecolumns{5}
  \tablecaption{\sc J2000 Positions of the Five Detected Images of SN Refsdal \label{tab:positions}}
  \tablehead{
    \colhead{Image} & \colhead{R.A.} & \colhead{Decl.}}
  \startdata
  S1 & 11$^{\rm h}$49$^{\rm m}$35.575$^{\rm s}$ & 22$^{\circ}$23$'$44.25$''$ \\
  S2 & 11$^{\rm h}$49$^{\rm m}$35.454$^{\rm s}$ & 22$^{\circ}$23$'$44.82$''$ \\
  S3 & 11$^{\rm h}$49$^{\rm m}$35.370$^{\rm s}$ & 22$^{\circ}$23$'$43.92$''$ \\
  S4 & 11$^{\rm h}$49$^{\rm m}$35.474$^{\rm s}$ & 22$^{\circ}$23$'$42.63$''$ \\
  SX & \rasx & \decsx \\
  \enddata
\end{deluxetable}


\begin{deluxetable}{ccccc}
\tablecaption{Photometry of SN Refsdal. 
\label{tab:photometry}
}
\tablehead{\colhead{Obj.} & \colhead{Filt.} & \colhead{MJD} & \colhead{Flux} & \colhead{Flux Err.}}
\startdata
S1 & F125W & 57004.89 & 0.240 & 0.012 \\
S1 & F125W & 57006.34 & 0.237 & 0.009\\
S1 & F125W & 57007.1 & 0.239 & 0.011 \\
S2 & F125W & 57013.746 & 0.200 & 0.011 \\
S2 & F125W & 57015.152 & 0.221 & 0.009 \\
S2 & F125W & 57015.96 & 0.218 & 0.010 \\
S3 & F125W & 56957.17 & 0.187 & 0.011 \\
S3 & F125W & 56958.68 & 0.206 & 0.010 \\
S3 & F125W & 56959.38 & 0.197 & 0.010 \\
S4 & F125W & 56994.383 & 0.042 & 0.010 \\
S4 & F125W & 56995.83 & 0.061 & 0.009 \\
S4 & F125W & 56996.59 & 0.059 & 0.011 \\
SX & F125W & 57004.527 & 0.009 & 0.010 \\
SX & F125W & 57006.027 & -0.004 & 0.008 \\
SX & F125W & 57006.74 & -0.002 & 0.009 \\
\enddata
\tablecomments{ A portion of the table is shown here. The full table is available online.}
\end{deluxetable}

\subsection{Photometry with {\tt DOLPHOT}} 
Fitting a model of the point-spread function (PSF) to an unresolved point source such as S1--SX should, in principle, yield a
measurement of the source's flux with an optimal signal-to-noise ratio (S/N). For a point source, pixels {\it closest} to the source's center have the greatest S/N.  When fitting a PSF
model to the data, one can assign greater weights to these central
pixels.  
By contrast, when measuring instead an aperture magnitude, the flux within a fixed radius is summed and an aperture correction is applied, and {\it all} pixels in the aperture are assigned effectively the same weight.

We have adapted the {\tt DOLPHOT} \citep{dolphin00} PSF-fitting package to be able to 
measure photometry from WFC3 IR difference imaging.
When applying {\tt DOLPHOT}, we use the default sky-fitting setting (FitSky=1) and model images S1--SX as point sources (Force1=1). 
When fitting images S1--SX, we instruct {\tt DOLPHOT} to assign additional weight to central pixels (PSFPhot=2).
Finally, we enforce the additional constraint that the flux of S1--SX acquired does not change between exposures taken within each imaging epoch, except for variation due to noise (ForceSameMag=0).

We estimate the uncertainty of the measured fluxes by using {\tt DOLPHOT} to inject fake 
point sources into individual exposures. We next measure the fluxes of the fake sources,
and estimate their uncertainty by computing the standard deviation of measured fluxes.

For the purpose of comparison, we use {\tt PythonPhot}\footnote{\url{https://github.com/djones1040/PythonPhot}}
\citep{jonesscolnicrodney15} to measure the light curves of images S1--SX of SN Refsdal
from coadded imaging. The {\tt PythonPhot} package includes an implementation of PSF-fitting photometry measurement
based on the {\tt DAOPHOT} package
\citep{stetson87}.  We have modified the code to inject fake stars at locations at which the
local background is similar to the locations of images S1--SX.

The measured photometry for all five images is shown in 
Figure~\ref{fig:each_image_light_curves}, along with a piecewise polynomial model fit (which will be described in Sec.~\ref{sec:fiducial_model}).  A full table of the measured photometry is available in the online version of the journal.  Table~\ref{tab:photometry} provides an excerpt. 

\subsection{Light Curves of Stellar Sources in Cluster Field}
To assess the performance of the photometry pipelines, we selected a set of stars from 
a coadded image of the field acquired with the ACS WFC F606W broadband filter
and required that the stars' F125W and F160W fluxes were similar to those of
S1--SX over the evolution of their observed light curves. We next ran {\tt DOLPHOT} in an image section around the location of
each star and determined the star's position in each image.
Using these positions, we then fit for the position, proper motion, and annual 
parallax. Finally, we ran {\tt DOLPHOT} and {\tt PythonPhot} on each of the comparison
stars, and calculated the standard deviation and reduced $\chi$-squared statistics of the flux measurements.

We estimate the photometric uncertainties by injecting 100  
fake point sources using {\tt DOLPHOT} and {\tt PythonPhot}. 
In the case of {\tt DOLPHOT}, we select areas near each star
without significant contamination. 
In the Appendix Figures~\ref{fig:stellarphot1}--\ref{fig:stellarphot5}, we plot the photometry of the stars, showing a
comparison between the results of the photometry codes.  From the {\tt DOLPHOT} tests we measure a reduced $\chi^2$ of 0.59$-$1.82 with a median of 1.06,  which affirms the accuracy of our uncertainty estimates for our {\tt DOLPHOT} photometry of the SN images. 
These test cases show that the {\tt DOLPHOT} photometry provides more reliable photometry and uncertainty estimates.  We therefore adopt the {\tt DOLPHOT} photometry for all SN images in the following analysis.

\subsection{The Fiducial Light-Curve Model}
\label{sec:fiducial_model}

To estimate the uncertainties and to evaluate and combine estimates of the time-delay fitting methods that we will apply to the SN Refsdal data (Sec.~\ref{sec:fitting}), we create simulated photometric data that approximate the real light-curve data.  To achieve this, we adopt a piecewise polynomial model as our ``fiducial'' light-curve model. 

Spectroscopic follow-up observations and the peculiar light curve of SN Refsdal have shown that it was a SN 1987A-like 
SN with a peak {\it B}-band luminosity
inferred to be a factor of 5--10 brighter than that of SN 1987A, given magnification favored by current cluster models \citep{kellybrammerselsing16}. 
As shown in Figure~\ref{fig:each_image_light_curves}, a piecewise polynomial function with separate, low-order polynomial functions for the photospheric and nebular phases can describe the light curves of both SN Refsdal and nearby SN 1987A-like SNe whose light curves have broadly similar shapes.
Indeed, \citet{rodneystrolgerkelly16} found that a low-order polynomial could describe the SN Refsdal data taken through the photospheric phase.

Table~\ref{tab:modelparams} describes fit statistics, including Akaike Information Criterion \citep[AIC;][]{aka74} values, for possible low-order models for each phase. While a fourth-order or fifth-order model for the photospheric phase, and a first-order model for the nebular phase, are preferred according to the AIC, we select a third-order model for the photospheric phase and a second-order model for the nebular phase. We choose the lower-order model for the photospheric phase, because time-variable microlensing magnification may introduce structure into the observed light curves that we do not want to include as part of the fiducial model. Moreover, as Figure~\ref{fig:each_image_light_curves} demonstrates, nearby SN 1987A-like SN light curves (which lack microlensing) in similar rest-frame bands are well-described by polynomials with this order. We select the higher, second-order model for the nebular phase, given that its evolution with time is expected to follow the non-linear, exponential rate of energy deposition from the decay of radioactive $^{56}$Co.

\begin{figure}[tbp]
\centering
\fig{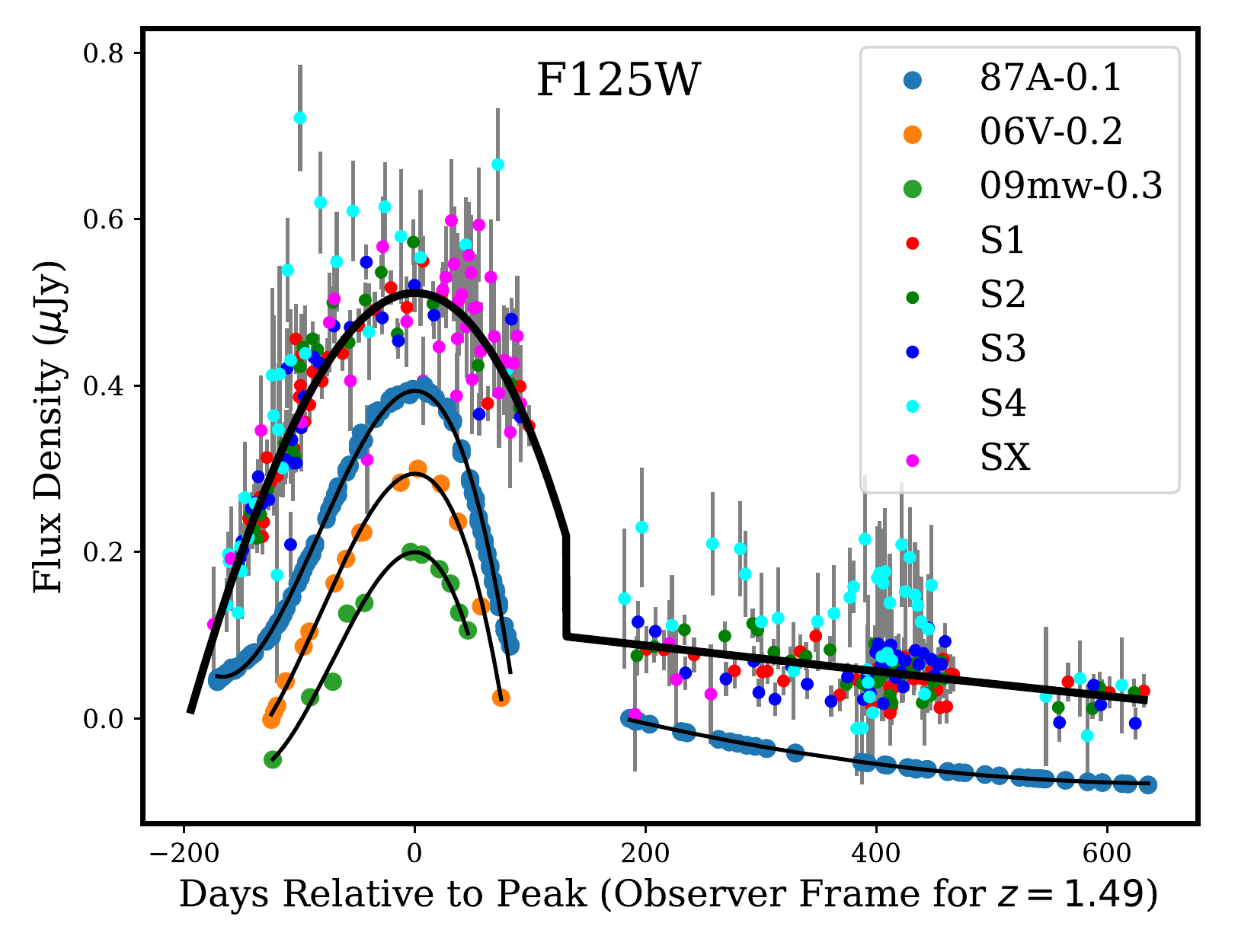}{0.49\textwidth}{}
\fig{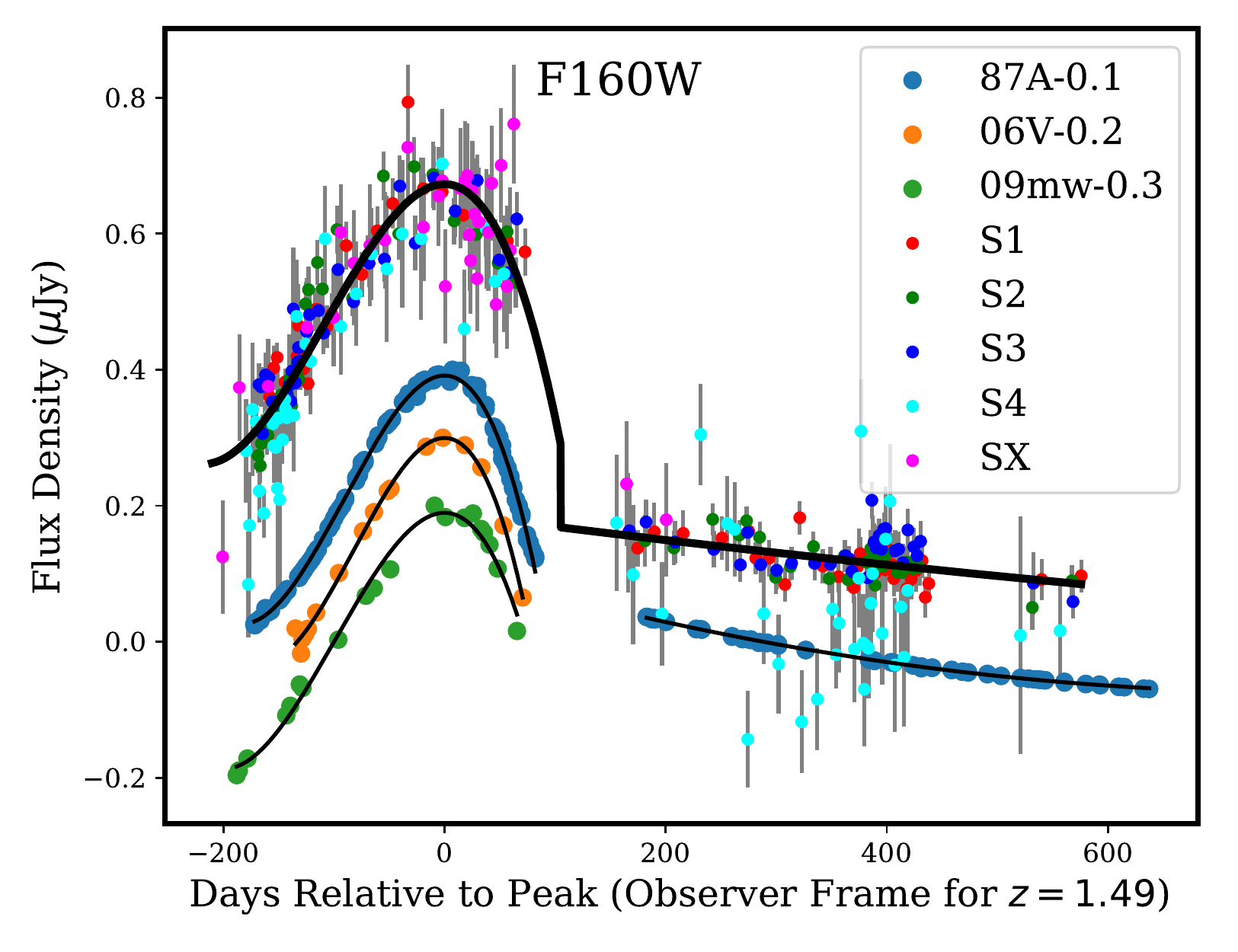}{0.49\textwidth}{}
\caption{The observer-frame ($z=1.49$) broadband light curves at 5000--7000\,\AA\ of both SN Refsdal and those of low-redshift SN 1987A-like SNe with a similar shape stretched in time by a factor of 2.49. All light curves are well fit by a piecewise polynomial function.
In the fits shown above, the photospheric phase light curve is modeled using a third-order polynomial and the nebular-phase light curve with a second-order polynomal. The uppermost points and best-fitting model are of SN Refsdal's F125W and F160W combined light curves constructed by shifting and rescaling the measured photometry of its five appearances according to the best-fitting time delays and magnifications relative to that of S1.
The F125W and F160W bandpasses correspond approximately to rest-frame {\it V} and {\it R} bandpasses at $z=1.49$, and we plot in respective panels above the measured {\it V} and {\it R} for the low-redshift examples of SN 1987A-like SNe. For visual clarity we rescale and then subtract constant flux values from the light curves of the nearby SNe. }
\label{fig:combined_light_curves}
\end{figure}

\section{Microlensing and Millilensing Simulations}
\label{sec:micromillilensing}

With a model for the SN light curve, we next simulate the effects of microlensing and millilensing on the light curves of SN Refsdal. 
In order to include these effects, we also need 
an estimate of the convergence ($\kappa$) and shear ($\gamma$) at the positions of the SN images.  Furthermore, an estimate for the stellar mass density and the substructure fraction along the line of sight to each of those images are needed.  Here we explain how we infer these and use them to simulate the effects of microlensing and millilensing.

\subsection{Convergence and Shear}

\begin{deluxetable}{ccccccc}
\centering
  \tablecolumns{7}
  \tablecaption{\sc Lensing Parameters}
  \tablehead{  \multicolumn{1}{c}{Name} & \multicolumn{1}{c}{$\kappa$}  & \multicolumn{1}{c}{$\gamma$} & \multicolumn{1}{c}{$\mu$} & \multicolumn{1}{c}{Parity} &
  \multicolumn{1}{c}{$\mu_t$} &
  \multicolumn{1}{c}{$\mu_r$} 
    }
  \startdata
S1 & $ 0.722$ & $0.093$ & $ 14.6$ & + & 5.4 & 2.7 \\
S2 & $ 0.733$ & $0.361$ & $ 16.8$ & - & $-10.6$ & 1.5 \\
S3 & $ 0.675$ & $0.230$ & $ 19.0$ & + & 10.5 & 1.8 \\
S4 & $ 0.727$ & $0.466$ & $ 7.0$ & - & $-5.2$ & 1.4 \\
SX & $ 0.966$ & $0.488$ & $ 4.2$ & - & $-2.2$ & 1.9
\enddata
\label{tab:lensparams}
\tablecomments{We measure these parameters from the v3 {\tt GLAFIC} \citep{kawamataoguriishigaki16,oguri10} maps posted online at \url{https://archive.stsci.edu/pub/hlsp/frontier/macs1149/models/glafic/v3/}.}
\end{deluxetable}

We adopt the {\tt GLAFIC} \citep{oguri10} estimates for the lensing parameter from a model that was developed for the MACS\,J1149 field, and updated to predict SN Refsdal's reappearance \citep{kawamataoguriishigaki16}.  This model was among the most accurate in matching the observed time delays and magnifications for the images S1--S4 \citep{rodneystrolgerkelly16}. After unblinding, we found that it best matches the full set of observations of SN Refsdal in our accompanying paper \citep{papertwo}. Many models have been developed of the cluster lens, and could also have been used for this purpose.  
Table~\ref{tab:lensparams} reports the relevant lens parameters from this model, used in the subsequent principal set of  simulations.  To investigate the importance of these parameters, we repeat our entire analysis using alternative lens parameters listed in Table~\ref{tab:lensparamsmore}.

\subsection{Density of Stellar Mass Along Sight Lines}

The stars along the line of sight to each images of SN Refsdal comprise multiple components. These can include the stellar populations in the early-type galaxy lens responosible for the Einstein-cross images S1--S4, (2) the MACS\,J1149 brightest cluster galaxy (BCG), and (3) the intracluster light (ICL).  We model the BCG and ICL together as a single underlying stellar population with the same properties (age, initial mass function, etc.).  Here we describe our models for these stellar populations, and we then describe the ray-tracing
calculations that we use to compute simulated magnification maps in the source plane at $z=1.49$. 
Finally, we briefly describe our simulations of millilensing by expected dark-matter subhalos.

\subsubsection{Models of the Early-Type Lens, BCG, and ICL}

To model the light distribution of the early-type lens and the BCG$+$ICL, we first extracted a 33$''$  $\times$ 39$''$ (East-West $\times$ North-South, respectively) subsection of the HFF coadded images. Using SExtractor, we next identified a set of pixels associated with each detected object (a ``segmentation'' image). We assembled a  merged mask of all sources except for the BCG, and used {\tt GALFIT} \citep{penghoimpey02} to fit a S\'{e}rsic model centered on the BCG to the WFC3 F125W image subsection using the mask.  During the fit, the sky is fixed to be zero, since the HFF imaging background is already removed. 
\citet{morishitaabramsontreu17} found that the background estimate used to construct the HFF coadded images was accurate for the MACS\,J1149 field.

\begin{figure}[tbp]
\centering
\subfigure{ \includegraphics[angle=0,width=0.48\textwidth]{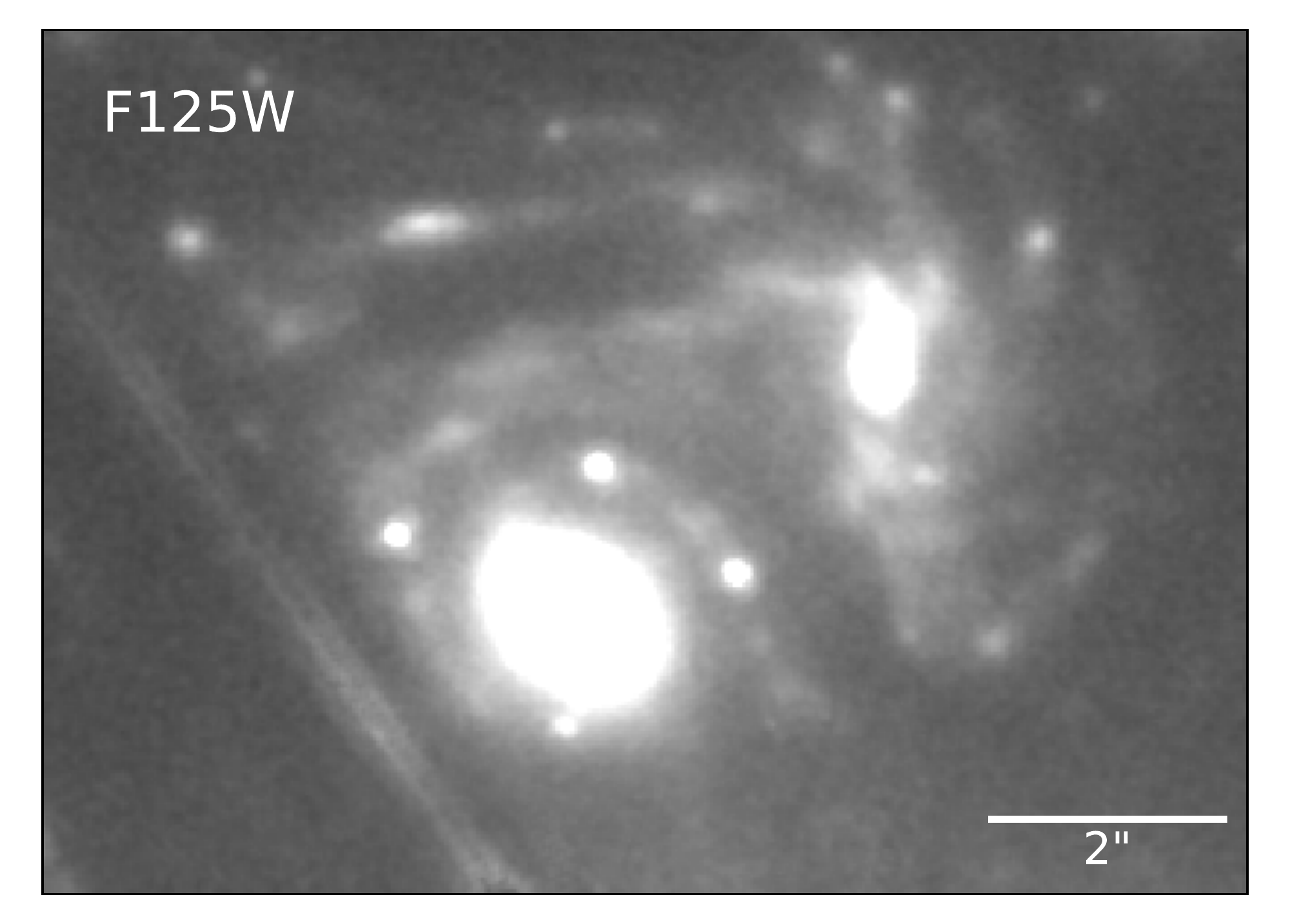} }
\subfigure{ \includegraphics[angle=0,width=0.48\textwidth]{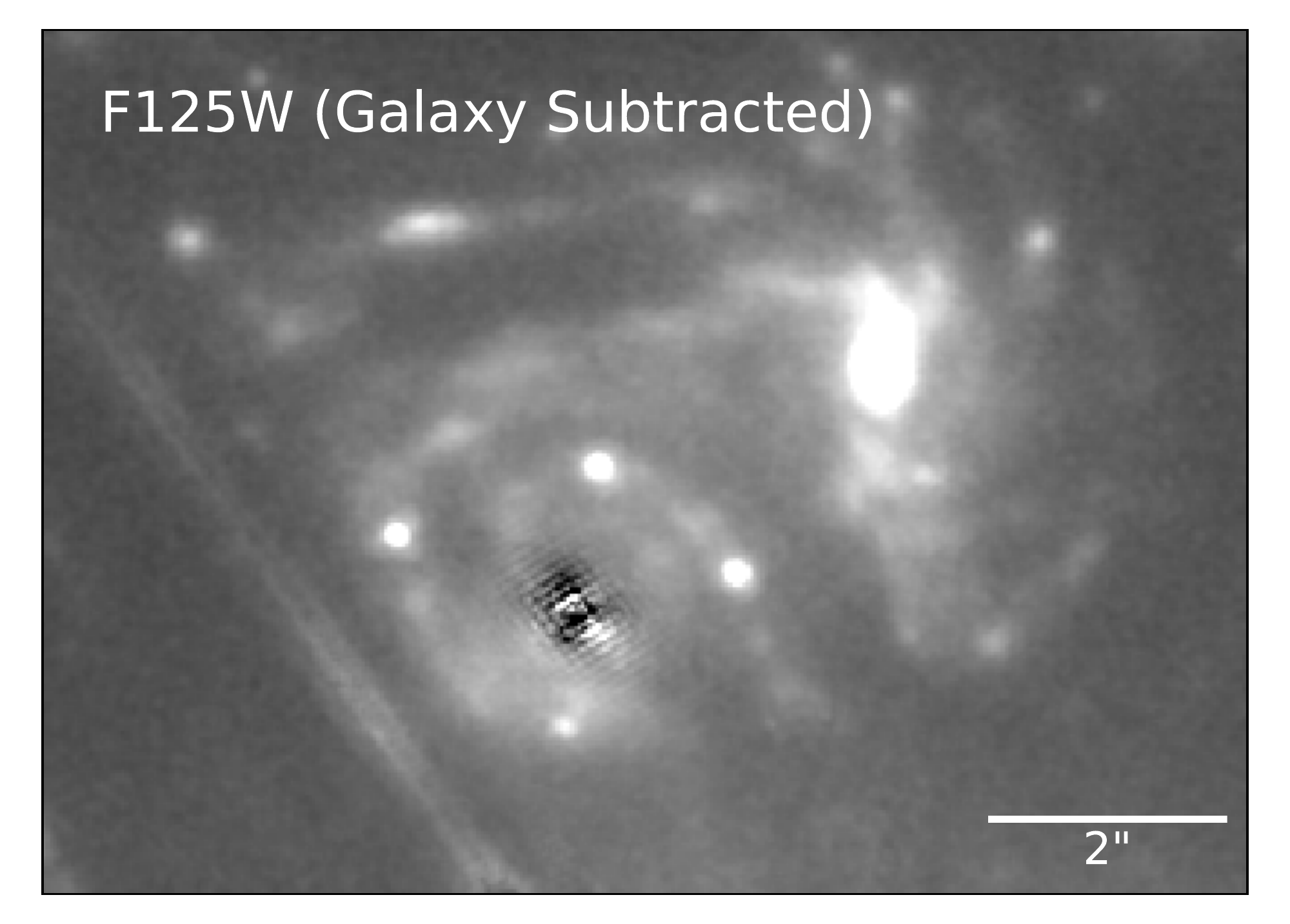} }
\caption{{\it HST} WFC3 F125W coadded image before and after best-fitting S\'{e}rsic model of the early-type galaxy lens ($z=0.54$) is subtracted. Upper panel shows image before subtraction of the early-type galaxy, and lower panel shows the result after  subtraction. In a previous and separate step, the cluster's BCG has already been modeled and subtracted to create the ``before'' image. The four point sources in the Einstein-cross configuration are images S1--S4 of SN Refsdal, and the underlying distorted spiral galaxy is the host galaxy of the SN ($z=1.49$).}
\label{fig:subtracted}
\end{figure}

The best-fitting S\'{e}rsic model for the single profile fit to the light distribution of the BCG, which transitions to that of the ICL, in the F125W bandpass has a S\'{e}rsic index of 4.86, an effective radius of 19.6$''$, and an F125W magnitude of 16.18\,AB.  Holding all parameters except for the total flux fixed, we next perform fits to subsections of the ACS WFC F435W, F606W, F814W, and WFC3 IR F105W, F125W, F140W, and F160W HFF coadditions. We subtract the best-fitting model for the BCG$+$ICL in each bandpass from the $33''\times 39''$ image subsection. The upper panel of Figure~\ref{fig:subtracted} shows the region around the SN images S1--S4 after subtracting this BCG$+$ICL model.  

The next step is to fit the early-type galaxy lens 
in the BCG$+$ICL-subtracted F125W image.  
We create a mask that excludes all pixels except those inside of an elliptical aperture with the same position angle, center, and ellipticity as the early-type galaxy. 
The aperture's semimajor axis is 1.8$''$.  Circular masks with radii of 
0.15$''$ are also placed over the locations of images of S1--S4 of SN Refsdal.   Our {\tt GALFIT} model consists of a S\'{e}rsic model for the early-type galaxy lens, and a background level that accounts for flux from the superposed lensed host galaxy. 

From the F125W image we find a S\'{e}rsic index of 2.22, an effective radius of 0.26$''$, and a total magnitude of F125W = 20.16\,AB for the early-type lens. The bottom panel of Figure~\ref{fig:subtracted} shows the S1--S4 region after this early-type galaxy has been subtracted.

\begin{figure}[tbp]
\centering
\includegraphics[width=0.49\textwidth]{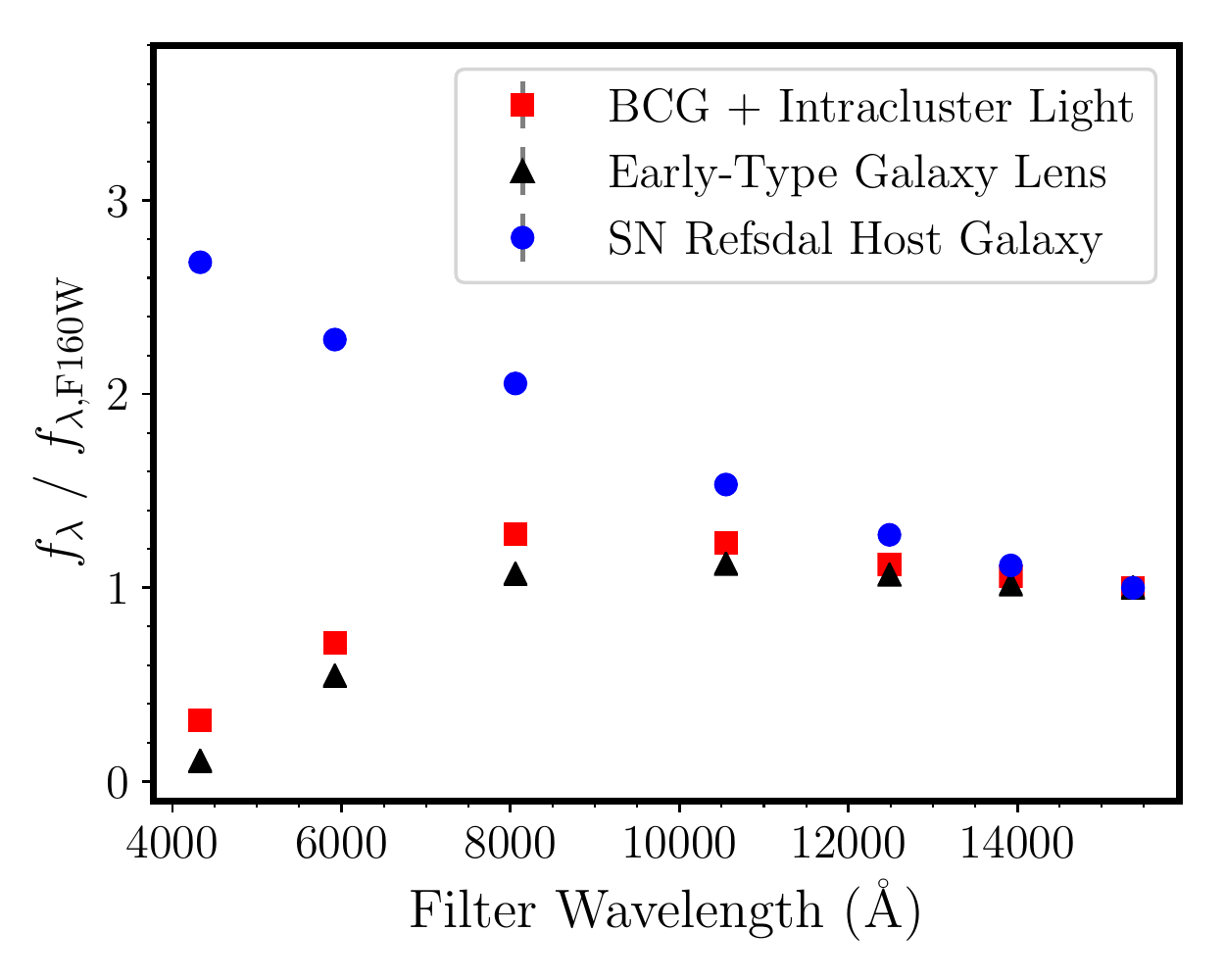}
\caption{Inferred SEDs of three separate underlying components at the positions of the images of SN Refsdal. Measured flux densities are shown relative to the F160W band.  The BCG$+$ICL are shown as red squares, the early-type galaxy that lenses the SN images S1--S4 are black triangles, and the background flux density that accounts for the SN Refsdal host galaxy are blue circles. 
All flux densities were measured using {\tt GALFIT} \citep{penghoimpey02} modeling, treating the BCG$+$ICL as a single S\'{e}rsic profile and the elliptical galaxy lens as a separate S\'{e}rsic profile. The light from the SN Refsdal host galaxy (the background spiral) is accounted for as a background flux level above the uniform sky brightness. Photometry is corrected for foreground Milky Way extinction using \citet{schlaflyfinkbeinerSFD11}. Measurement uncertainties are shown, but are smaller than the size of the points.  See the main text for further details of the fitting procedure. 
 }
\label{fig:seds}
\end{figure}

We next repeat the fit for the BCG$+$ICL-subtracted coadded image sections in all other available {\it HST} bandpasses from 4000\AA\ to $1.6 \mu$m,  holding all parameters fixed except for the galaxy's total flux and the local surface brightness.  
Figure~\ref{fig:seds} shows the resulting measured SEDs for the BCG$+$ICL, the early-type galaxy lens, and the host galaxy of SN Refsdal.  The inferred SED of the BCG$+$ICL matches that of the early-type galaxy. 

\subsubsection{SED Fitting to Estimate the Stellar Mass Density at the SN Image Positions}

\begin{deluxetable*}{cccccc}
\label{tab:starmassdens}
  \tablecolumns{6}
  \tablecaption{\sc Stellar and Total Mass Density Along Multiple Lines of Sight to SN Refsdal}
  \tablehead{  \multicolumn{1}{c|}{SN} & \multicolumn{3}{c|}{$\Sigma_\ast / (10^7\,{\rm M}_{\odot}\,{\rm kpc}^{-2})$ } & \multicolumn{1}{c|}{$\Sigma_{\rm tot} / (10^7\,{\rm M}_{\odot}\,{\rm kpc}^{-2})$} & \multicolumn{1}{c}{$\kappa_\star / \kappa_{\rm tot}$} \\
   \multicolumn{1}{c|}{image}  & \multicolumn{1}{c}{BCG$+$ICL} & \multicolumn{1}{c}{Early-Type Gal.} & \multicolumn{1}{c|}{Total}  & \multicolumn{1}{c|}{} & \multicolumn{1}{c}{}
    }
  \startdata
S1 & $ 0.434 \pm 0.005$ & $0.724 \pm 0.008$ & $1.158 \pm 0.009$ & $185.7 \pm 9.9$ & $0.0062 \pm 0.0003$ \\
S2 & $ 0.403 \pm 0.004$ & $0.484 \pm 0.005$ & $0.887 \pm 0.007$ & $188.3 \pm 11.4$ & $0.0047 \pm 0.0003$ \\
S3 & $ 0.297 \pm 0.003$ & $0.419 \pm 0.005$ & $0.716 \pm 0.006$ & $176.7 \pm 12.5$ & $0.0041 \pm 0.0003$ \\
S4 & $ 0.273 \pm 0.003$ & $1.520 \pm 0.017$ & $1.793 \pm 0.017$ & $190.4 \pm 11.4$ & $0.0094 \pm 0.0006$ \\
SX & $ 1.392 \pm 0.022$ & $0.000 \pm 0.000$ & $1.392 \pm 0.022$ & $231.5 \pm 13.4$ & $0.0060 \pm 0.0004$
\enddata
\end{deluxetable*}

We next want to model the measured SEDs to derive estimates for the physical properties of the BCG+ICL and the elliptical galaxy lens.
To that end, we correct the fluxes for Galactic extinction using \citet{schlaflyfinkbeinerSFD11}, 
and then use the Fitting and Assessment of Synthetic Templates (FAST) software tool \citep{kriekvandokkumlabbe09} and 
the \citet[][hereafter BC03]{bc03} stellar population model to model the SEDs shown in Figure~\ref{fig:seds}. 
Our fitting uses the BC03 isochrones, a \citet{cha03} initial-mass function (IMF), and an exponentially declining star-formation history; moreover, it assumes
no extinction due to dust associated with the BCG, ICL, or early-type lens.
The stars are set to have initial masses that are between 0.1\,M$_{\odot}$ and 100\,M$_{\odot}$, and we include models with metallicities $Z$ of 0.004, 0.008, 0.02, and 0.05. 
We measure $\log(M_{\odot} / f_{\rm F125W})$, where $ f_{\rm F125W}$ has units of $\mu$Jy, to be 9.01 and 9.15 dex for the combined BCG and ICL model fit and the early-type galaxy lens, respectively.  
Table~\ref{tab:starmassdens} reports the inferred stellar mass and total mass (i.e., including dark matter) along the line of sight to each of the five SN Refsdal images.

\subsection{Simulated Microlensing Magnification Maps}\label{sec:MicrolensingSimulations}

\begin{figure*}[tbp]
\centering 
\fig{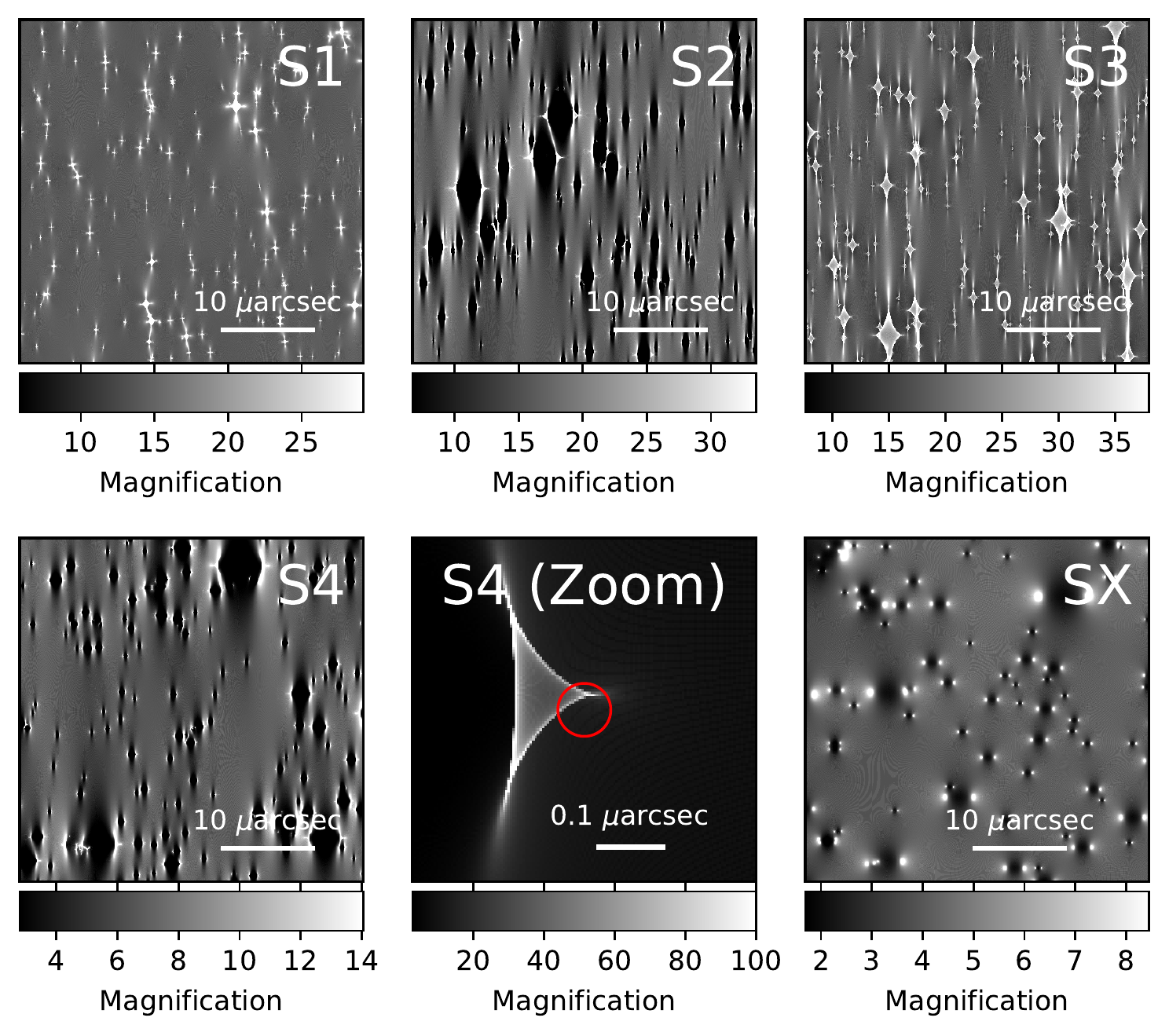}{\textwidth}{}
\caption{Simulated source-plane magnification maps for random realizations of stellar microlenses for SN Refsdal. Here we show simulations we have generated using the $\kappa$ and $\gamma$ values listed in Table~\ref{tab:lensparams} from the {\tt GLAFIC} v3 model \citep{kawamataoguriishigaki16,oguri10}. The red circle shows the size of a photosphere after 10 days of expansion with a velocity of 6000\,km\,s$^{-1}$. 
All panels (except for the zoomed-in panel) is 36 microarcseconds (0.31\,pc) on a side. The Einstein radius of a solar-mass star at the lens' redshift is 0.015\,pc.
 }
\label{fig:microlens}
\end{figure*}

\begin{figure}[tbp]
\centering
\fig{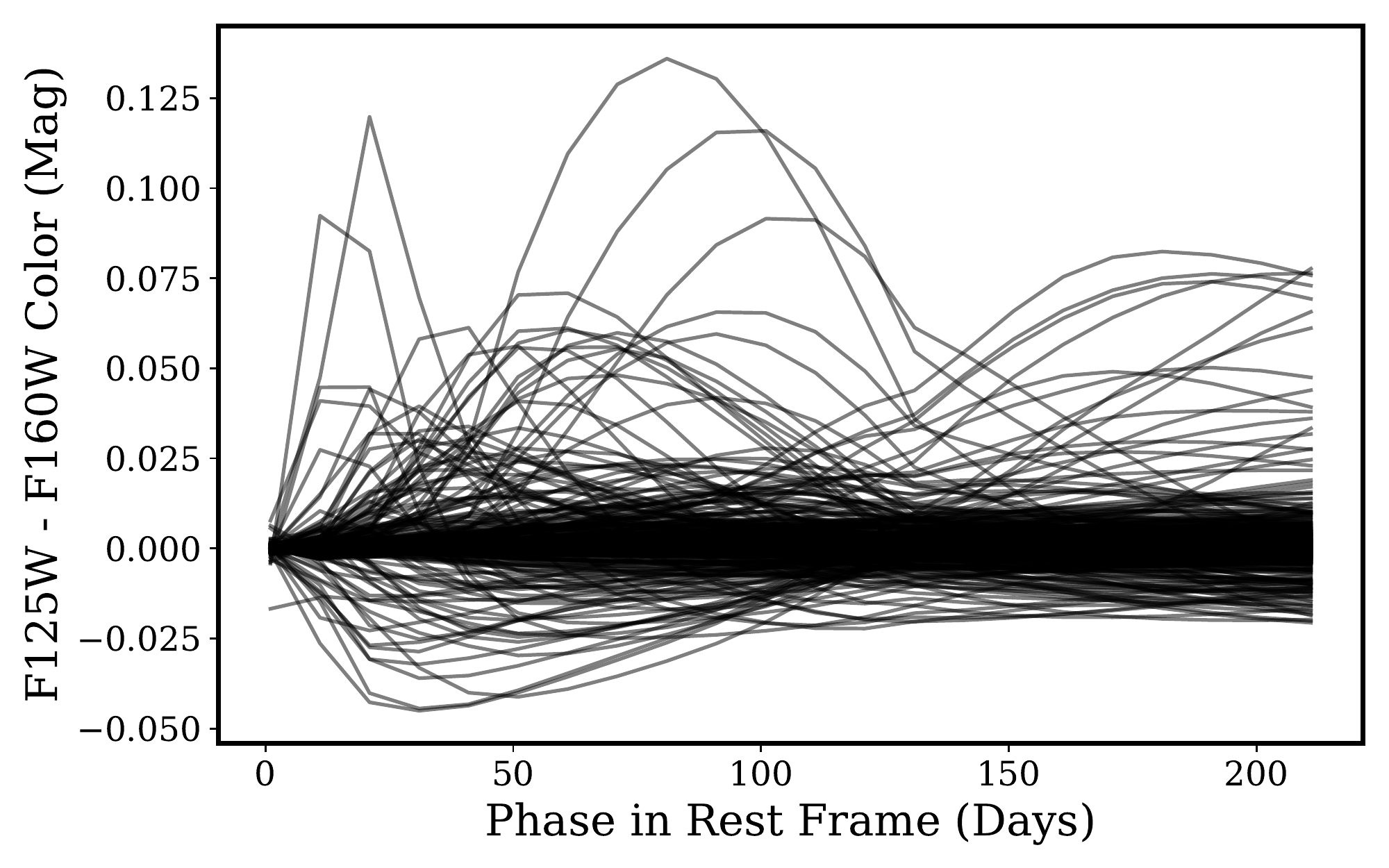}{0.49\textwidth}{}
\caption{Effect of microlensing on the apparent F125W--F160W color ($-2.5 \times \log_10 ({\rm F125W} / {\rm F160W})$) of SN Refsdal from simulations, as a function of rest-frame time since explosion.   Shown are 1000 random realizations of a SN 1987A-like photosphere expanding linearly in a stellar microlensing field.  These curves demonstrate that microlensing magnification will affect the total F125W and F160W flux from the photosphere almost equally, resulting in color changes of $<$0.05 mag.}
\label{fig:ml_achromatic}
\end{figure}

\begin{figure}[tbp]
\centering
\fig{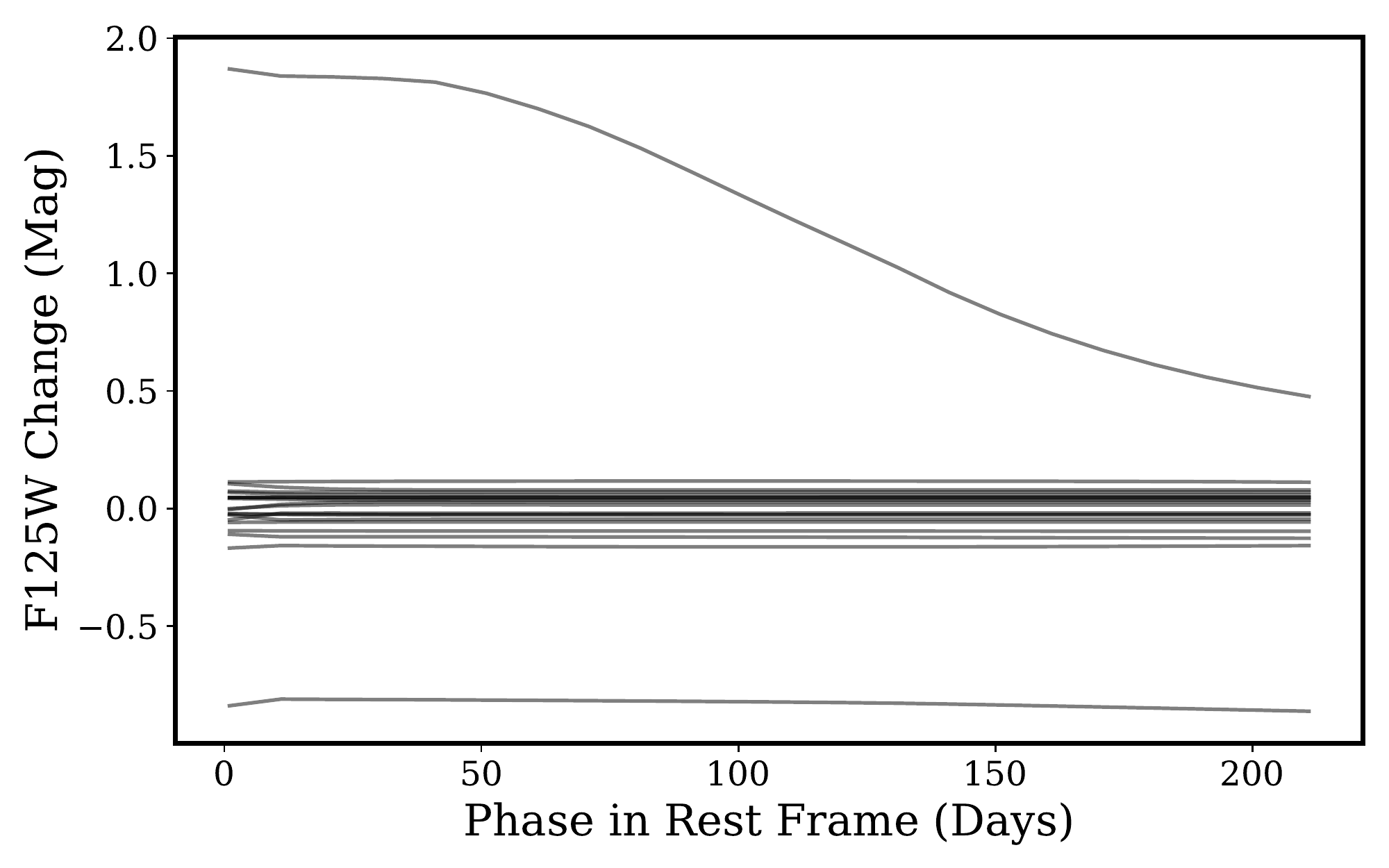}{0.49\textwidth}{}
\caption{Simulated magnification of SN photosphere in the F125W bandpass due to microlensing as a function of SN phase. Here we show light curves corresponding to different random configurations of masses and positions of foreground microlenses.  These simulations demonstrate that, in almost all cases, expected microlensing will not appreciably affect the shape of the light curve. }
\label{fig:ml_shape}
\end{figure}

\begin{figure*}[tbp]
\centering
\fig{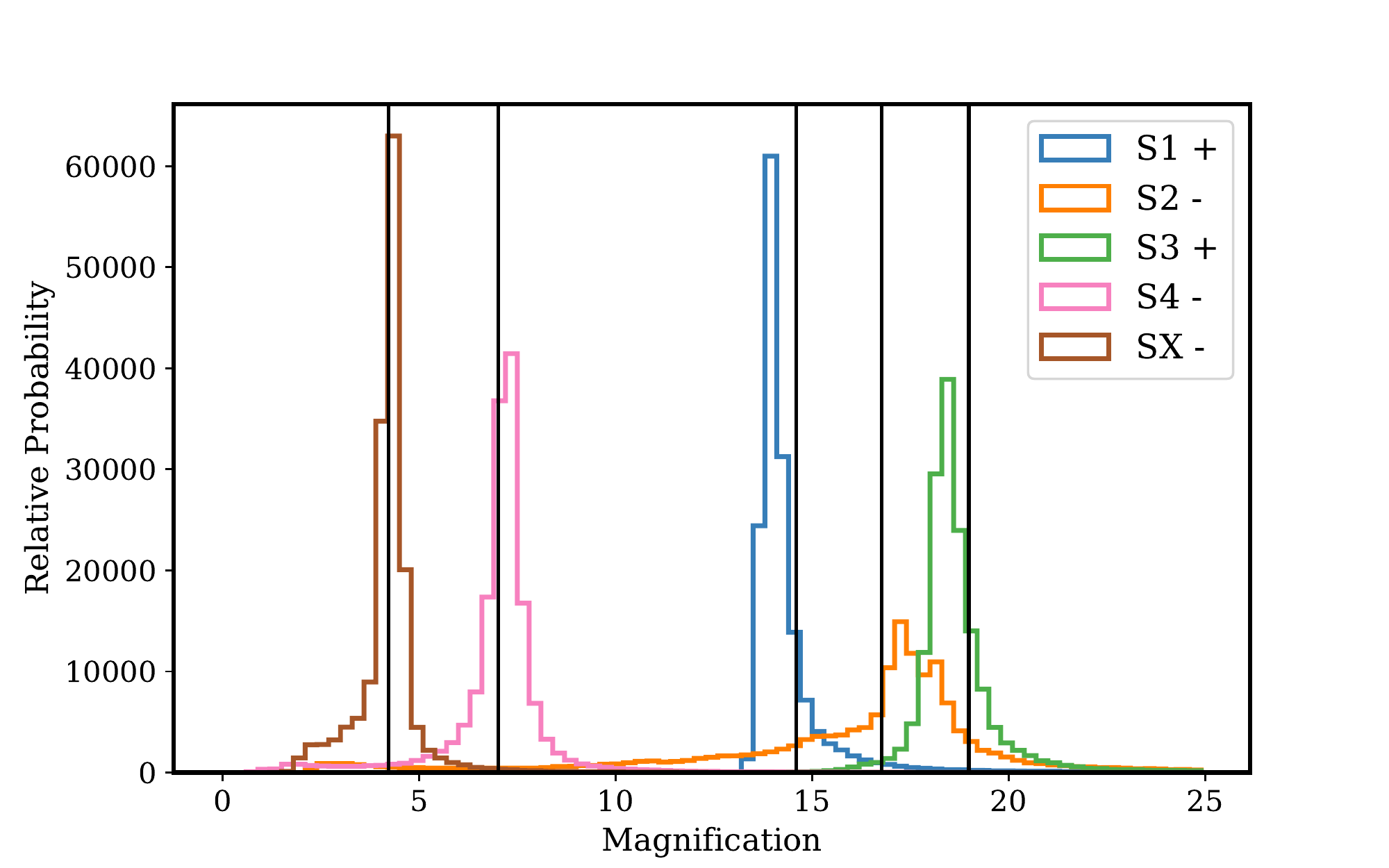}{\textwidth}{}
\caption{ Distributions of total magnification of images S1--S4 and SX of SN Refsdal from microlensing simulations. The shape of the distributions depends on the parity of the SN image, the shear $\gamma$ and convergence $\kappa$ of the local lensing potential, and the mass density of the stellar microlenses listed in Tables~\ref{tab:starmassdens} and \ref{tab:lensparams}. The respective vertical lines show the magnification for the parameters in Table \ref{tab:lensparams} that we use to carry out our simulation. The negative-parity images (S2, S4, and SX) have significant tails extending to low magnification. We have convolved the simulated magnification maps with a circular kernel with a radius of 0.0017\,pc equal to that expected for a photosphere with a velocity of 6000 km\,s$^{-1}$ that has expanded for 100 days in its rest frame.
}  
\label{fig:microhist}
\end{figure*}

Using the estimated shear and convergence along the lines of sight
to each image, we generated caustic magnification patterns in the source plane with an adapted version of the 
{\tt MULE} inverse ray-tracing code \citep{doblerfassnachttreu15}.
However, we obtain similar results instead also using the {\tt microlens} code \citep{wambsganss90,wambsganss99}.
We next use a spherically symmetric model for SN 1987A \citep{dessarthillier19} at phases of 10.9, 21.3, 31.1, 41.5, 50.1, 66.8, 80.8, 97.8, 117.0, 177.0, 220.0, and 300.0 days to 
generate the microlensing simulations. 
We rescale the radius of the photosphere models linearly with time and interpolate between the rescaled model to approximate the effect of the expansion of the SN during its photospheric phase.

In our simulation, each pixel has a length of 0.002 Einstein radii for a solar-mass lens.
For the photospheric expansion velocity of $-8356 \pm 1105$\,km\,s$^{-1}$
at a phase of $-47 \pm 8$ days relative to maximum brightness
measured by \citet{kellybrammerselsing16}, the SN photosphere would
expand in radius by a single pixel each 1.25 days in the rest frame at $z=1.49$. 
Each ray-tracing simulation creates a map of the magnification across a grid of $1024 \times 1024$ pixels, and
we place the simulated SN photosphere at the center of the pixel grid. 
Figure~\ref{fig:microlens} plots example source-plane magnification maps at the positions of
S1--SX. 
The fact that we use spherically symmetric explosion model 
may present a limitation of the 
simulation. A more sophisticated treatment would use a two-dimensional or even three-dimensional explosion model \citep[e.g.,][]{goldsteinnugentprecisetimedelays18,bonvintihhonovamillon19}.
To approximate an asymmetric explosion, we perturb the 
circular isophotes of the symmetric model to become elliptical with a ratio of the major axes of $b/a = 0.8$. The isophotal brightness of each elliptical isophote matches that of the circular isophote whose diameter is equal to the length of the ellipse's major axis. As shown in Table~\ref{tab:sxdelayvariants}, the elliptical photosphere yields almost identical constraints on the SX--S1 time delay.

In Figure~\ref{fig:ml_achromatic}, we show the simulated impact of microlensing on the F125W--F160W color.  This  
indicates that stellar microlensing of the photosphere of a SN 1987A-like SN affects its 
apparent F125W and F160W magnitudes equally within $\lesssim0.05$\,mag.
Furthermore, Figure~\ref{fig:ml_shape} shows that microlensing is not expected to strongly affect the 
shape of the light curve of SN 1987A-like SNe during the photospheric phase. 
Microlensing can also impact the overall magnification of each of the images, causing either increases or decreases to the apparent total magnification of individual SN images.  Figure~\ref{fig:microhist} shows the distributions of total magnifications inferred from our simulations, including effects from both ``macro'' lensing and microlensing.

\subsection{Millilensing Simulations}
\label{sec:millilensing}

\begin{figure*}[ptb]
\gridline{\fig{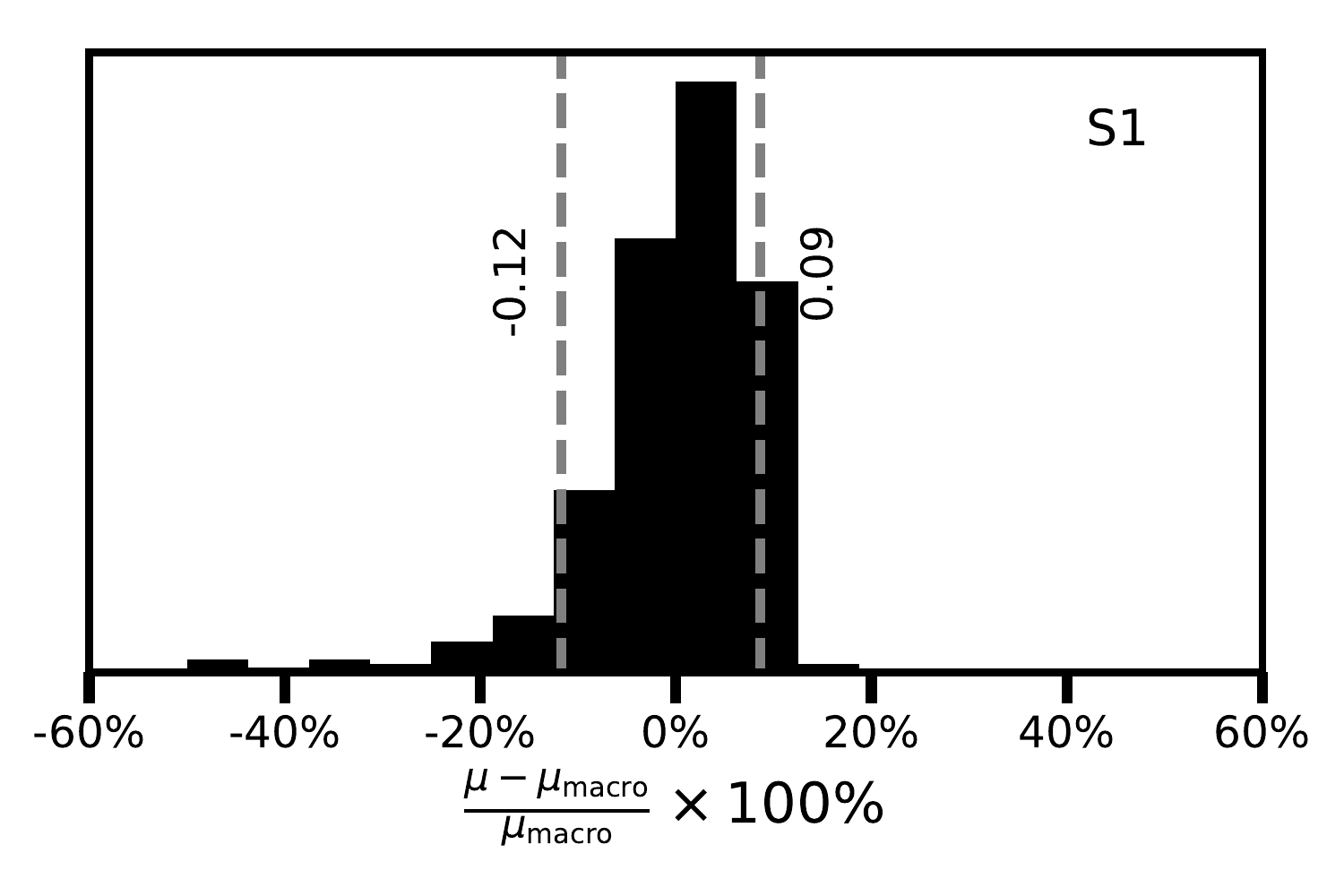}{0.49\textwidth}{}
\fig{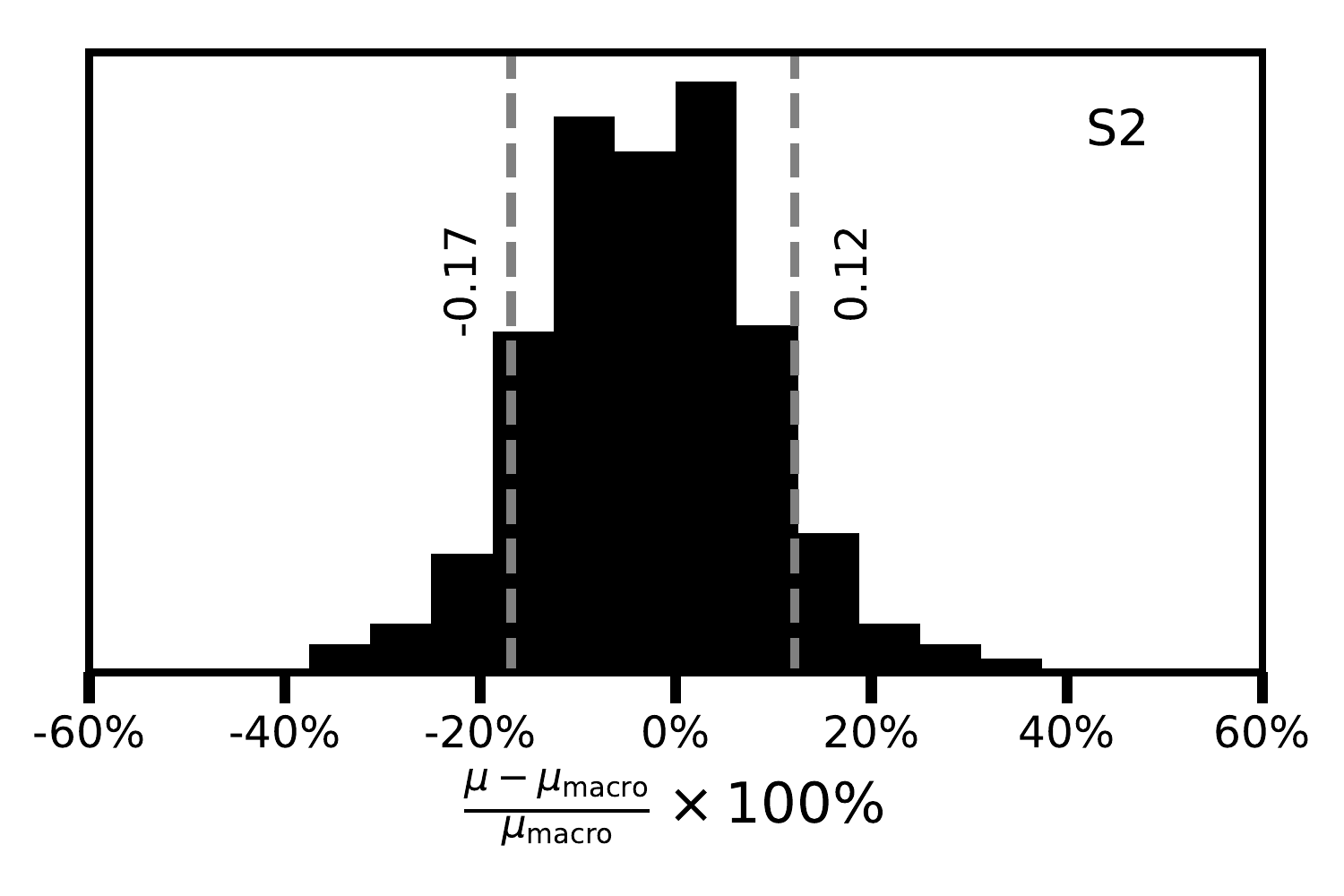}{0.49\textwidth}{}}
\vspace{-3em}
\gridline{\fig{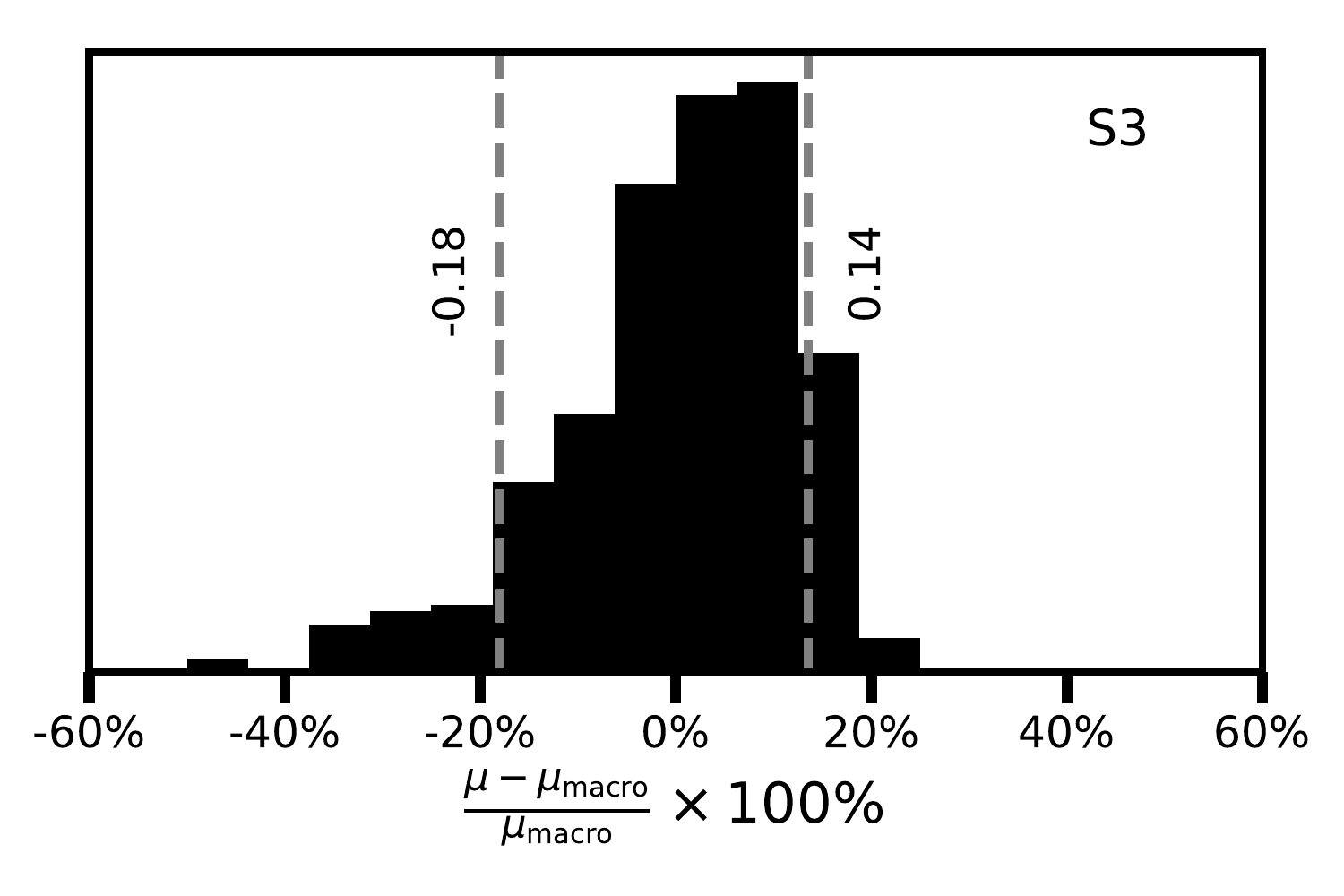}{0.49\textwidth}{}
\fig{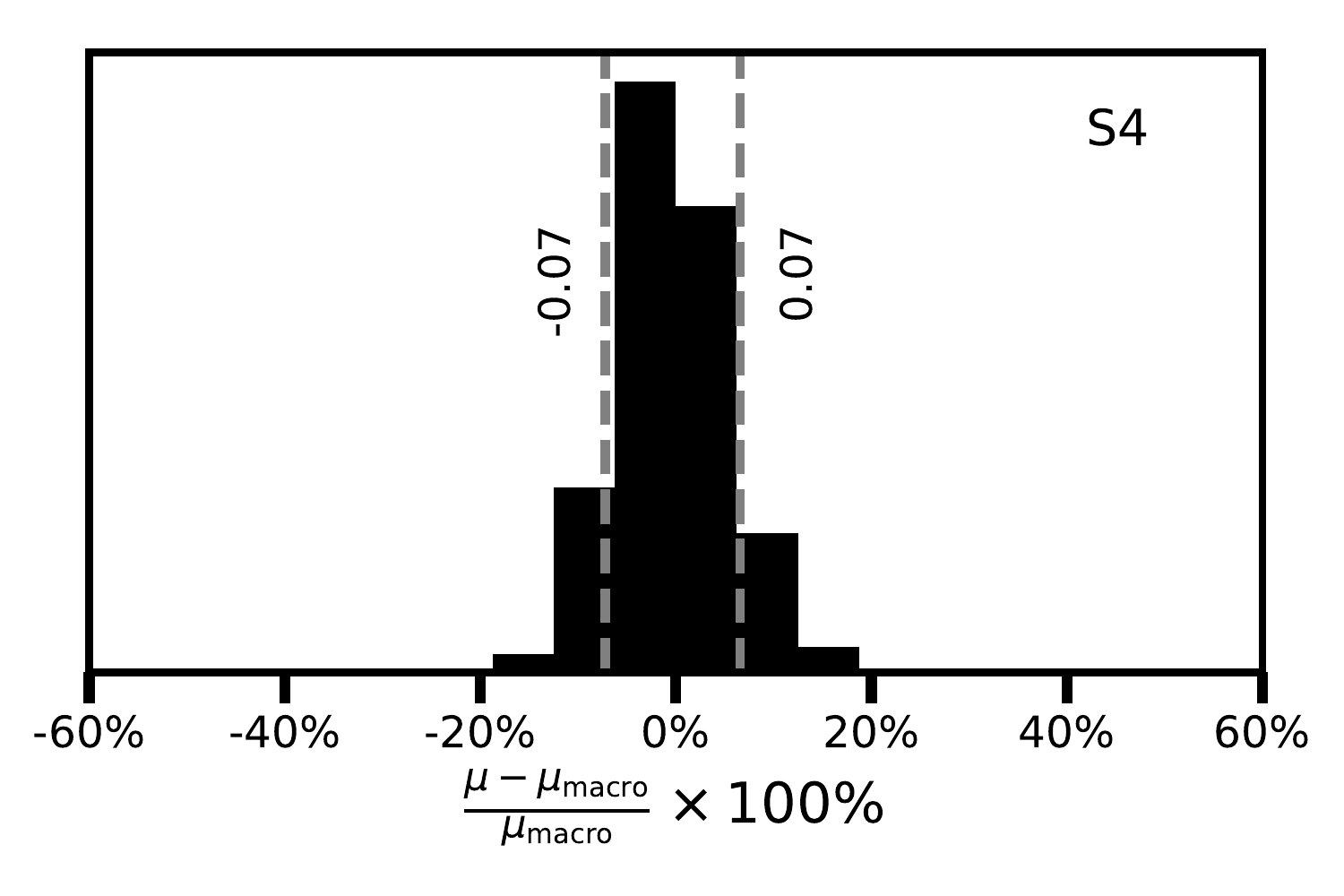}{0.49\textwidth}{}}
\vspace{-3em}
\fig{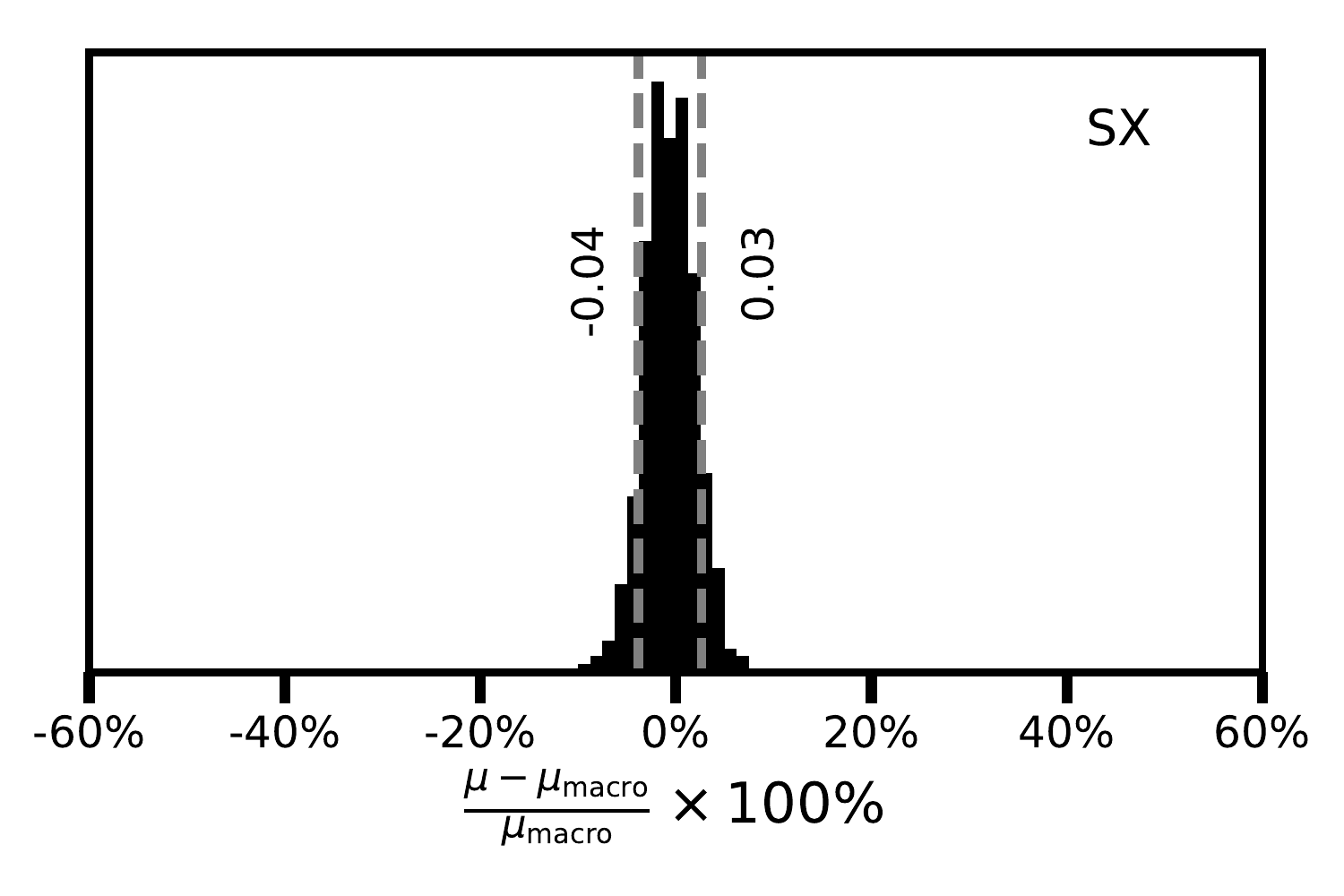}{0.49\textwidth}{}
\vspace{-2em}
\caption{Simulation of effect of millilensing dark-matter subhalos on the magnification of images S1--SX.  Here we show results where 0.5\% of the cluster dark matter is in the form of substructure, but Table~\ref{tab:millilensing} shows results where 2\% of the cluster dark matter  consists of subhalos.
}
\label{fig:millilensing}
\end{figure*}

Using the ray-tracing technique presented by \citet{gilmanbirrertreu19}, we compute the expected millilensing due to dark-matter subhalos for a projected mass density in substructure bracketing the range of expected values based on theoretical and observational arguments \citep{gilmanbirrernierenberg20}. In practice, we explore $10^7$\,M$_{\odot}$\,kpc$^{-2}$ in the main lens plane, corresponding to 0.5\% substructure mass fraction, as well as $4 \times 10^7$\,M$_{\odot}$\,kpc$^{-2}$, or a 2\% mass fraction in substructure. We model the line-of-sight structure using a Sheth-Tormen halo mass function \citep{shethmotormen01}. In these simulations, halos and subhalos are rendered in a range $10^6$--$10^{10}$\,M$_{\odot}$. Masses outside of this range are either massive enough to contain a visible counterpart, or too small to matter statistically \citep{gilmanbirrernierenberg20}. Table \ref{tab:millilensing} lists the 16\% and 84\% quantiles of the fractional change in magnification due to 0.5\% and 2\% substructure fractions, and Figure \ref{fig:millilensing} plots the distributions we measure for the case where the substructure fraction is 0.5\%. These show that magnification due to millilensing ranges from a $\sim$4\% to  $\sim$18\% effect among the five SN images.

\begin{deluxetable}{ccccc}
  \tablecolumns{5}
  \tablecaption{\sc Millilensing Simulations}
  \tablehead{  \multicolumn{1}{c}{Image} & \multicolumn{2}{|c}{0.05\%} of Dark Matter  & \multicolumn{2}{|c}{2\%} of Dark Matter \\ 
 & \multicolumn{1}{|c}{16\%}  & \multicolumn{1}{c|}{84\%}  & \multicolumn{1}{c}{16\%}  & \multicolumn{1}{c}{84\%}  }
\startdata
S1 & -0.12 & 0.09 & -0.14 & 0.09 \\
S2 & -0.17 & 0.12 & -0.17 & 0.14 \\
S3 & -0.18 & -0.14 & -0.08 & 0.08 \\
S4 & -0.07 & 0.07 & -0.08 & 0.08 \\
SX & -0.04 & 0.03 & -0.04 & 0.04 \\
\enddata
\label{tab:millilensing}
\tablecomments{Effect of millilensing by populations of dark-matter subhalos on magnification of images S1--SX. Here we list the 16\% and 84\% quantiles of $(\mu - \mu_{\rm macro}) / \mu$, where $\mu$ is the magnification measured from each simulation, and $\mu_{\rm macro}$ is the magnification that the SN images would have due to the galaxy cluster and cluster members in the absence of millilensing or microlensing. The columns at left show the results of simulations where 0.5\% of the dark matter along the line of sight consists of substructure, and the columns at right show results where substructure accounts for 2\% of the dark matter along the line of sight. }
\end{deluxetable}

\subsection{Modeling Correlation Structure in Light Curves}
\label{sec:correlationstructure}
The light curve of any SN will not, of course, be described exactly as a piecewise polynomial. 
\citet{baklanovlyskovablinnikov21} modeled the light curves of SN Refsdal measured through 2015 presented by \citet{rodneystrolgerkelly16} and \citet{kellyrodneytreu16} using a radiation hydrodynamics code. We first fit the  \citet{baklanovlyskovablinnikov21} F125W and F160W model using the same piecewise polynomial model that we apply to the data. We next model these residuals as a realization of a Gaussian process \citep[GP; see][]{rasmussen06} with a squared exponential covariance function, and estimate correlated deviations from the piecewise polynomial with a timescale of 10 days. We then model residuals of the piecewise model from the actual data as a combination of a correlated component common to all images, and an uncorrelated (``white noise'') component whose variance differs between images. The correlated component is modeled as a realization of a GP with a squared exponential kernel and fixed timescale of 10~days. For each filter, we estimate the standard deviations of the correlated and uncorrelated perturbations, which are listed in Table~\ref{tab:perturbationcoeffs}. We use our GP model to generate realizations of perturbations to the best-fitting piecewise polynomial model that are informed both by the \citet{baklanovlyskovablinnikov21} model, and the residual variation seen in the real data.

\begin{deluxetable*}{c|c|ccccc}
  \tablecolumns{10}
  \tablecaption{\sc Parameters of Residual Model}
  \tablehead{ Filter & Correlated & \multicolumn{5}{c}{Uncorrelated} \\
  & $K$ & S1 & S2 & S3 & S4 & SX }
\startdata
\hline
F125W & 0.00670 & 0.00576 & 0.01463 & 0.02097 & 0.03788 & 0 \\
F160W & 0.01066 & 0.00998 & 0.00416 & 0 & 0.02492 & 0 \\
\hline
\enddata
\label{tab:perturbationcoeffs}
\tablecomments{Standard deviations of correlated and uncorrelated components of the model used to generate residuals from the underlying piecewise polynomial.}
\end{deluxetable*}

\subsection{Simulated SN Refsdal Light Curves}
\label{sec:SimulatedLightCurves}

\begin{figure*}[htp!]
\fig{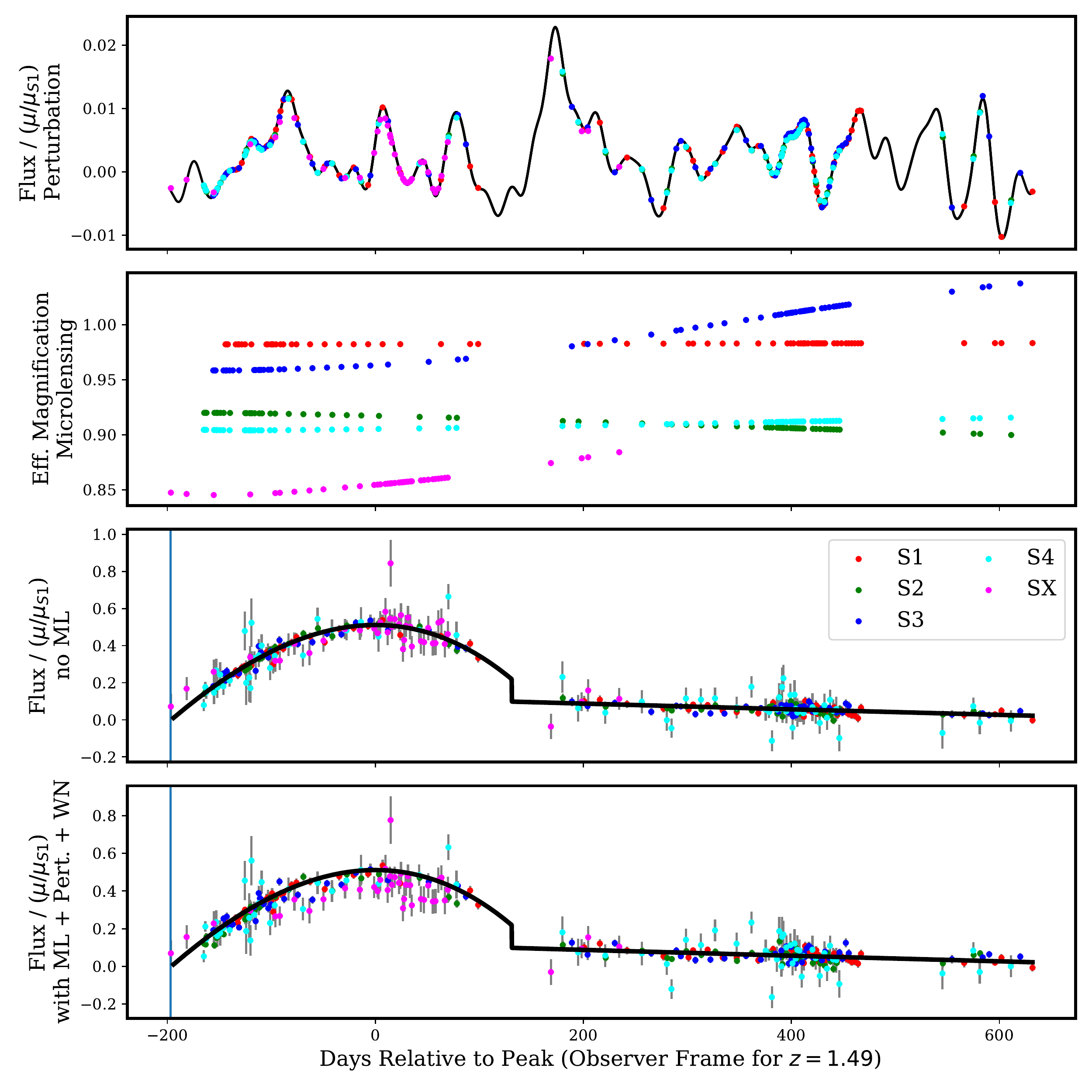}{0.8\textwidth}{}
\caption{Components of a single set of simulated light curves of the five appearances of SN Refsdal. The upper panel shows the perturbation applied to the fiducial piecewise polynomial model. The perturbation is a realization of a Gaussian process model of the residuals of the \citet{baklanovlyskovablinnikov21}  light curves  from the best-fitting piecewise polynomial model. The next panel plots the change in apparent magnification in the F125W band 
due to microlensing by a randomly placed set of stars and stellar remnants. The middle panel plots the generative light-curve model without microlensing (ML) included.
Motivated by fits to the light curves of low-redshift SN 1987A-like SNe with light curves similar to that of SN Refsdal (see Fig.~\ref{fig:combined_light_curves}), the generative model consists of separate polynomials for the photospheric and nebular SN phases. In the bottom panel, we show the simulated light curves after adding the microlensing plotted in the upper panel.  In both the middle and bottom panels, we divide the observed flux by $\mu / \mu_{\rm S1}$ to remove the relative magnification ratios among the light curves due to smoothly distributed matter (i.e., in the absence of microlensing) and we remove their relative offsets. 
}
\label{fig:fake_construction}
\end{figure*}

\begin{figure*}
\gridline{\fig{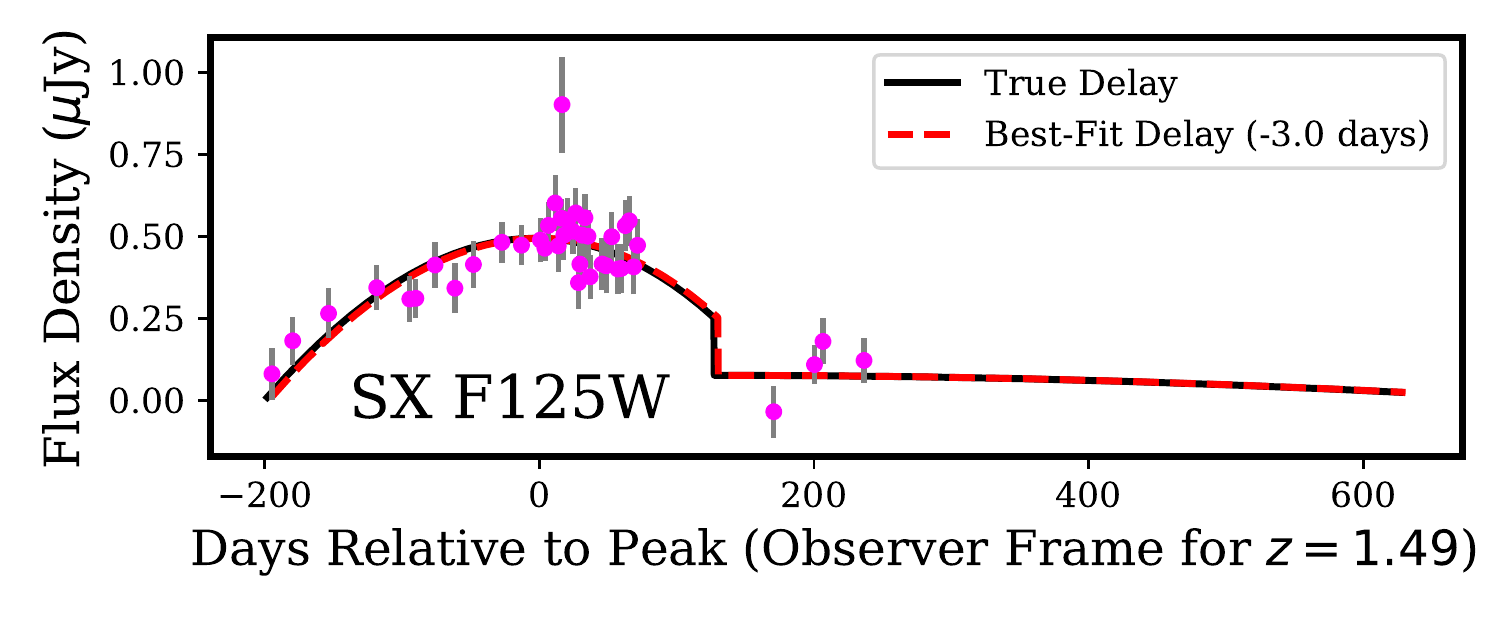}{0.49\textwidth}{}
\fig{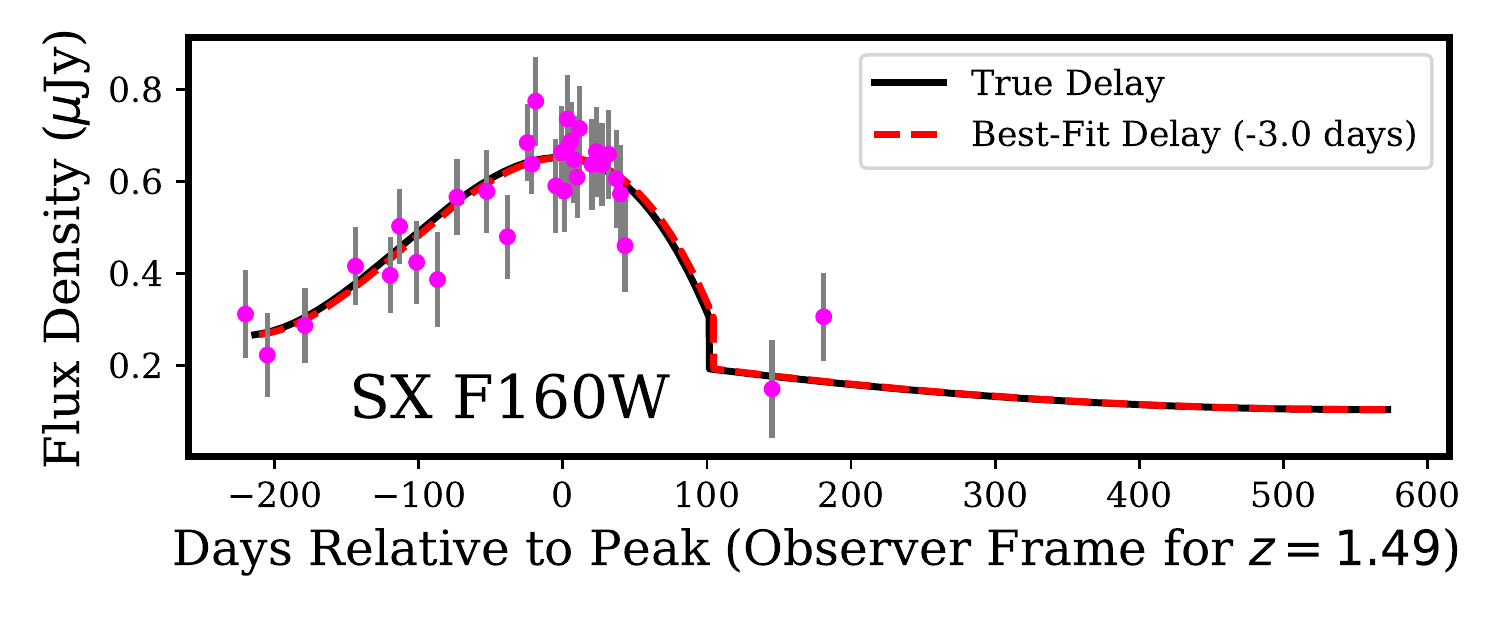}{0.49\textwidth}{}}
\caption{Comparison between recovered and actual delay of SX relative to S1 for the example simulation shown in Fig.~\ref{fig:fake_construction}. We show the best-fitting piecewise polynomial model (red dashed) and the same fit after shifting the model according to the true delay used to create the simulated light curve, leaving its normalization unchanged. For this realization, the time delay is recovered within three days. Estimated time delays are only affected by time-varying changes in total magnification arising from microlensing that change the shape of the light curve \citep[e.g.,][]{doblerkeeton06}. As a consequence, the inferred time delay require strong microlensing events, which should not occur frequently, given the relatively small value of $\kappa_s$. }
\label{fig:each_image_fake}
\end{figure*}

We now have all the components we believe are needed to assemble simulated light curves that mimic the SN Refsdal data.  These simulations will then be used to evaluate our light-curve fitting algorithms (Sec.~\ref{sec:fitting}) and define weights when combining the inferences from all models together (Sec.~\ref{sec:Combining}).

Figure \ref{fig:fake_construction} shows the steps to produce a single simulated dataset, which was chosen at random from the thousand primary simulated light curves. 
We use the fiducial light-curve model (a piecewise polynomial, described in Sec.~\ref{sec:fiducial_model})  to generate 1000 simulated light curves with the same 
cadence and photometric uncertainties as for SN Refsdal, as well as similar apparent magnitudes to those of SN Refsdal. We apply additional perturbations from our GP model of the expected variation around the fiducial model.
We next apply our simulations of the microlensing magnifications derived from the inferred stellar-mass densities, shear, and convergence along the lines of sight to each image (as described in Sec.~\ref{sec:MicrolensingSimulations}). 
Since the uncertainties in our measurements are background-dominated, we 
do not change photometric uncertainties when rescaling by microlensing fluctuations.
We include the effects of millilensing (Sec.~\ref{sec:millilensing}) until after fitting each microlensed light curve, since its effect is only a constant change in magnification.  

In Figure~\ref{fig:each_image_fake}, for the F125W and F160W light curves of SX, we compare the piecewise polynomial model that best fits the simulated light curves in Figure~\ref{fig:fake_construction}, against the same model shifted to reproduce the actual relative time delay. In the case of this simulated light curve, the best-fitting time delay is within three days of the true delay. Indeed, the estimated time delay can only be substantially affected by microlensing if the apparent {\it shape} of the light curve is altered, which requires a significant change in magnification with time. Given the low values of $\kappa_{\star}$ for the images of SN Refsdal, the optical depth for microlensing is small, so these events should be comparatively rare.

Four separate teams attempted to recover the relative time 
delays and magnification ratios from the one-thousand sets of simulated light 
curves, using the methods described in
Section~\ref{sec:fitting}. 
The fitting was fully blind, meaning the true values of the 
simulated time delays and magnification ratios were hidden from 
the light-curve fitting teams.  Section~\ref{sec:Combining} 
presents the analysis of these simulation tests and their 
application when we combine their results.

\section{Time Delay and Magnification Measurement Methods}
\label{sec:fitting}

In this Section, we describe the four algorithms that we have applied to 
recover the relative time delays and magnifications of images S1--SX.  We then explain how we used the simulated light curves to derive probability distributions for each of the observable lensing properties of SN Refsdal, and then use those probability distributions to combine the estimates from all models.

\subsection{Light-Curve Fitting Algorithms}

\subsubsection{Piecewise Photospheric and Nebular Polynomial Model}

\begin{deluxetable}{ccccccc}
  \tablecolumns{7}
  \tablecaption{\sc Piecewise Model Comparison}
  \tablehead{Phot. & Neb. & $\chi^2$ & $n_{\rm par}$ & AIC & $\chi^2_{\nu}$ & $\Delta t_{\rm X1}$}
\startdata
2 & 1 & \chisqtwoonedolphot & \paramstwoonedolphot & \aictwoonedolphot  & \redchisqtwoonedolphot & \delaysxsonetwoonedolphot \\
3 & 1 & \chisqthreeonedolphot & \paramsthreeonedolphot & \aicthreeonedolphot  & \redchisqthreeonedolphot & \delaysxsonethreeonedolphot\\
4 & 1 & \chisqfouronedolphot & \paramsfouronedolphot & \aicfouronedolphot  & \redchisqfouronedolphot & \delaysxsonefouronedolphot \\
5 & 1 & \chisqfiveonedolphot & \paramsfiveonedolphot & \aicfiveonedolphot  & \redchisqfiveonedolphot & \delaysxsonefiveonedolphot \\
6 & 1 & \chisqsixonedolphot & \paramssixonedolphot & \aicsixonedolphot  & \redchisqsixonedolphot & \delaysxsonesixonedolphot \\
\hline
2 & 2 & \chisqtwotwodolphot & \paramstwotwodolphot & \aictwotwodolphot  & \redchisqtwotwodolphot & \delaysxsonetwotwodolphot \\
3 & 2 & \chisqthreetwodolphot & \paramsthreetwodolphot & \aicthreetwodolphot  & \redchisqthreetwodolphot & \delaysxsonethreetwodolphot \\
4 & 2 & \chisqfourtwodolphot & \paramsfourtwodolphot & \aicfourtwodolphot  & \redchisqfourtwodolphot & \delaysxsonefourtwodolphot \\
5 & 2 & \chisqfivetwodolphot & \paramsfivetwodolphot & \aicfivetwodolphot  & \redchisqfivetwodolphot & \delaysxsonefivetwodolphot \\
6 & 2 & \chisqsixtwodolphot & \paramssixtwodolphot & \aicsixtwodolphot  & \redchisqsixtwodolphot & \delaysxsonesixtwodolphot
\enddata
\tablecomments{Results from fitting piecewise models with photospheric and nebular components each of varying order.}
\label{tab:modelparams}
\end{deluxetable}

Our first light-curve-fitting algorithm uses the same piecewise polynomial apparatus that we used for generating the simulated light curves, as shown in  Figure~\ref{fig:combined_light_curves} and described in Section~\ref{sec:SimulatedLightCurves}. For the reasons described in Section~\ref{sec:fiducial_model}, we adopt a third-order polynomial for the photospheric phase and a second-order polynomial for the nebular phase.
We include separate models for the F125W and F160W light curves (i.e., assuming no fixed rule for color variation), but simultaneously fit the light curves for all images S1--SX. Table~\ref{tab:modelparams} presents the best-fitting parameters of this model for various polynomial orders, as well as the Akaike Information Criterion \citep[AIC;][]{aka74} values and the inferred time delays from image S1 to SX.

\subsubsection{Bayesian Gaussian Process Model}
We construct a Bayesian GP model for the light curves, motivated by the approaches of \citet{tak17} and \citet{hojjati13} for modeling lensed quasar time delays.  Our framework describes the intrinsic SN light-curve shape, perturbations due to microlensing, and relative time delays and magnification due to strong lensing, within a single probabilistic model.  In this model, the time series data of each image are treated as noisy measurements of the same underlying light curve, shifted in time and magnitude and modulated by a microlensing curve. The latent light curve, assumed to be the same for all images, is modeled as a realization of a GP for smooth functions \citep[see][for a background on Gaussian processes]{rasmussen06}. The microlensing perturbations are also modeled as GP realizations, but with no correlation between the multiple images. Our overall model has twelve parameters -- two latent light-curve hyperparameters, two microlensing hyperparameters, four time delays, and four magnitude shifts.  We sample the joint posterior distribution of these parameters, conditioning simultaneously on the light-curve observations of all the images using the \texttt{dynesty} nested sampling package \citep{speagle20}.  We obtain posterior distributions of the time delays by marginalizing over all other parameters.

\subsubsection{{\tt Supernova Time Delays} Parametric Model}
We fit the light-curve data using version 2 of the {\tt Supernova Time
Delays (SNTD)} software package \citep{pierelrodney19}.
Here we model SN Refsdal's flux as a function of time using
\begin{equation}
\label{eq:bazin_model}
F(t)=A\frac{e^{-(t-t_0)/\tau_{\rm{fall}}}}{1+e^{-(t-t_0)/\tau_{\rm{rise}}}}+B,
\end{equation}
presented by \citet{bazinpalanquedelarouillerich09} as a generic model for SN light curves, 
where $\tau_{\rm{rise}}$ and $\tau_{\rm{fall}}$
characterize the relative rise and decline times of the light
curve (respectively), and $A$ and $B$ set its normalization. 
While this model is flexible, its simple shape minimizes sensitivity to
microlensing features in the light
curves of S1--SX.

In order to measure the time delays of the simulated light curves described
in Section \ref{sec:SimulatedLightCurves}, we employed {\tt SNTD}'s 
parallel method according to which the light curves of S1--SX are first fit separately
before deriving joint likelihood estimates for the light-curve model parameters \citep[see ][]{pierelrodney19}.
We note that {\tt SNTD} was designed to use spectrophotometric models of SNe,
to leverage, when available, prior knowledge of the intrinsic color and
shape of well-studied SN species including Type Ia SNe or common types of core-collapse SNe.
SN Refsdal's light curve, however, was peculiar and similar to that of SN 1987A \citep{kellyrodneytreu16},
the explosion of a blue supergiant, and a useful spectrophotometric model was unfortunately not available.  

\begin{figure*}
\gridline{\fig{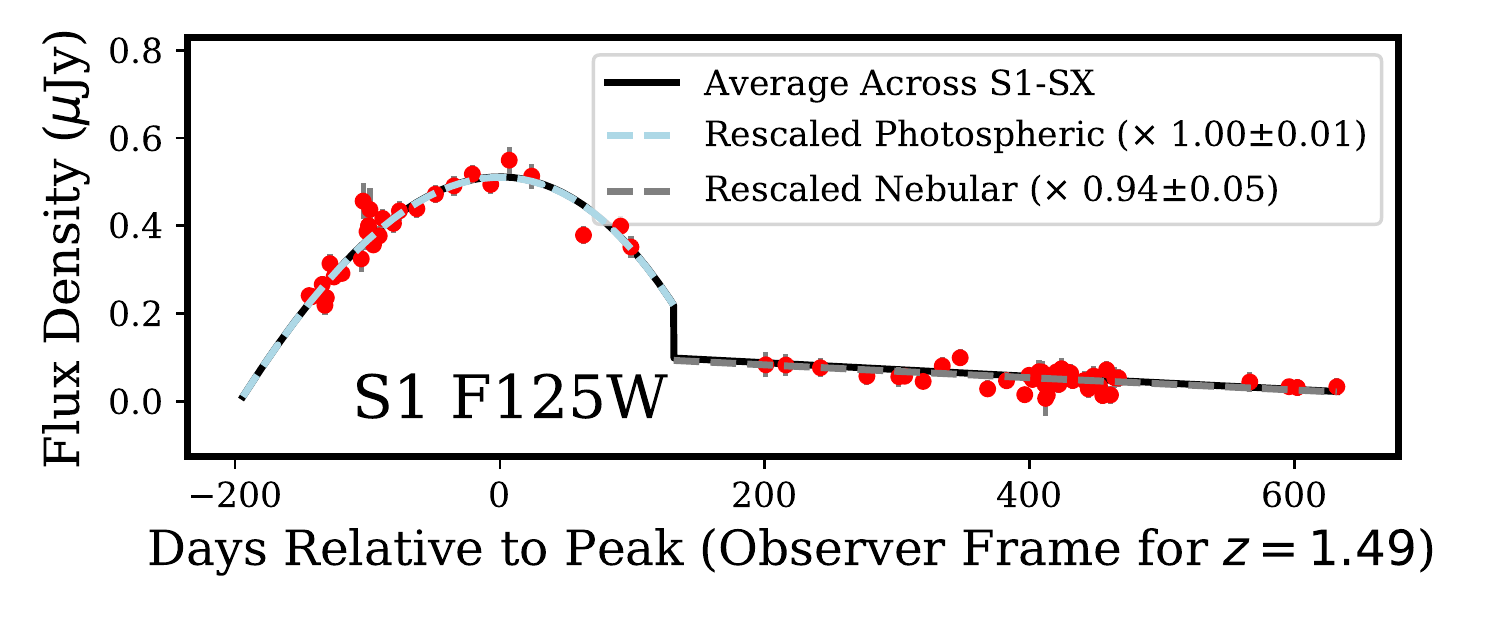}{0.49\textwidth}{}
\fig{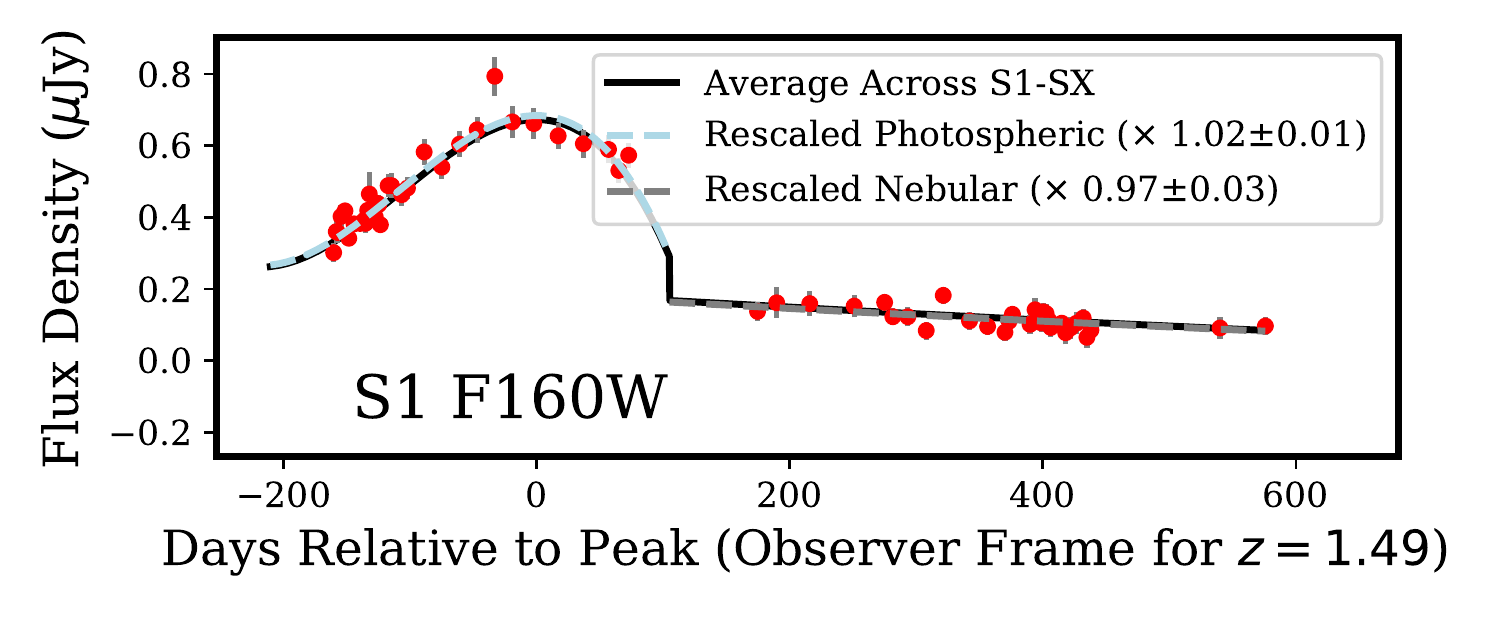}{0.49\textwidth}{}}
\vspace{-3em}
\gridline{\fig{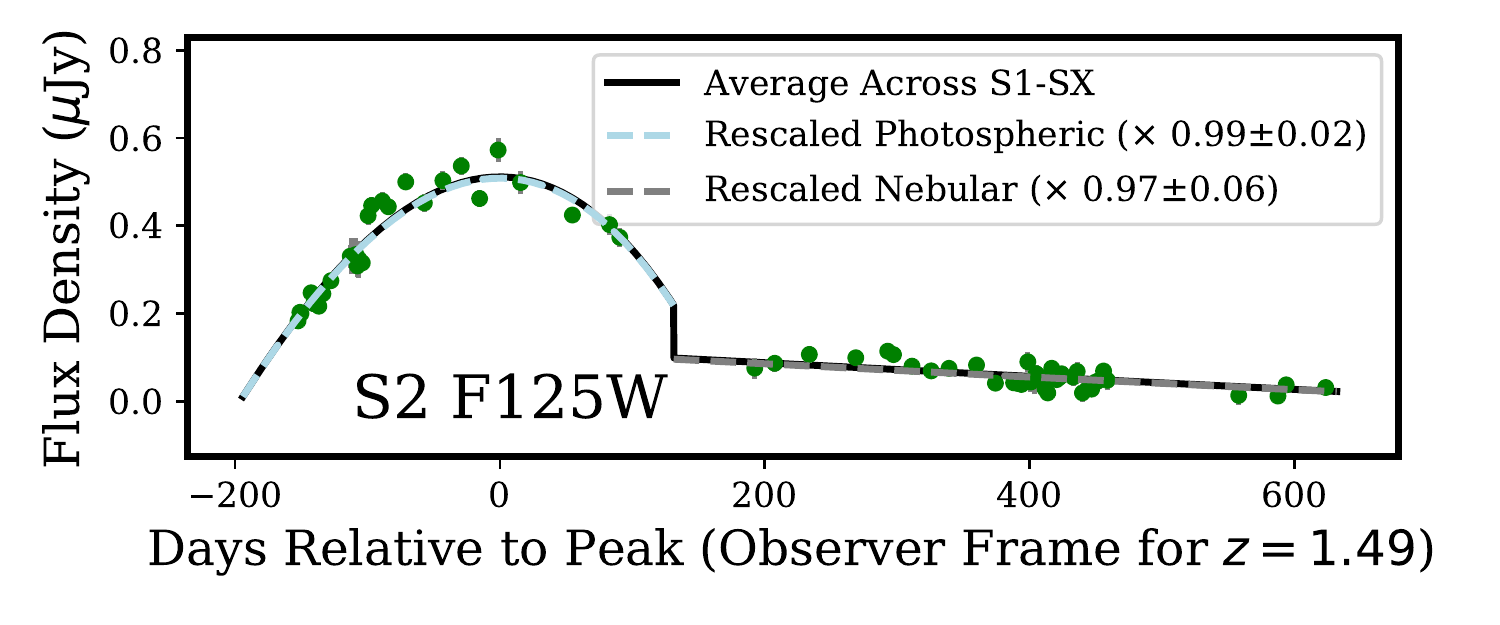}{0.49\textwidth}{}
\fig{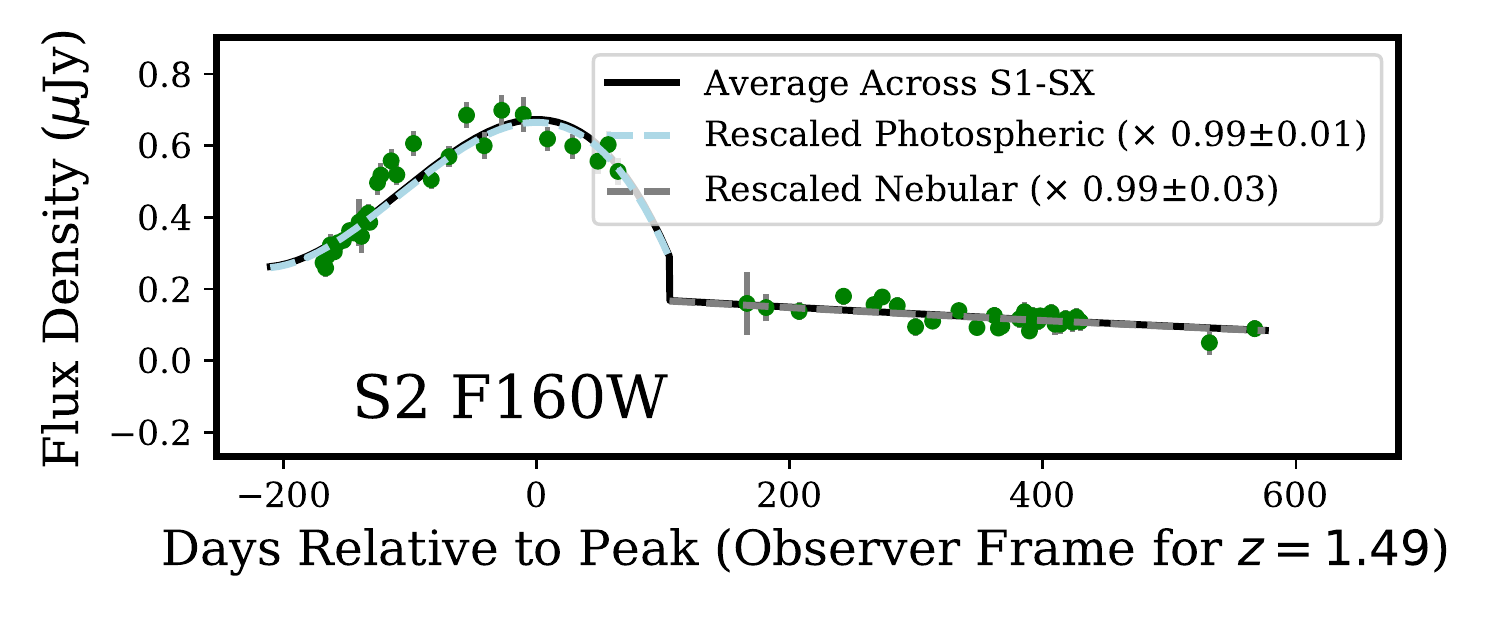}{0.49\textwidth}{}}
\vspace{-3em}
\gridline{\fig{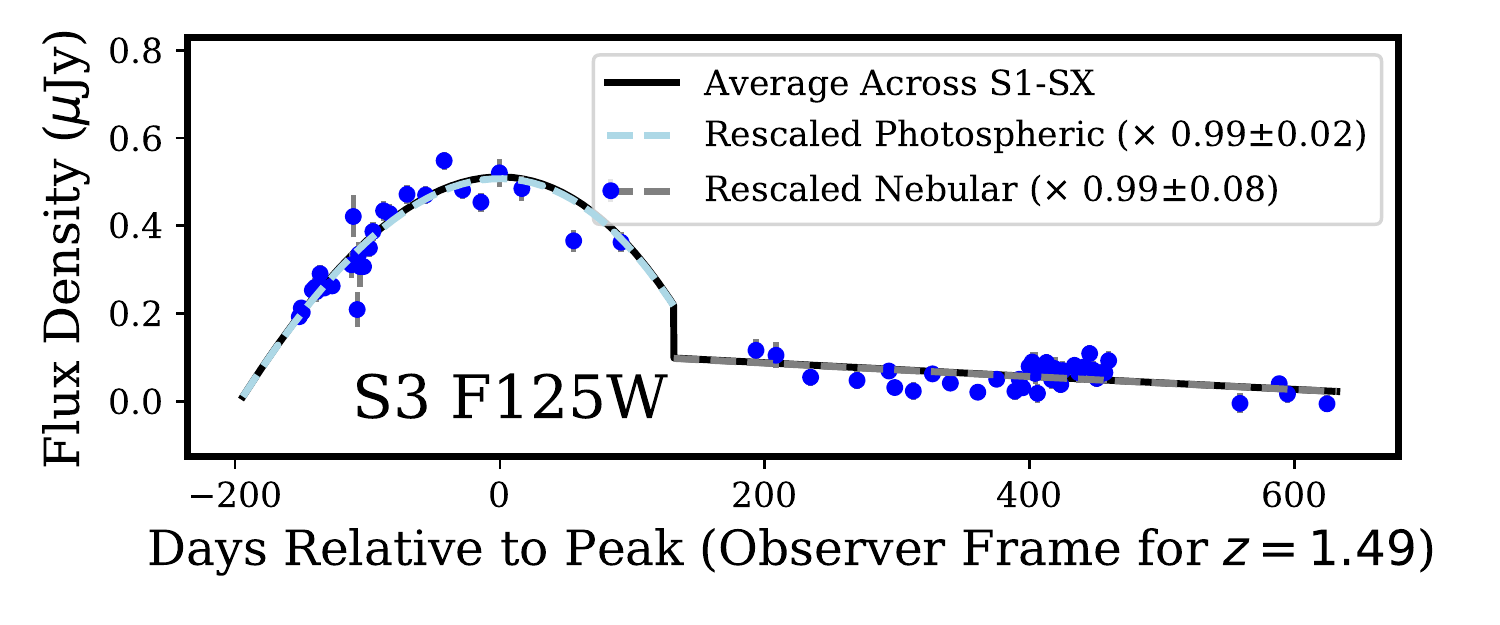}{0.49\textwidth}{}
\fig{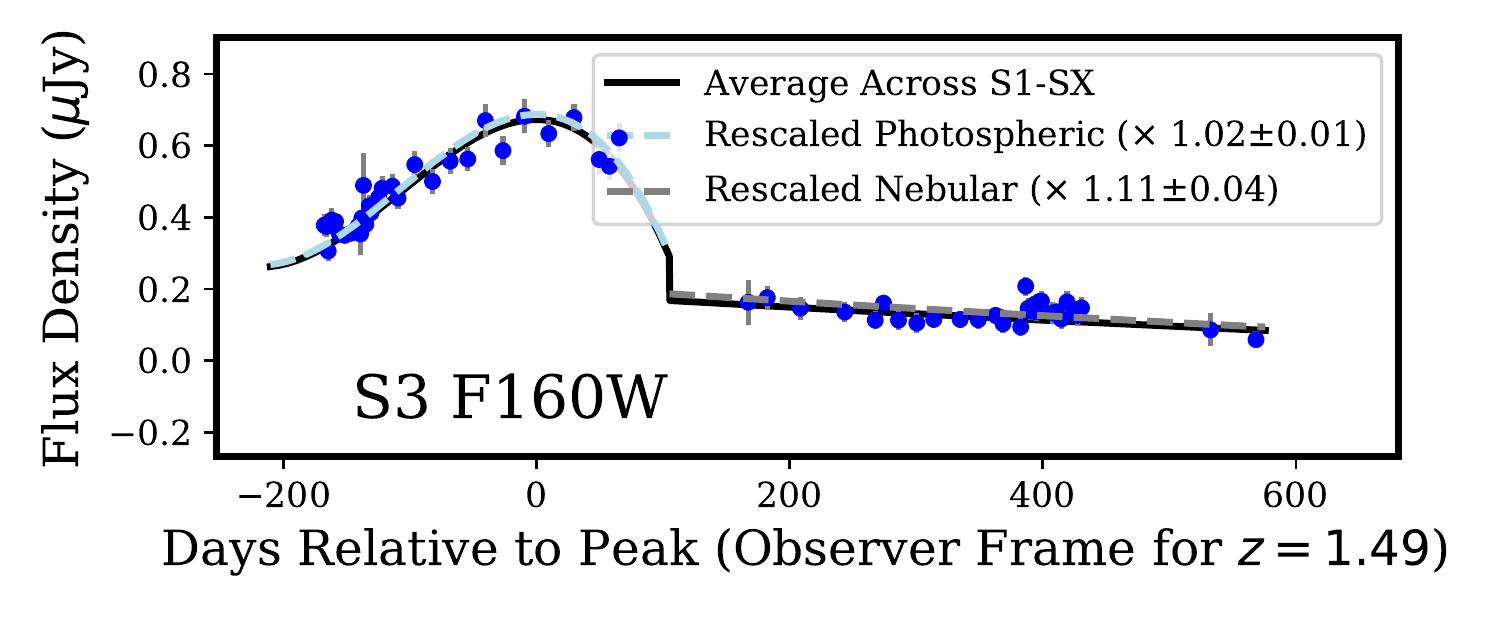}{0.49\textwidth}{}}
\vspace{-3em}
\gridline{\fig{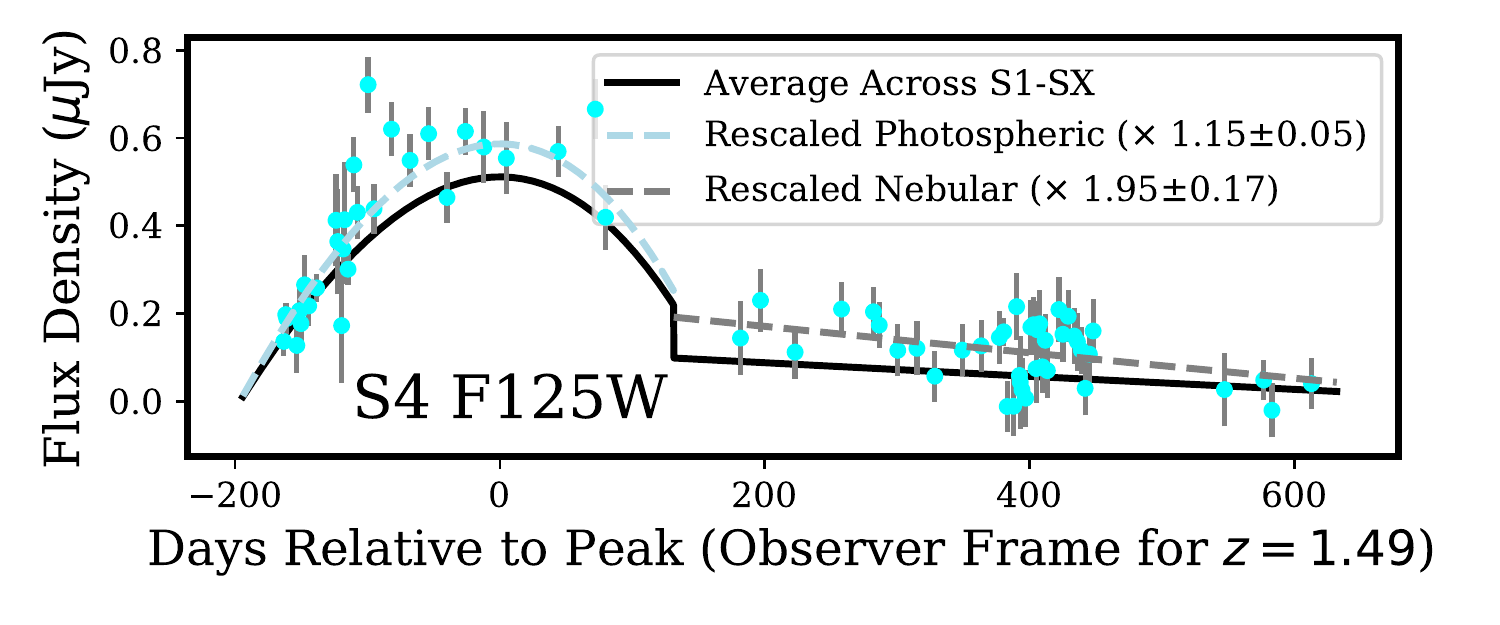}{0.49\textwidth}{} 
\fig{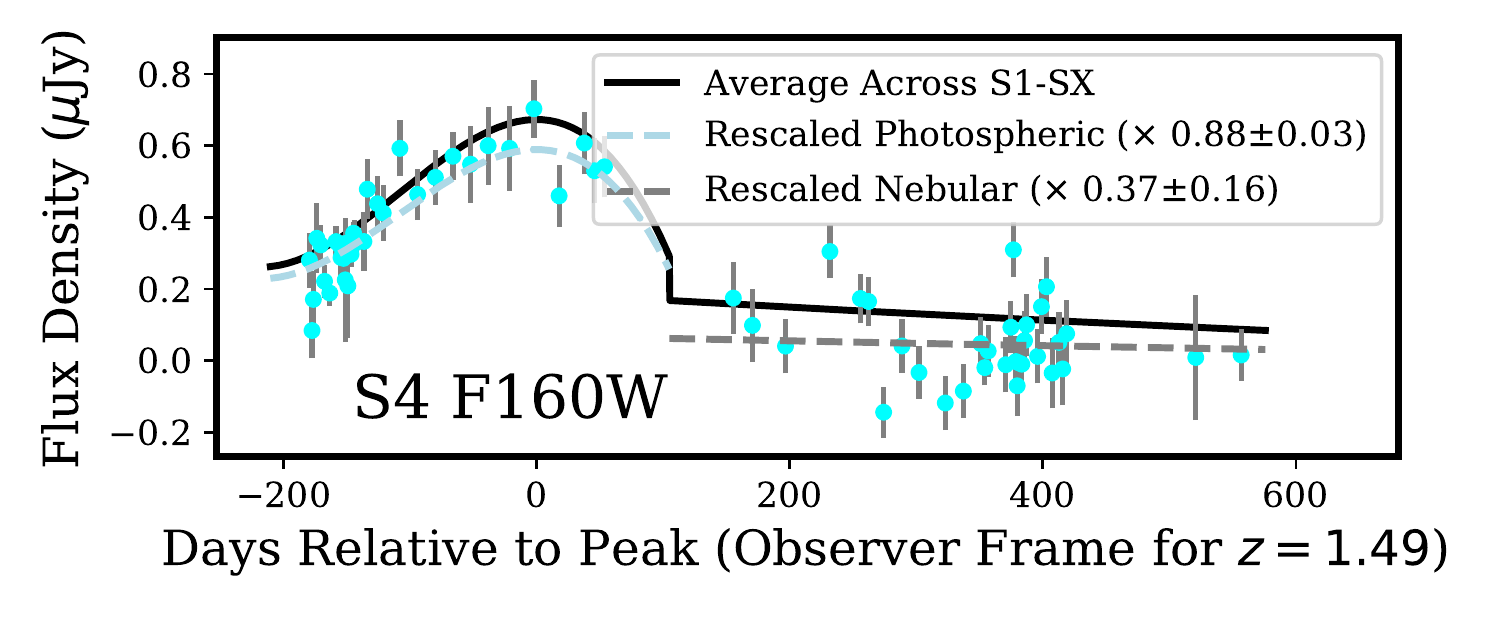}{0.49\textwidth}{}}
\caption{Photometry of the five images of SN Refsdal and light-curve fits using the fiducial light-curve model.  The fiducial piecewise polynomial model is shown fit simultaneously to the F125W and F160W (rest-frame approximately {\it V} and {\it R}) light curves of images S1--SX (solid).  The dashed line shows the same model fit after rescaling to match the photospheric and nebular light curves of each image separately. 
Each legend lists the respective best-fitting factor used to rescale each light-curve component, and we list the ratios between the photospheric- and nebular-phase factors in Table~\ref{tab:fluxratios}.  Microlensing that is not achromatic could cause the apparent color of each image to become comparatively bluer or redder. Although not expected given that the galaxy-scale lens is an early-type galaxy, extinction by dust would cause the color to become redder.  The color of image S4 is much more {\it blue} than the average color of Refsdal images.
}
\label{fig:each_image_light_curves}
\end{figure*}

\begin{table}
\small
\centering
\begin{tabular}{cccc}
\multicolumn{1}{c}{Object} &  \multicolumn{1}{c}{Filter} &  \multicolumn{1}{c}{Ratio} & \multicolumn{1}{c}{Uncertainty}  \\
\hline
S1 & F125W & 0.968 & 0.031 \\
S2 & F125W & 0.982 & 0.035 \\
S3 & F125W & 0.991 & 0.046 \\
S4 & F125W & 1.561 & 0.110 \\
SX & F125W & 0.979 & 0.035 \\
\hline
S1 & F160W & 0.996 & 0.017 \\
S2 & F160W & 0.989 & 0.016 \\
S3 & F160W & 1.063 & 0.023 \\
S4 & F160W & 0.709 & 0.066 \\
SX & F160W & 1.016 & 0.028 \\
\hline
\end{tabular}
\caption{\label{tab:fluxratios} Anomalous color of image S4 of SN Refsdal. Ratios between measured fluxes and model light curve 
that best fits the light curves of all images.  The best-fitting model accounts for the fluxes of all images only by changing 
the relative magnifications of the image.  }
\end{table}

\begin{figure*}
\gridline{\fig{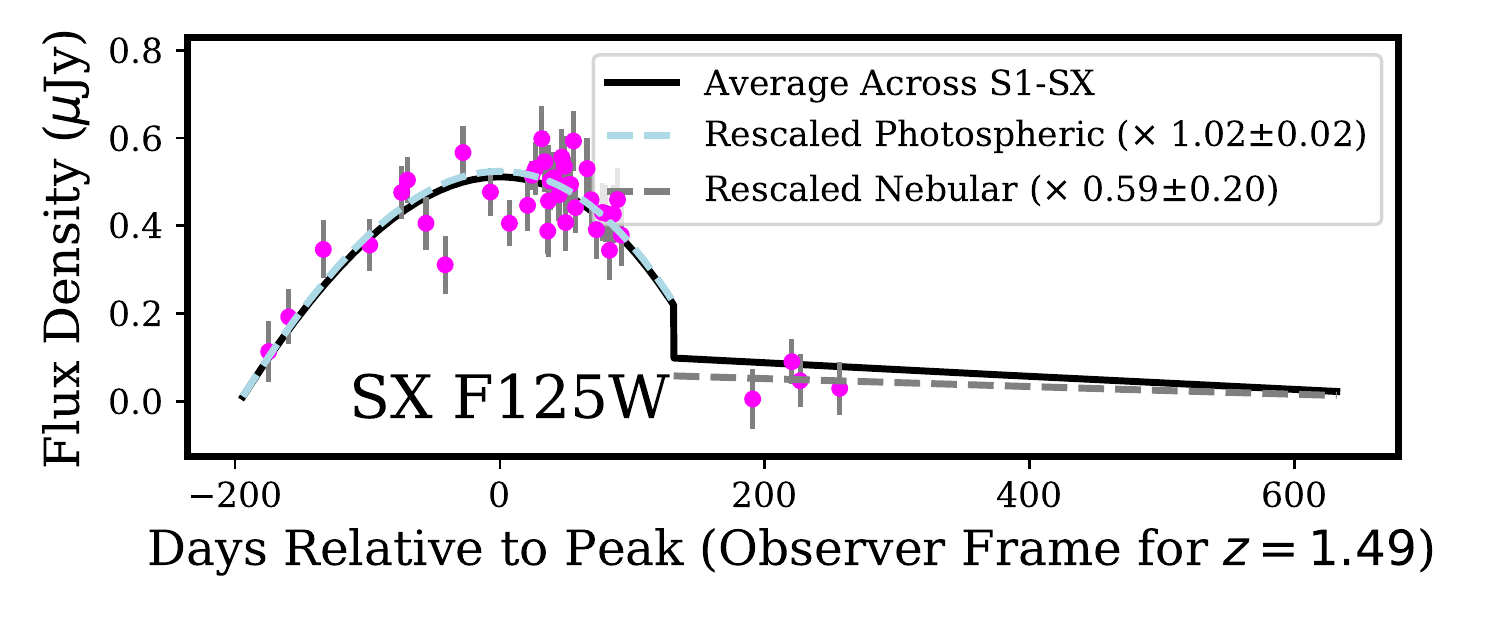}{0.49\textwidth}{}
\fig{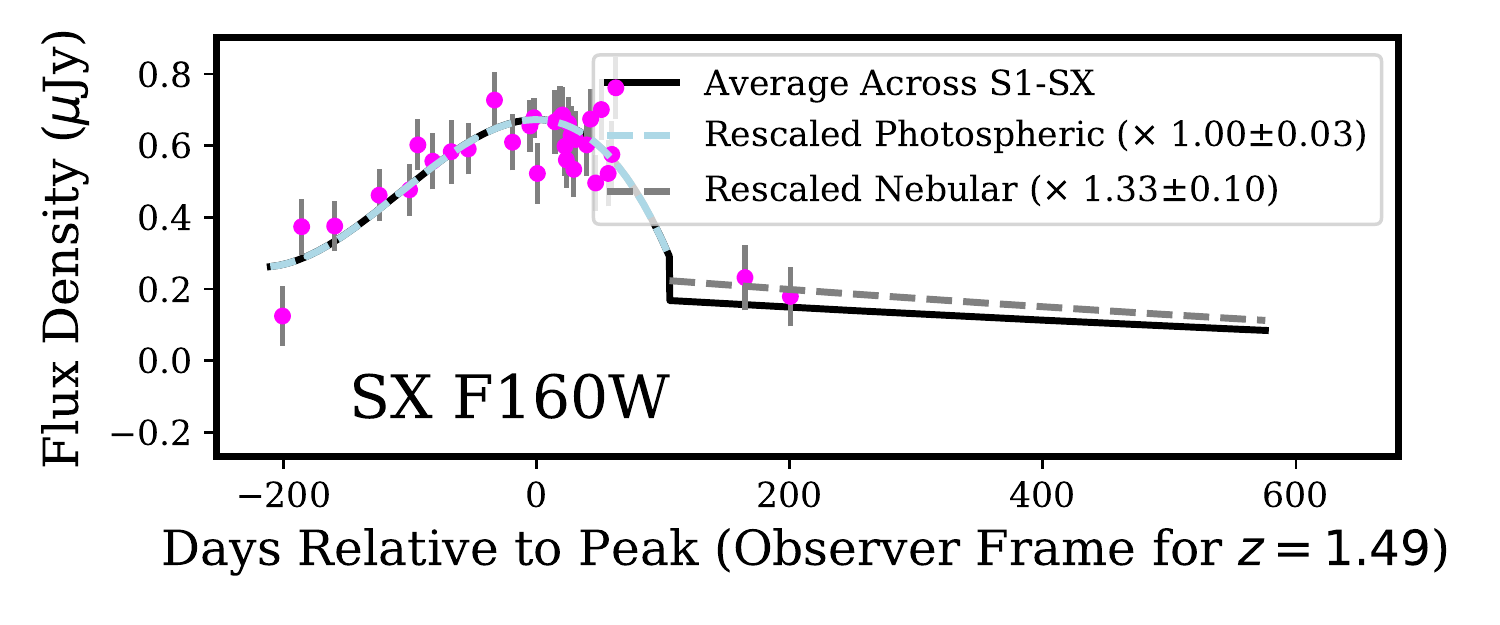}{0.49\textwidth}{}}
\caption{Continuation of Figure~\ref{fig:each_image_light_curves}.
}
\label{fig:each_image_light_curves_continued}
\end{figure*}

We applied the parallel algorithm to the F125W and F160W light
curves separately, since Equation~\ref{eq:bazin_model} only describes the 
flux in a single band.  For each filter, we did two iterations.  We first identified the time $\tau_{\rm pk}$ as the epoch with the maximum observed flux,
and carried out a``coarse" fit subject to the following bounds on the fitting parameters:
\begin{eqnarray*}
 \tau_{\rm pk}-150 < t_0 < \tau_{\rm pk}+150\\
20 < t_{\rm rise} < 80\\
60 < t_{\rm fall} < 300 \\
0.001 < A < 4 \\
0 < B < 2.
\end{eqnarray*}
From this coarse run we identified the set of parameters that
returned the best $\chi^2$ as ($t_{0,{\rm c}}$, $t_{\rm rise,c}$,
$t_{\rm fall,c}$, $A_{\rm c}$, $B_{\rm c}$).  
We adopted the first three of those parameter values as estimates for the time of peak, rise time, and fall time, respectively.

Our next step was to carry out a ``fine'' fit using the same bounds for A and B as in the coarse run while applying the 
following constraints:
\begin{eqnarray*}
 t_{0,{\rm c}}-50 < t_0 < t_{0,{\rm c}}+50\\
t_{\rm rise,c}-3 < t_{\rm rise} < t_{\rm rise,c}+3\\
t_{\rm fall,c}-10 < t_{\rm fall} < t_{\rm fall,c}+10. 
\end{eqnarray*}
After completing this procedure for the F125W and F160W bands
separately, we defined the final time delay and magnification
estimates from the band preferred by the Bayesian
evidence for the fine-run fitting results.

During the testing phase with simulated light curves (i.e., before
applying any fitting method to the actual Refsdal data), we
experimented with more aggressive constraints on the parameter
values. Specifically, we experimented with tighter bounds on the model
parameters when fitting to the light curves of S2, S3, and S4, and then using these
parameters to model the light curves of S1 and SX. 
However, this procedure resulted in a $\sim7$-day bias in the estimated
time of peak for the S1 and SX light curves, and we discarded this approach.

\subsubsection{{\tt Python Curve Shifting3} Free-Knot Spline Model}
{\tt Python Curve Shifting3 (PyCS3)} is a new version of the PyCS package, which was originally designed for application to lensed quasar light curves, and includes several independent methods for measuring time delays and estimating uncertainties \citep{tewescourbinmeylen13, milloncourbinbonvin20b}. For application to the SN Refsdal data, we use the free-knot splines estimator, which fits third-order splines to the light-curve data. The intrinsic variations of the source, common to all lensed images, are fit with a single spline. The variability of the spline is controlled by the initial spacing of the knots. The position of the knots, coefficients of the spline, and time and magnitude shifts of all the light curves are iteratively adjusted to minimize a global $\chi^2$, following the bounded optimal knots algorithm  \citep{molinaridurandsabatier04}. The details of the fitting process are given by \citet{tewescourbinmeylen13}, and assessments of the robustness of the method when applied to lensed quasars are presented by 
\citet{liaotreumarshall15} and \citet{bonvintewescourbin16}.

We explore a set of plausible knot spacing values with the automated pipeline described by \citet{milloncourbinbonvin20a} and select the value providing the most precise results, according to the PyCS3 uncertainties estimation procedure. Note that we do not use this procedure for our final estimates of the uncertainties. Instead, we use the free-knot splines fitting to estimate a single value for the relative time delays, and we compute the uncertainty using the residuals of the {\tt PyCS3} estimates, compared to the simulated light curves as described in Section~\ref{sec:Combining}. Finally, we obtain our final time-delay estimates by measuring the time delays and flux ratios separately in the F125W and F160W band and by taking the weighed mean, with the weight being proportional to the inverse of the estimated uncertainties in each band.

\subsection{Probability Distributions for Each Observable, from Each Method}

\begin{figure*}
\centering
\gridline{\fig{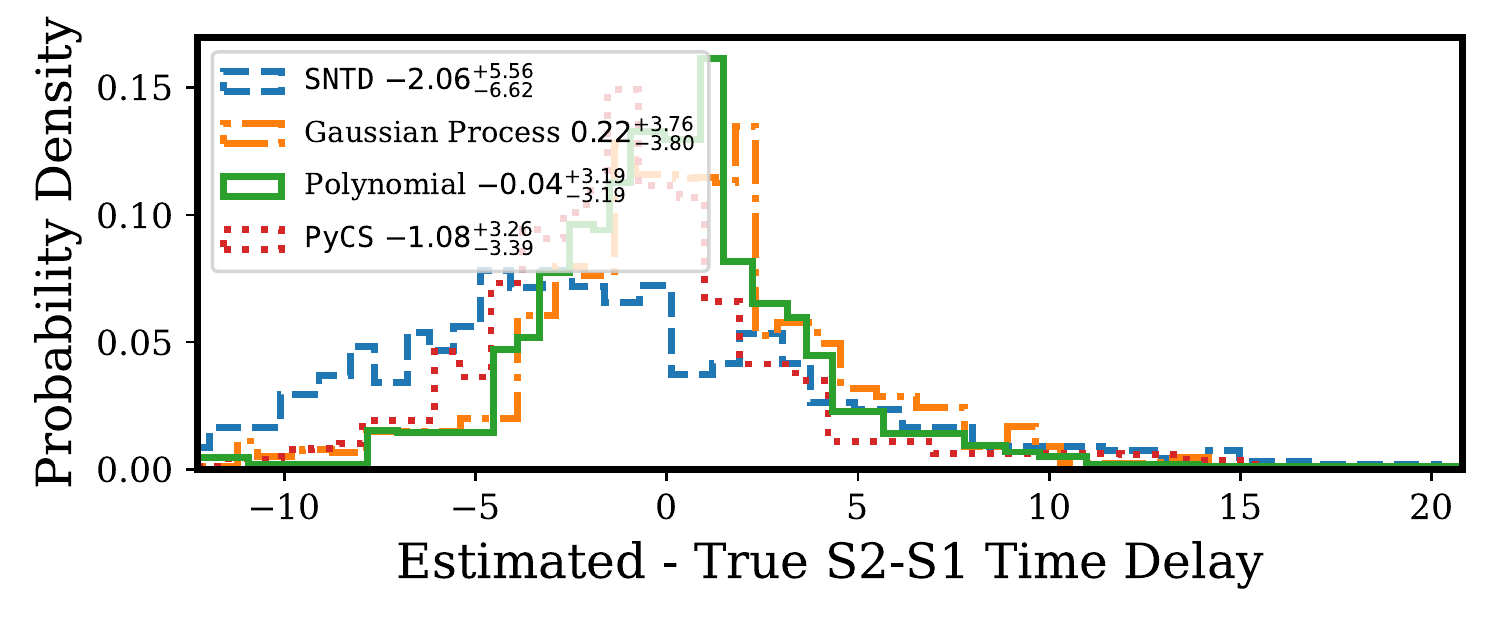}{0.49\textwidth}{}  
\fig{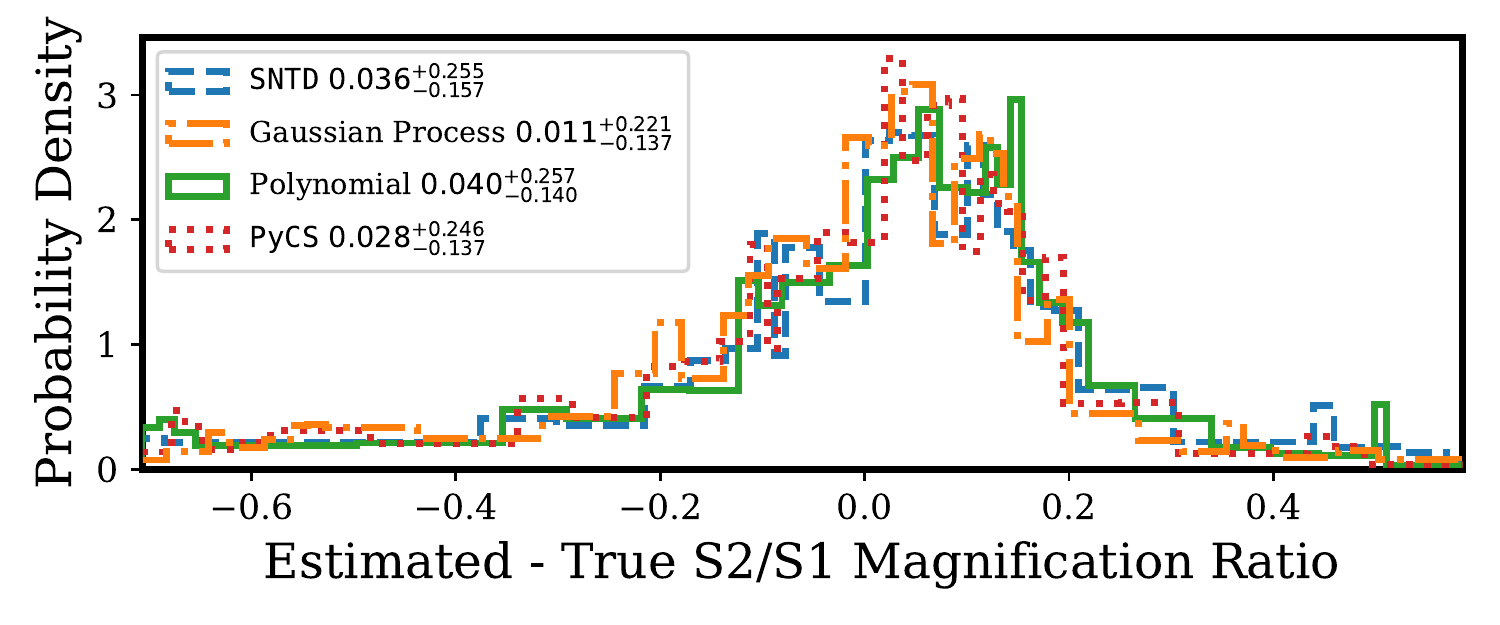}{0.49\textwidth}{}}
\vspace{-3em}
\gridline{\fig{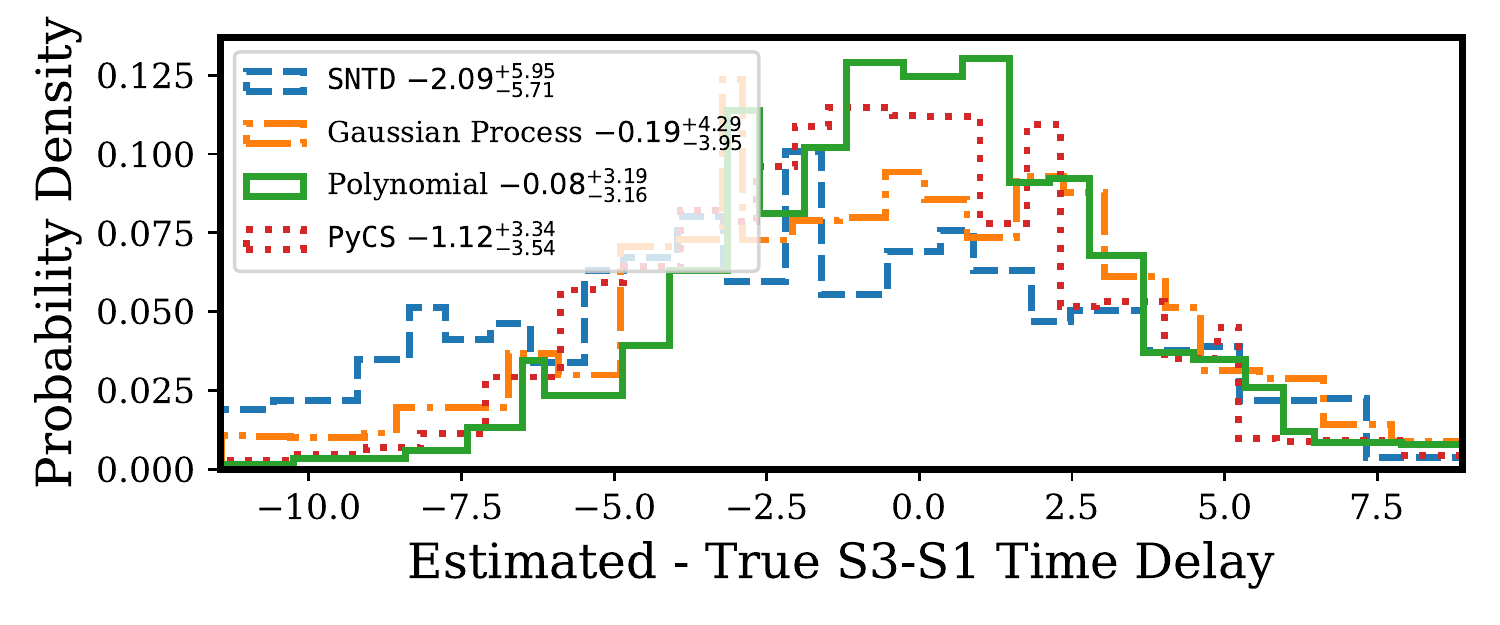}{0.49\textwidth}{}  
\fig{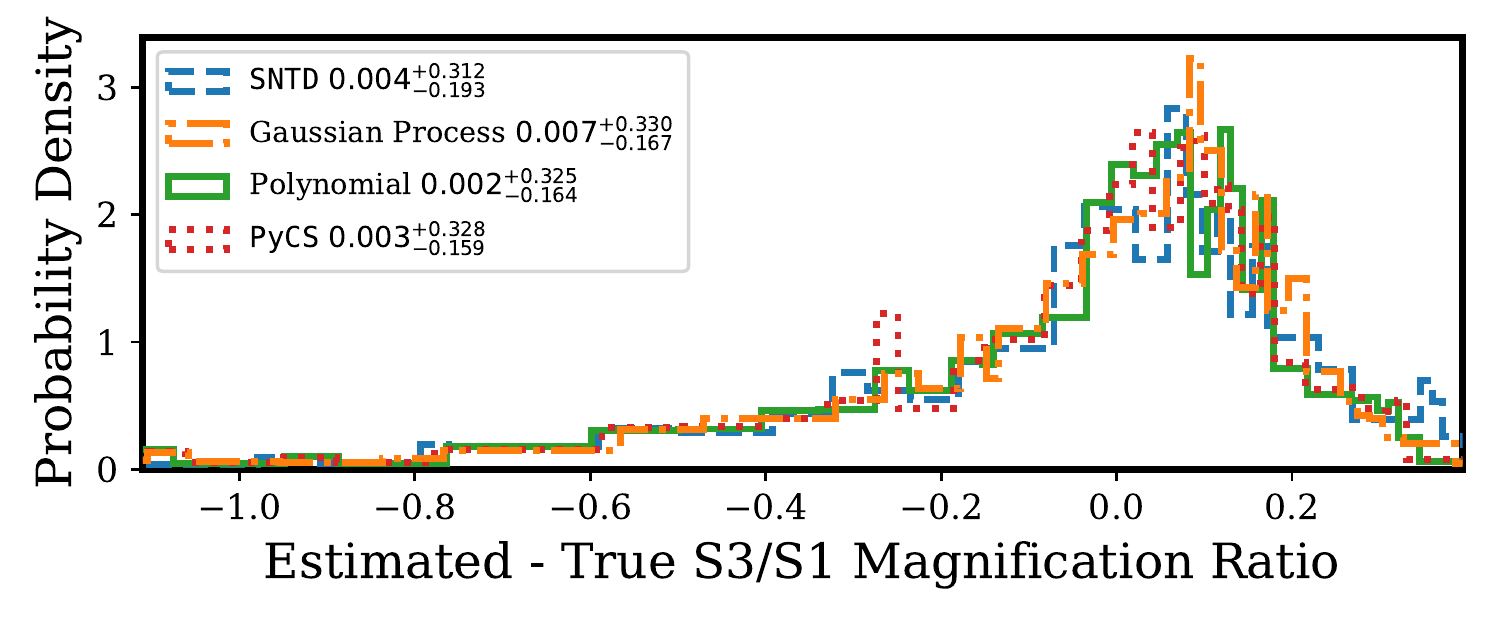}{0.49\textwidth}{}}
\vspace{-3em}
\gridline{\fig{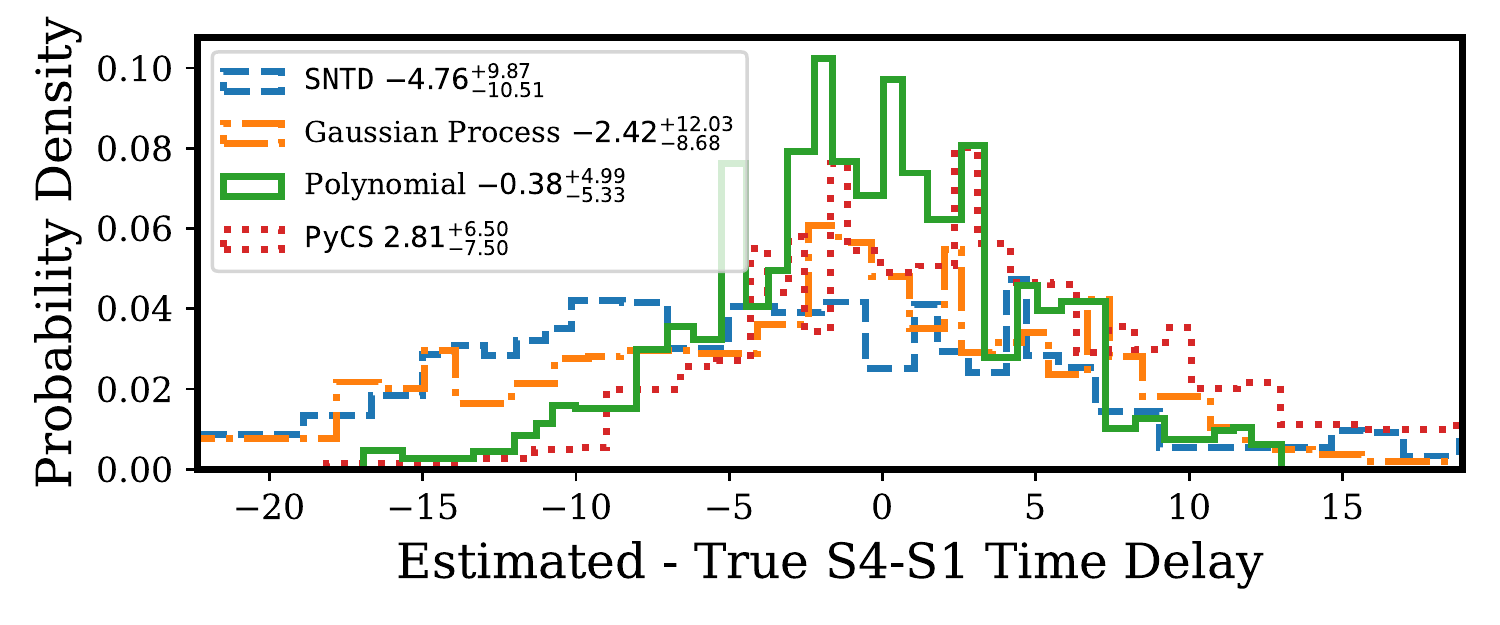}{0.49\textwidth}{}  
\fig{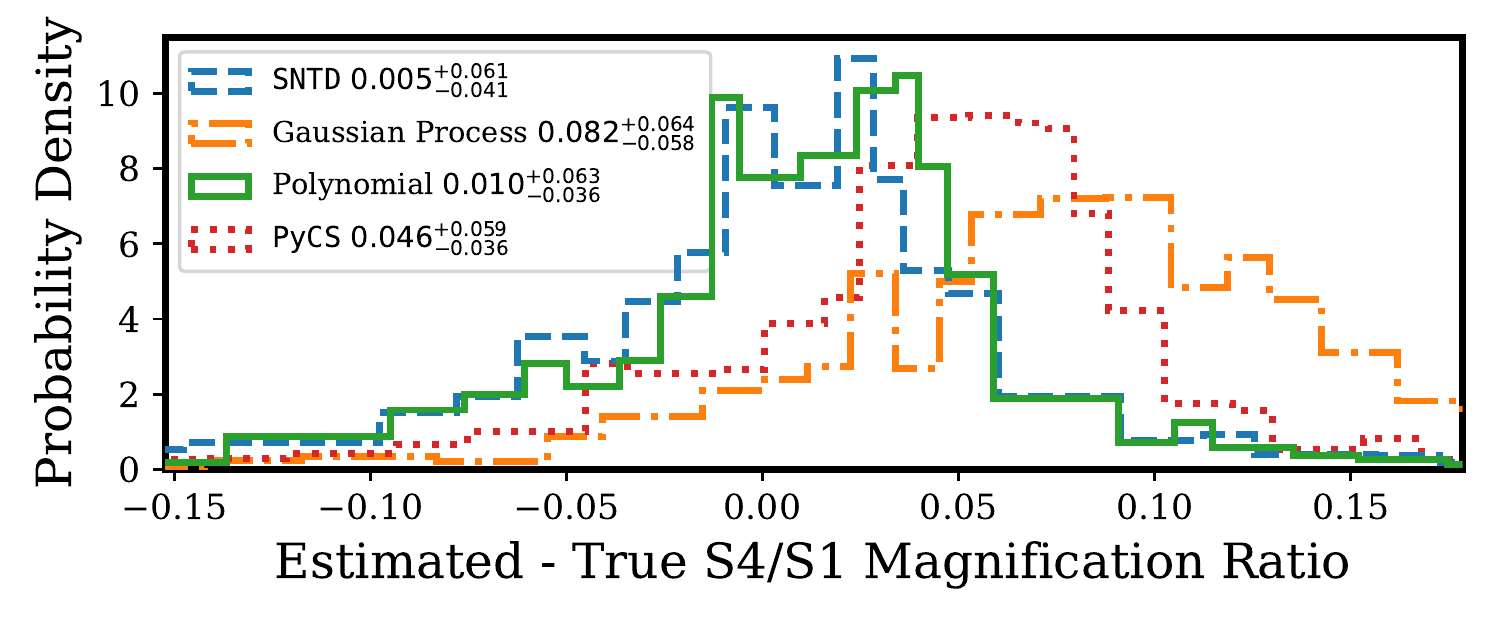}{0.49\textwidth}{}}
\vspace{-3em}
\gridline{\fig{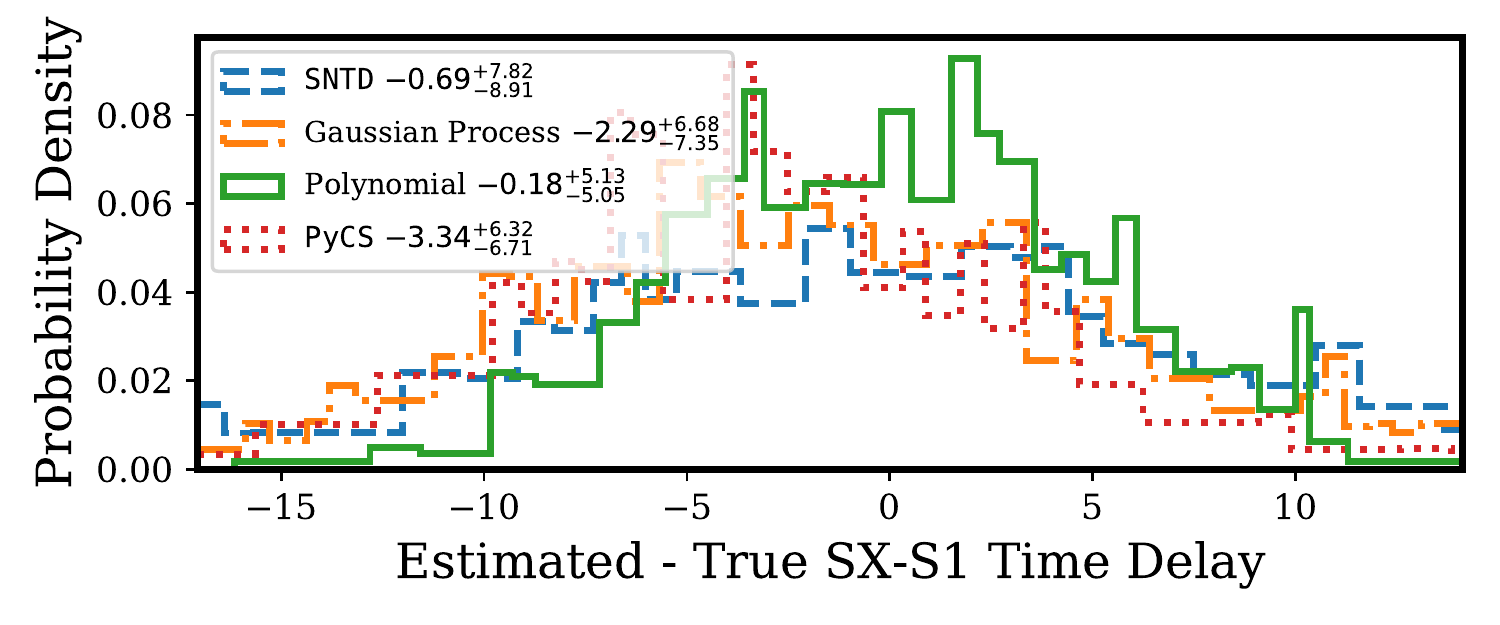}{0.49\textwidth}{}  
\fig{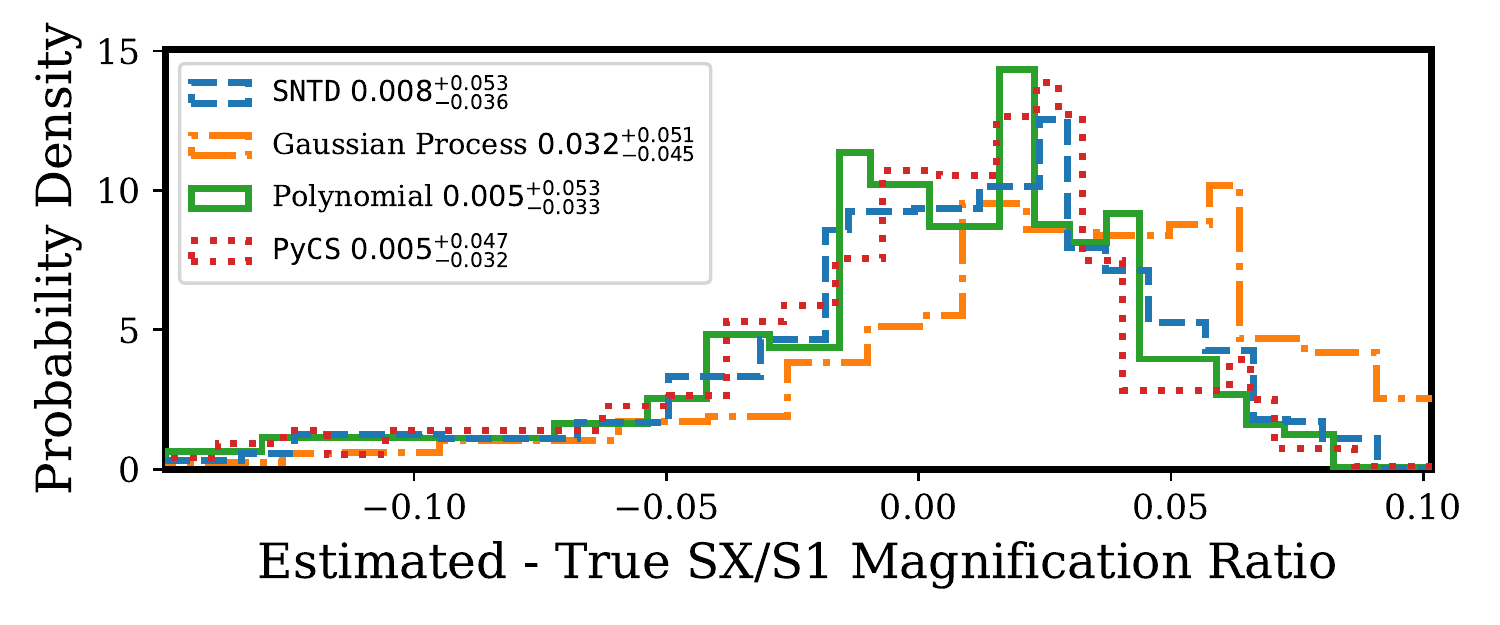}{0.49\textwidth}{}}
\vspace{-3em}
\caption{Distributions of residuals of time delay (left) and magnification ratio (right) estimated from simulated light curves by four separate algorithms. As indicated, each row shows results from the simulated data that mimics one pair of images, from S2:S1 in the top tow to SX:S1 in the bottom row.
The 16th, 50th, and 84th percentiles are listed in the legends.
}
\label{fig:residuals}
\end{figure*}

Each of the four methods described in Section~\ref{sec:fitting}, when
given the light curves of images S1--SX,
can yield a single estimate for eight lensing observables $\mathcal{O}$. These observables consist of the four time delays between images S2, S3, S4, and SX relative to
S1  ($\Delta t_{2,1}, ..., \Delta t_{x,1}$), and the four respective magnification ratios ($\mu_2 / \mu_1 , ..., \mu_X / \mu_1$).  The choice of image S1 as the
reference image is arbitrary. 
In order to combine the estimates from multiple methods with appropriate weighting for each method, we need a consistent way to define a probability distribution $p(\mathcal{O} | D; M_k)$ 
for each method $M_k$ given the light-curve data $D$.
We address this by performing fits to the 1000 simulated light curves, which were described in Section~\ref{sec:SimulatedLightCurves}.

\begin{deluxetable*}{c|cccc|cccc}
\label{tab:RMSscatter}
  \tablecolumns{9}
  \tablecaption{\sc RMS Scatter}
 \tablehead{ Method & \multicolumn{4}{c|}{Time Delay} & \multicolumn{4}{c}{Magnif. Ratio}  \\  & S2--S1 & S3--S1 & S4--S1 & SX--S1 & 
 S2/S1 & S3/S1 & S4/S1 & SX/S1  }
  \startdata
Polynomial &  \rmsstwosonedelayKellymedianchabrier & \rmssthreesonedelayKellymedianchabrier & \rmssfoursonedelayKellymedianchabrier & \rmssxsonedelayKellymedianchabrier & \rmsstwosonemuKellymedianchabrier & \rmssthreesonemuKellymedianchabrier & \rmssfoursonemuKellymedianchabrier & \rmssxsonemuKellymedianchabrier\\
Gaussian Process & \rmsstwosonedelayThorpmedianchabrier & \rmssthreesonedelayThorpmedianchabrier & \rmssfoursonedelayThorpmedianchabrier & \rmssxsonedelayThorpmedianchabrier & \rmsstwosonemuThorpmedianchabrier & \rmssthreesonemuThorpmedianchabrier & \rmssfoursonemuThorpmedianchabrier & \rmssxsonemuThorpmedianchabrier  \\
{\tt SNTD} & \rmsstwosonedelayRPthreerdmedianchabrier & \rmssthreesonedelayRPthreerdmedianchabrier & \rmssfoursonedelayRPthreerdmedianchabrier & \rmssxsonedelayRPthreerdmedianchabrier & \rmsstwosonemuRPthreerdmedianchabrier & \rmssthreesonemuRPthreerdmedianchabrier & \rmssfoursonemuRPthreerdmedianchabrier & \rmssxsonemuRPthreerdmedianchabrier  \\
{\tt PyCS} & \rmsstwosonedelayMillonsplmedianchabrier & \rmssthreesonedelayMillonsplmedianchabrier & \rmssfoursonedelayMillonsplmedianchabrier & \rmssxsonedelayMillonsplmedianchabrier & \rmsstwosonemuMillonsplmedianchabrier & \rmssthreesonemuMillonsplmedianchabrier & \rmssfoursonemuMillonsplmedianchabrier & \rmssxsonemuMillonsplmedianchabrier  \\
\enddata
\tablecomments{The RMS scatters in the time delay columns are given in units of days.  The magnification ratio columns  are unitless, and report the scatter of flux magnification ratios (i.e., no conversion into magnitudes has been done). When computing the scatter, we have excluded $>30$ day outliers for the time delays, and 1.5 for the magnification.}
\end{deluxetable*}

\begin{deluxetable*}{c|cccc|cccc}
\label{tab:OutlierFraction}
  \tablecolumns{9}
  \tablecaption{\sc $>3\sigma$ Outlier Fraction}
 \tablehead{ Method & \multicolumn{4}{c|}{Time Delay} & \multicolumn{4}{c}{Magnif. Ratio}  \\  & S2--S1 & S3--S1 & S4--S1 & SX--S1 & 
 S2/S1 & S3/S1 & S4/S1 & SX/S1  }
  \startdata
Polynomial &  \outlierfractionstwosonedelayKellymedianchabrier & \outlierfractionsthreesonedelayKellymedianchabrier & \outlierfractionsfoursonedelayKellymedianchabrier & \outlierfractionsxsonedelayKellymedianchabrier & \outlierfractionstwosonemuKellymedianchabrier & \outlierfractionsthreesonemuKellymedianchabrier & \outlierfractionsfoursonemuKellymedianchabrier & \outlierfractionsxsonemuKellymedianchabrier\\
Gaussian Process & \outlierfractionstwosonedelayThorpmedianchabrier & \outlierfractionsthreesonedelayThorpmedianchabrier & \outlierfractionsfoursonedelayThorpmedianchabrier & \outlierfractionsxsonedelayThorpmedianchabrier & \outlierfractionstwosonemuThorpmedianchabrier & \outlierfractionsthreesonemuThorpmedianchabrier & \outlierfractionsfoursonemuThorpmedianchabrier & \outlierfractionsxsonemuThorpmedianchabrier  \\
{\tt SNTD} & \outlierfractionstwosonedelayRPthreerdmedianchabrier & \outlierfractionsthreesonedelayRPthreerdmedianchabrier & \outlierfractionsfoursonedelayRPthreerdmedianchabrier & \outlierfractionsxsonedelayRPthreerdmedianchabrier & \outlierfractionstwosonemuRPthreerdmedianchabrier & \outlierfractionsthreesonemuRPthreerdmedianchabrier & \outlierfractionsfoursonemuRPthreerdmedianchabrier & \outlierfractionsxsonemuRPthreerdmedianchabrier  \\
{\tt PyCS} & \outlierfractionstwosonedelayMillonsplmedianchabrier & \outlierfractionsthreesonedelayMillonsplmedianchabrier & \outlierfractionsfoursonedelayMillonsplmedianchabrier & \outlierfractionsxsonedelayMillonsplmedianchabrier & \outlierfractionstwosonemuMillonsplmedianchabrier & \outlierfractionsthreesonemuMillonsplmedianchabrier & \outlierfractionsfoursonemuMillonsplmedianchabrier & \outlierfractionsxsonemuMillonsplmedianchabrier  \\
\enddata
\tablecaption{Outliers are defined as time-delay measurements that differ from the mean by more than 30 days, or measured magnification
ratios that deviate by more than 1.}
\end{deluxetable*}

For the $i$th simulated light curve, we calculate the residual of each {\it input} value $\mathcal{O}_{{\rm input}; i }$ from the {\it estimated} value $\mathcal{O}_{{\rm est}; i}$,
\begin{equation}
\delta \mathcal{O}_{{\rm resid}; i} = \mathcal{O}_{{\rm est}; i} - \mathcal{O}_{ {\rm input}; i}.
\label{eq:prob_dist}
\end{equation}
We use a histogram of these residuals to construct a probability distribution 
$P_{\rm resid}$.

We next compare the ability of the four fitting methods to recover the input relative time delays. As shown in Figure~\ref{fig:residuals}, the polynomial method has the least bias, and the 16th, 50th, and 84th percentiles of its residual distribution are most closely separated. The root-mean-square (RMS) scatter of the polynomial method, as listed in Table~\ref{tab:RMSscatter}, is also smallest. In some cases, the performance of the Gaussian Process and {\tt PyCS} fitting methods approaches that of the polynomial method. The light curve of S4 has a lower signal-to-noise ratio than those of the other images, however, and in that case, both methods are not as effective in recovering its relative time delay as the polynomial method.

Considering the codes' ability to recover magnification ratios, Figure~\ref{fig:residuals} and Table~\ref{tab:RMSscatter} demonstrate that all fitting codes yield comparably accurate and precise magnification ratios S2/S1 and S3/S1. In the case of the S4/S1 ratio, the Gaussian Process and {\tt PyCS} methods show a greater bias than the polynomial and {\tt SNTD} methods. 
In Table~\ref{tab:OutlierFraction}, we report the fraction of simulated light curves for which the difference between measurement and truth ($\delta\mathcal{O}_{{\rm resid};i}$) is more than 30 days (for time delays) or $>1$ (for magnification ratios).

The IMF of the stellar populations in the early-type galaxy lens \citep[e.g.,][]{treu10,conroyvandokkum12} and the intracluster medium may have a \citet{salpeter55} IMF instead of the \citet{cha03} IMF that we have assumed for our simulations. 
In Tables~\ref{tab:salpeterRMSscatter} and \ref{tab:salpeterOutlierFraction}, we report the same set of statistics instead using microlensing simulations computed for a \citet{salpeter55} IMF. 
In Figure~\ref{fig:residuals_salpeter}, we plot the residual distributions given a \citet{salpeter55} IMF.
The residual distributions for simulations that adopt a Salpeter IMF show a very modest increase ($\lesssim 10$\%) in the difference between the 16th and 84th percentiles in the case of the SX--S1 time delay, and larger ($\sim 30$\%) increases for the S2--S1, S3--S1, and S4--S1 images.

We next want to compute the probability distribution for a property of the light curve $\mathcal{O}$, given a single-point estimate of the same observable $\mathcal{O}_{{\rm est;D};k}$, derived from the actual light-curve data $D$ using method $M_k$, to compute this probability distribution.
To accomplish this, we use the distribution of residuals $P(-\delta \mathcal{O}_{{\rm resid}})$
assembled from fits to the simulations.
This probability distribution $P(\mathcal{O} | D; M_k)$ is given by

\begin{equation}
P(\mathcal{O} | D; M_k) = P(\mathcal{O} | \mathcal{O}_{{\rm est;D};k}) = P_{\rm resid}(\mathcal{O} - \mathcal{O}_{{\rm est;D};k}).
\end{equation}

\subsection{Results from Bootstrapping Samples}
\label{sec:Bootstraps}
We computed the relative time delays and magnifications of the SN light curves after creating one thousand light curves after performing bootstrapping with replacement. Since the uncertainties we measured were smaller than those we computed from the simulated light curves, which included perturbations from the piecewise polynomial model, microlensing, and millilensing, we adopt the results from the simulations. 

\subsection{Combination of Estimates from Distinct Light-Curve Algorithms}
\label{sec:Combining}

For each observable, we sought to identify an optimal combination of the set of four probability distributions $P(\mathcal{O} | D; M_k)$ corresponding to the respective methods. 
We find the non-negative linear combination of probability distributions $P(\mathcal{O} | D; M_k)$ that {\it maximizes the likelihood} of the input observable values used to construct the simulations,
\begin{equation}
\max_{w \in \mathcal{S}_1^K} \frac{1}{n} \sum_{i=1}^{n} \log \sum_{k=1}^K w_k P(\mathcal{O}_{{\rm input}; i} | D_i; M_k),
\label{eq:opt}
\end{equation}
subject to $\mathcal{S}_1^K = \{ w \in [0,1]^K : \sum_{k=1}^K w_k = 1\}$.
When computing these weights using fits to the simulated light curves, we use a modified probability function $P_{\rm penal}(\mathcal{O} | D; M_k)$ that has the effect of penalizing bias. The penalized probability distribution is given by

\begin{equation}
P_{\rm penal}(x) = \\
P_{\rm resid}(x - {\rm median}(\Delta \mathcal{O}_{{\rm resid}})).
\end{equation}
For a distribution where the median residual, ${\rm median}(\delta \mathcal{O}_{{\rm resid}})$, is equal to zero, then $P_{\rm penal} = P_{\rm resid}$. 
Given best-fitting weights $w_k$, the probability distribution for the actual light curve becomes
\begin{equation}
P(\mathcal{O} | D ) = \sum_{k=1}^{K} w_k P(\mathcal{O}|D; M_k).
\label{eq:weighted_sum}
\end{equation}

We note that maximizing the likelihood of the relative time delays and magnifications used to construct the simulated light curve  corresponds to minimizing the Kullback-Leibler divergence between a single-valued distribution where the residual $\delta \mathcal{O}_{{\rm resid}; i} = 0$ and the stacked set of probability distributions $p(\mathcal{O} | D )$.  The use of the Kullback-Leibler divergence as a means to identify an optimal combination instead of {\it predictive} models was developed by \citet{yaoyulingvehtari18}.

The Kullback-Leibler divergence allows us to consider full probability distributions, as opposed to single-point estimates. For this analysis, we use the residual distribution for all simulated light curves to construct the probability for every light curve as described in Equation~\ref{eq:prob_dist}, the framework we describe could accommodate a separate probability distribution for each simulated light curve.

To calculate the likelihood of the input light-curve parameters
in Equation~\ref{eq:opt}, 
we need to choose how to construct a probability density function from the set of residuals $\delta \mathcal{O}_{{\rm resid}; i}$. The probability density must always be nonzero, with the bin widths sufficiently large to avoid small-number statistics. We create bins such that each bin includes three residuals (i.e., each contains three values) in each of the $<5$\% and $>95$\% quantiles, 15 residuals in the 5--30\% and 70--95\% quantiles, and 30 residuals in the 30--70\% quantiles. Moreover, we adjust these bins so that the bin width is at least 0.5 days for the time delays, or 0.01 for the magnification ratios.

For each observable, we carry out an optimization using Equation~\ref{eq:opt} to identify the weights that maximize the likelihood of the parameters used to construct the light curves.  Table~\ref{tab:methodweights} lists the weights that we obtain. The piecewise polynomial method receives the majority of the weight for all of the relative time-delay measurements, while multiple codes receive majority weight for the estimates of magnification ratios.
Marginal probability distributions from the combination of four methods are shown as a ``corner plot” in Figure~\ref{fig:corner}.

In Figure~\ref{fig:tension}, we show the separate constraints from all the methods. 

\begin{deluxetable*}{c|cccc|cccc}
  \tablecolumns{9}
  \tablecaption{\sc Weights of Methods to Maximize Likelihood of True Parameters}
 \tablehead{ Method & \multicolumn{4}{c|}{Time Delay} & \multicolumn{4}{c}{Magnif. Ratio}  \\  & S2--S1 & S3--S1 & S4--S1 & SX--S1 & 
 S2/S1 & S3/S1 & S4/S1 & SX/S1  }
  \startdata
Polynomial & \weightstwosonedelayKellymedianchabrier & \weightsthreesonedelayKellymedianchabrier & \weightsfoursonedelayKellymedianchabrier & \weightsxsonedelayKellymedianchabrier & \weightstwosonemuKellymedianchabrier & \weightsthreesonemuKellymedianchabrier & \weightsfoursonemuKellymedianchabrier & \weightsxsonemuKellymedianchabrier  \\
Gaussian Process & \weightstwosonedelayThorpmedianchabrier & \weightsthreesonedelayThorpmedianchabrier & \weightsfoursonedelayThorpmedianchabrier & \weightsxsonedelayThorpmedianchabrier & \weightstwosonemuThorpmedianchabrier & \weightsthreesonemuThorpmedianchabrier & \weightsfoursonemuThorpmedianchabrier & \weightsxsonemuThorpmedianchabrier  \\
{\tt SNTD} & \weightstwosonedelayRPthreerdmedianchabrier & \weightsthreesonedelayRPthreerdmedianchabrier & \weightsfoursonedelayRPthreerdmedianchabrier & \weightsxsonedelayRPthreerdmedianchabrier & \weightstwosonemuRPthreerdmedianchabrier & \weightsthreesonemuRPthreerdmedianchabrier & \weightsfoursonemuRPthreerdmedianchabrier & \weightsxsonemuRPthreerdmedianchabrier  \\
{\tt PyCS} & \weightstwosonedelayMillonsplmedianchabrier & \weightsthreesonedelayMillonsplmedianchabrier & \weightsfoursonedelayMillonsplmedianchabrier & \weightsxsonedelayMillonsplmedianchabrier & \weightstwosonemuMillonsplmedianchabrier & \weightsthreesonemuMillonsplmedianchabrier & \weightsfoursonemuMillonsplmedianchabrier & \weightsxsonemuMillonsplmedianchabrier  \\
\enddata
\label{tab:methodweights}
\tablecomments{Weight assigned to each method determined by maximizing the likelihood of each input time delay or magnification.}
\end{deluxetable*}

\section{Relative Time Delays and Magnification Ratios}
\label{sec:td_mr}

With a procedure for optimally combining estimates from the light-curve fitting methods and estimating uncertainty, we turn to measurements of the relative time delays and magnification ratios from the SN Refsdal light-curve data. 
We used {\tt DOLPHOT} photometry, restricted to data acquired after 24 December 2013 (MJD = 56650) to fit the 
light curves of S1--S4,  
and 10 October 2015 (MJD = 57300) for SX.  To enable the time-delay fitting to be blind, we generated a random number for each light curve, and added that random number to the Modified Julian Date corresponding to each photometric measurement.  Each of the four fitting methods was applied to the shifted light curves, and combined using the weighted sum described in Equation~\ref{eq:weighted_sum}.

\begin{figure*}
\centering
\gridline{
\fig{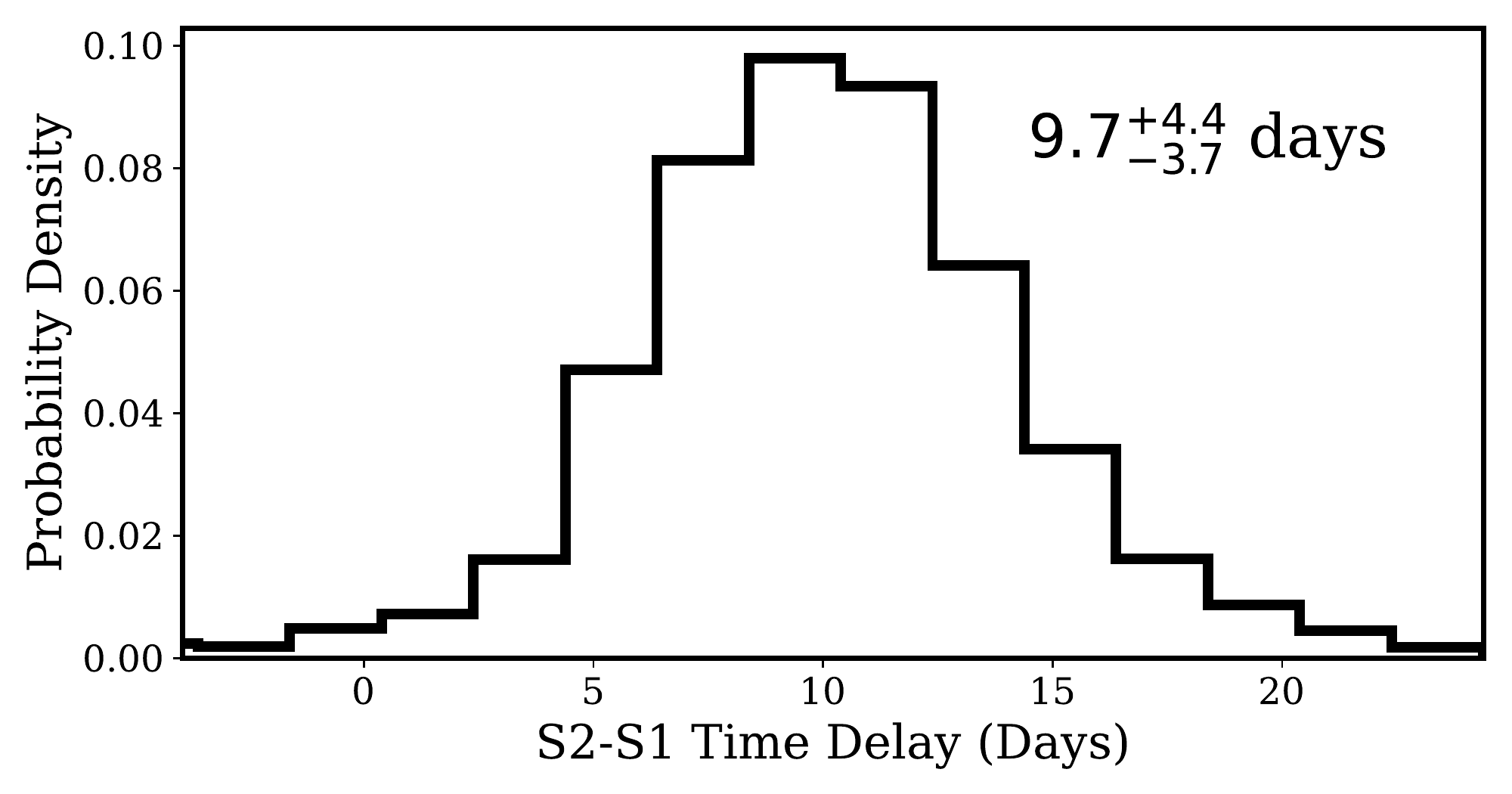}{0.49\textwidth}{}
\fig{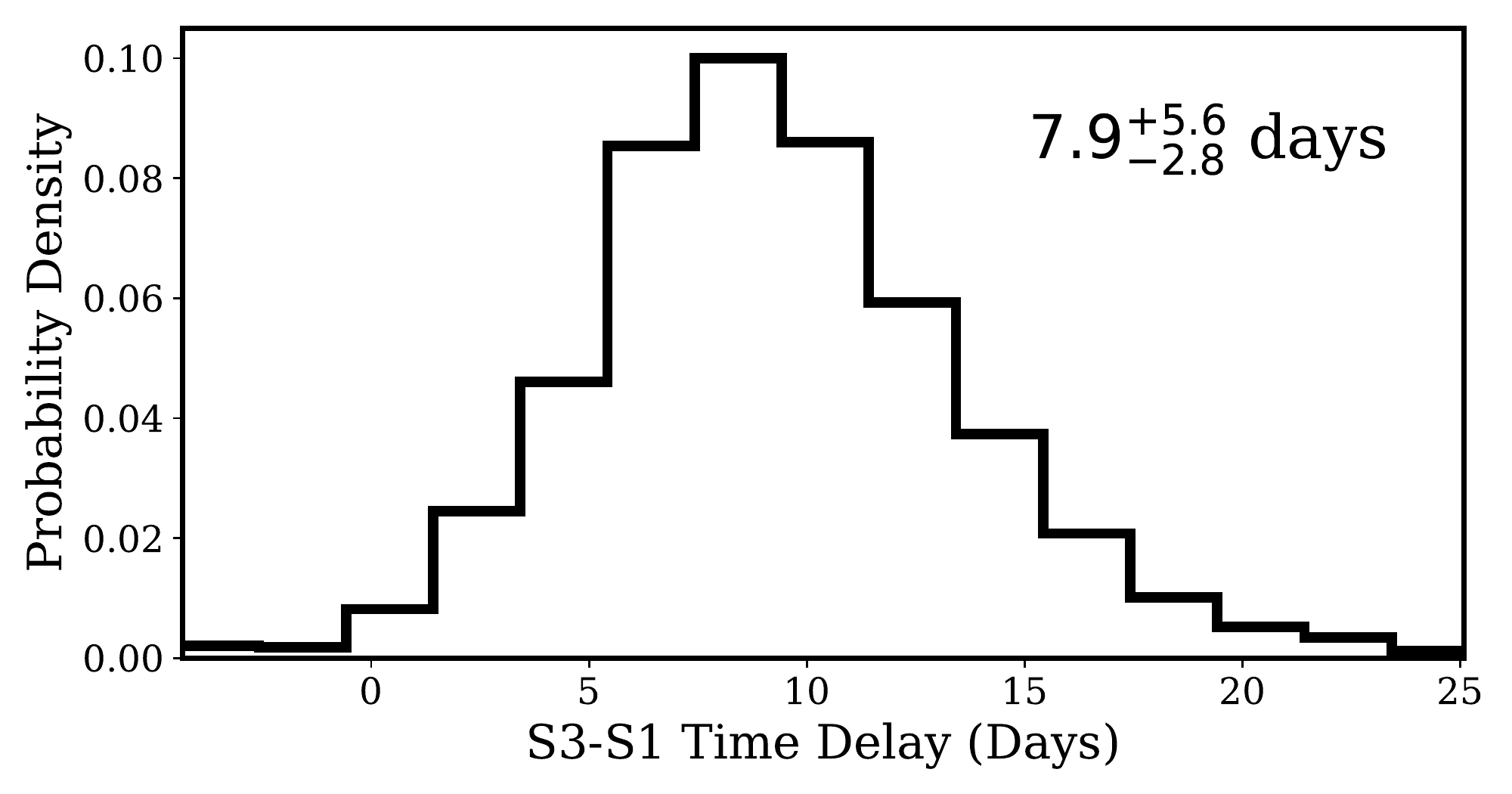}{0.49\textwidth}{}}
\vspace{-3em}
\gridline{
\fig{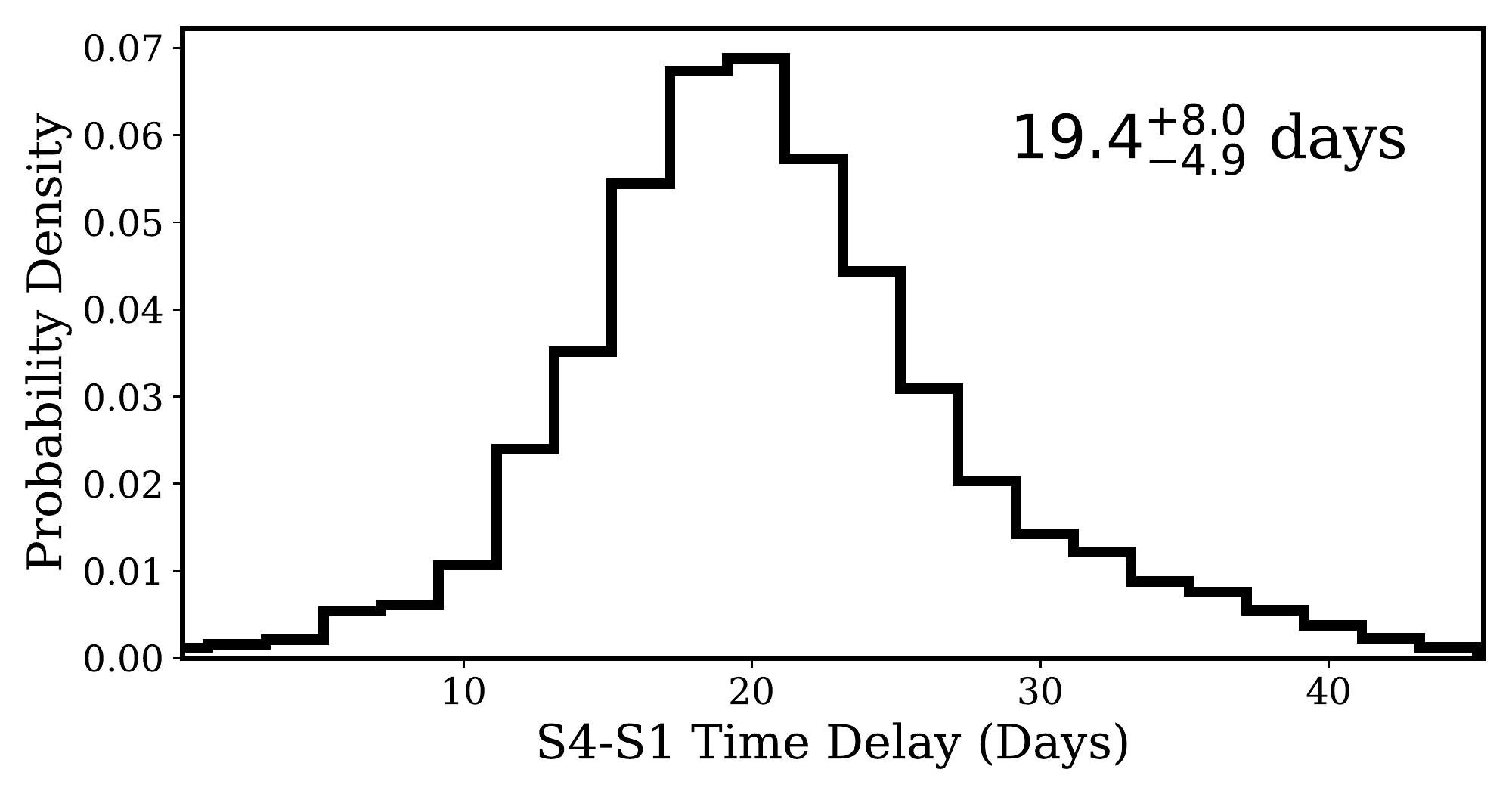}{0.49\textwidth}{}
\fig{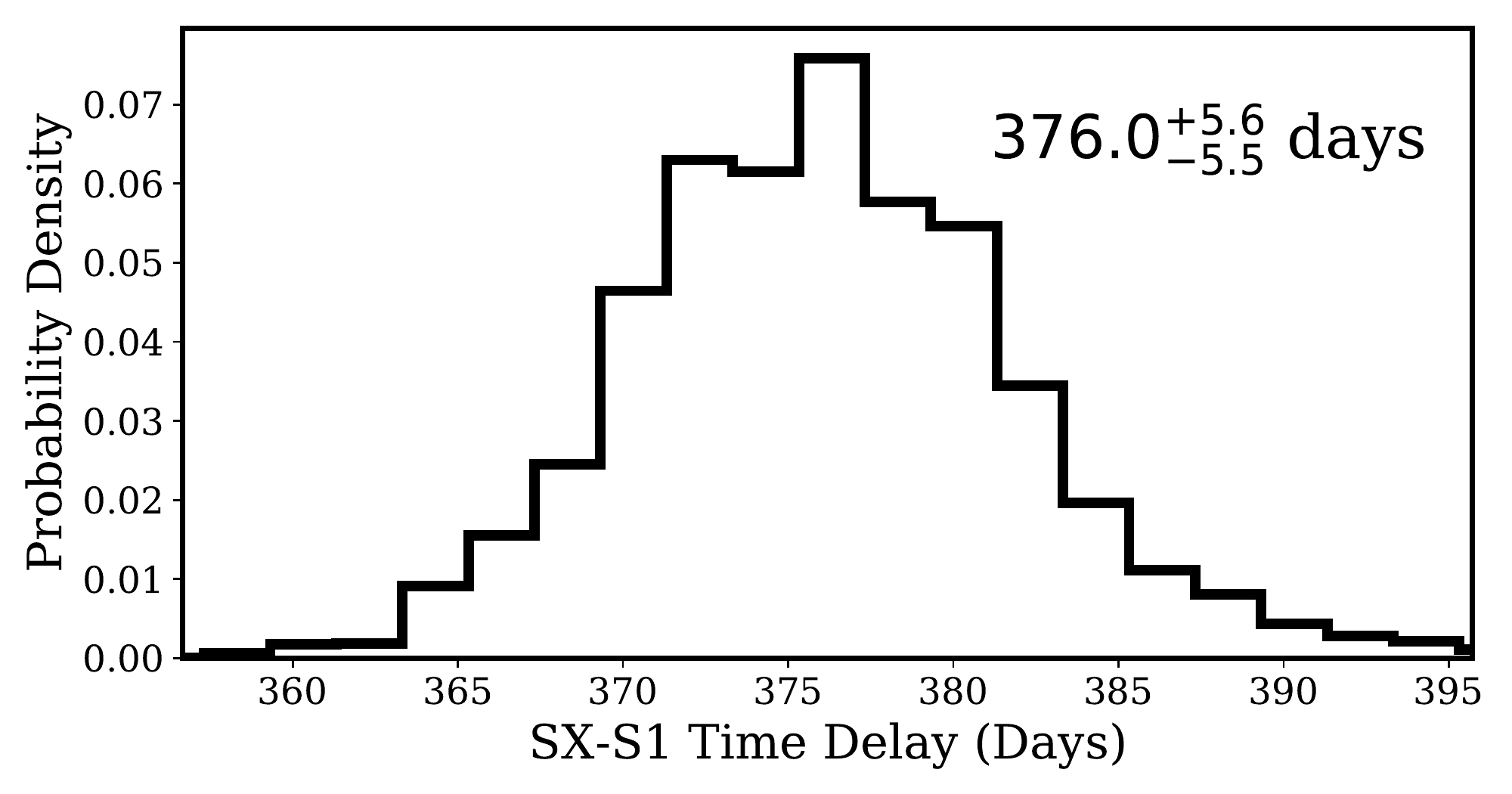}{0.49\textwidth}{}}
\vspace{-2em}
\caption{Relative time delays between images S2--SX and S1. Constraints shown are the weighted combination of results from four different algorithms where weights are determined by maximizing the likelihood of the actual time delays in simulated light curves. The values in the upper right-hand corner are the 16th percentile confidence level, maximum-likelihood value, and 84th percentile confidence level, as listed in Table~\ref{tab:measurementsdelaycorr}. }
\label{fig:delay_combined}
\end{figure*}

\begin{figure*}[tbp]
\centering
\gridline{\fig{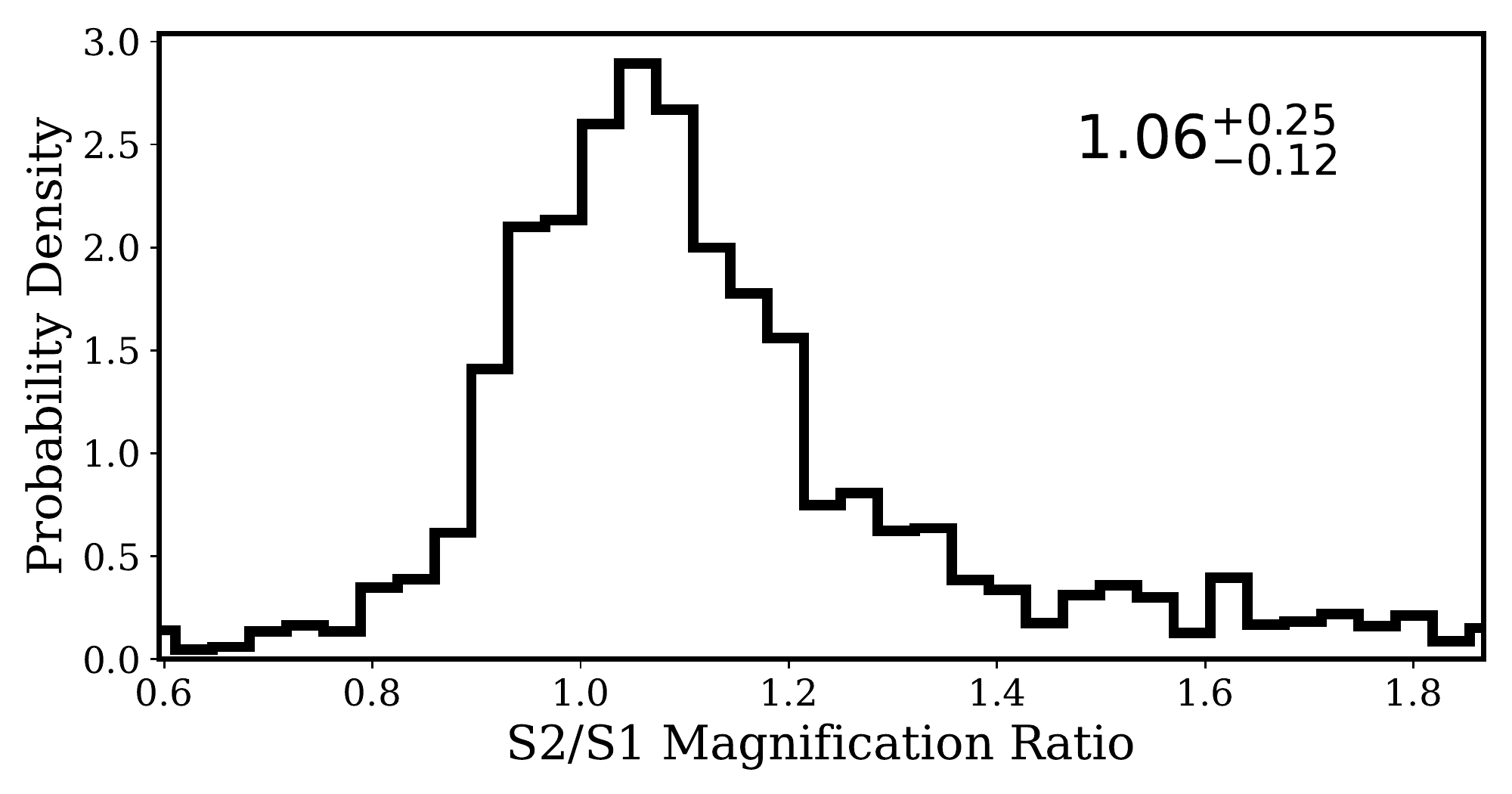}{0.49\textwidth}{}
\fig{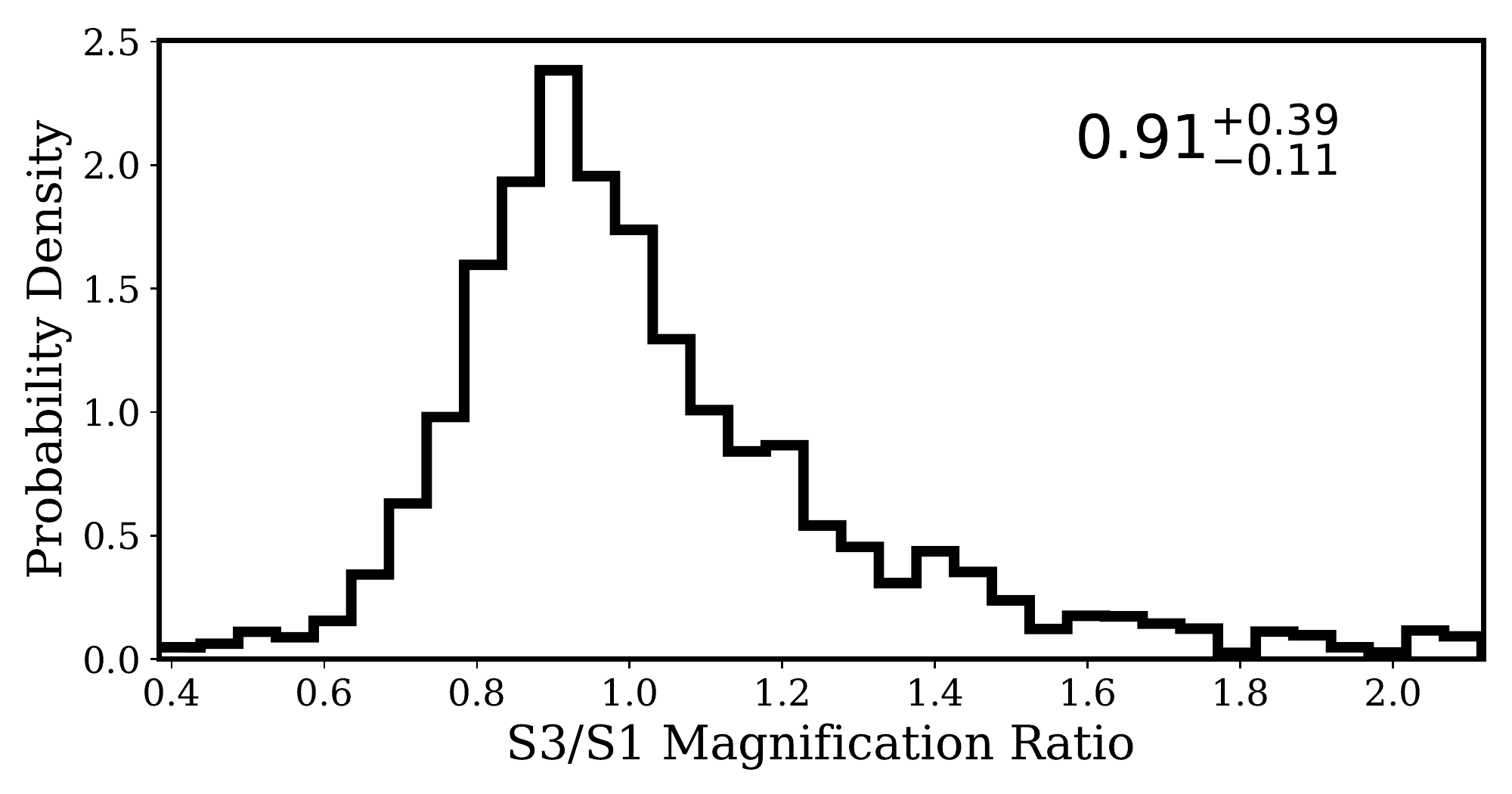}{0.49\textwidth}{}}
\vspace{-3em}
\gridline{
\fig{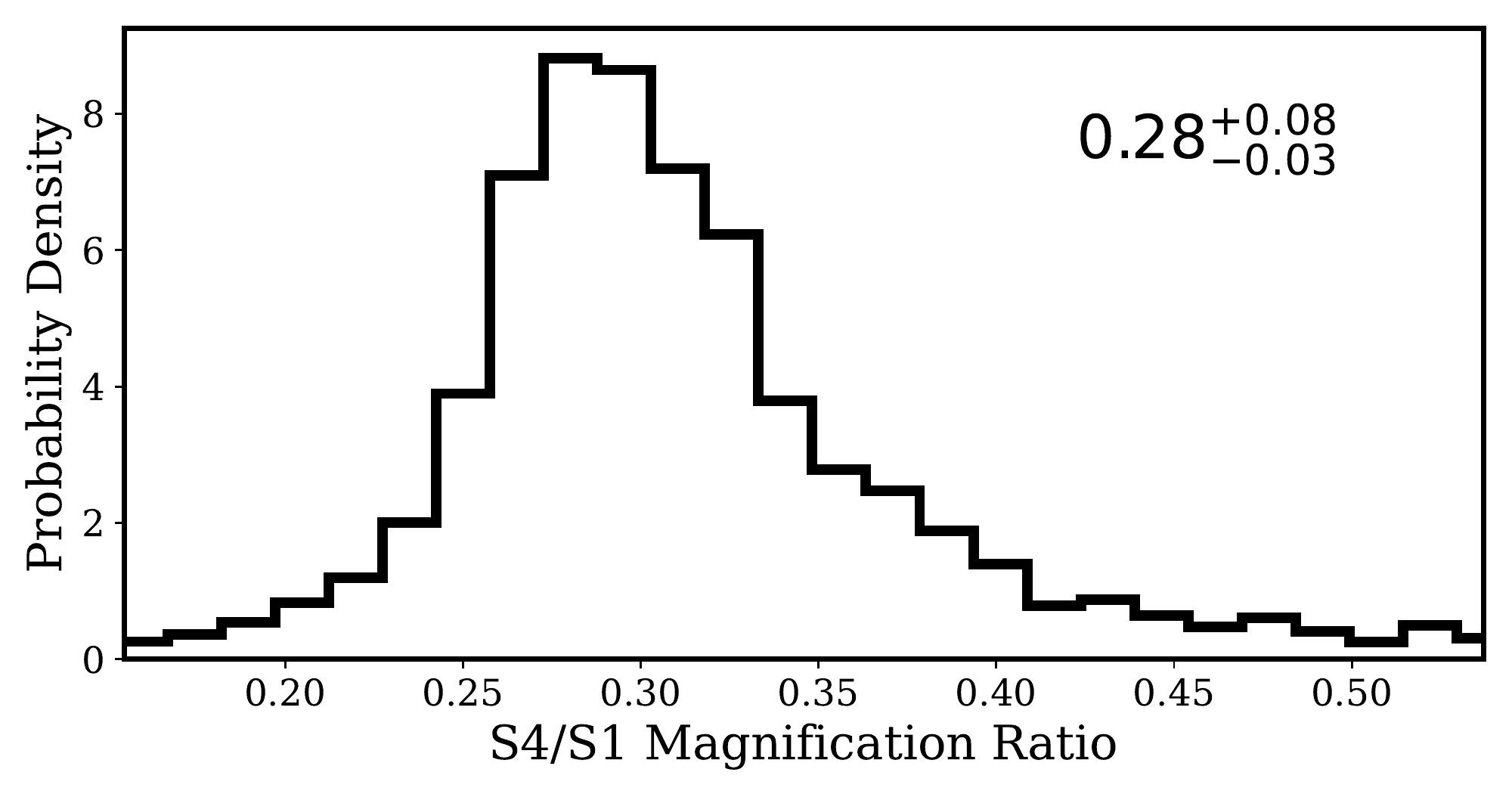}{0.49\textwidth}{}
\fig{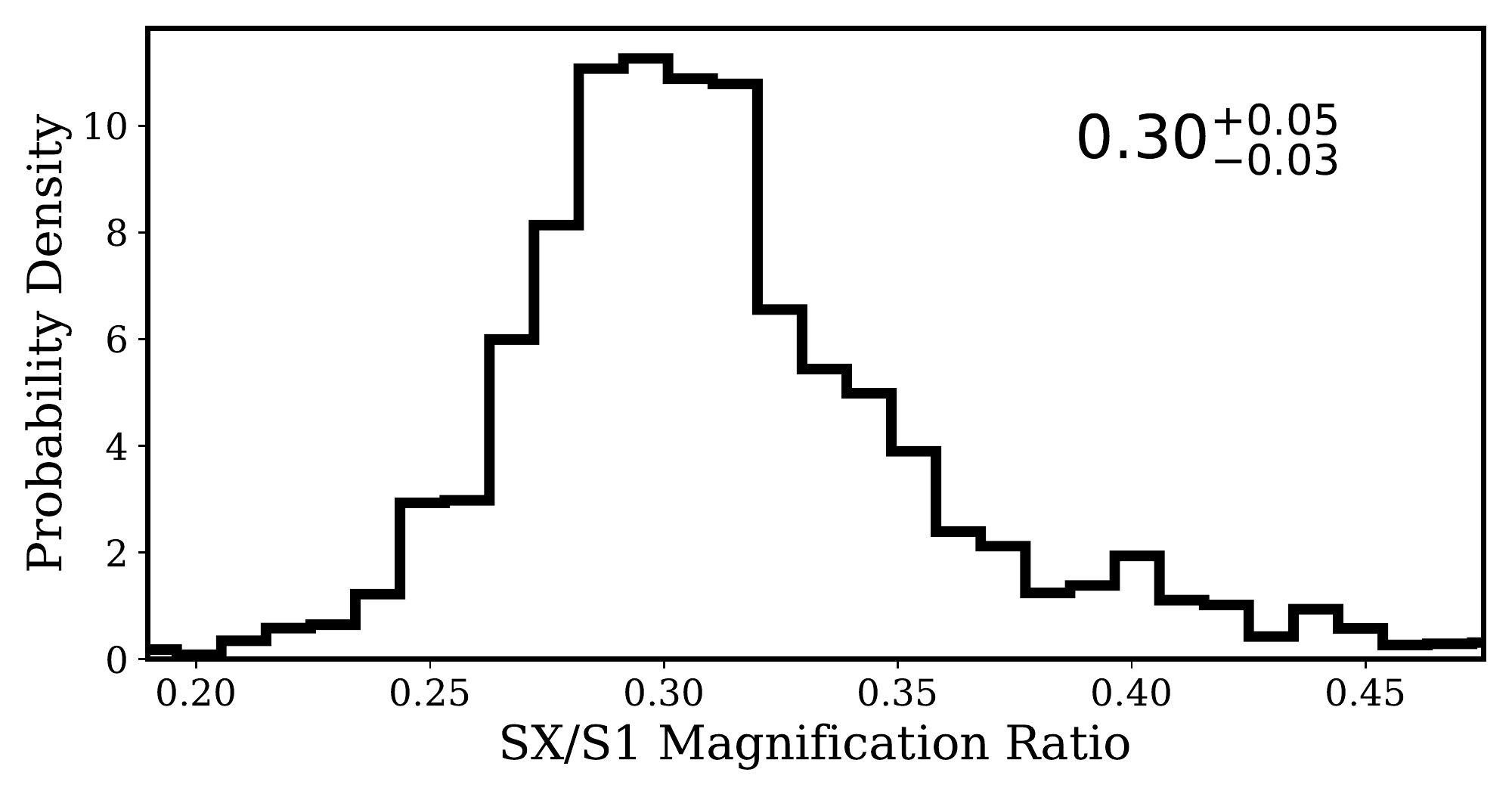}{0.49\textwidth}{}}
\vspace{-2em}
\caption{Magnification ratios between images S2--SX and S1. Constraints shown are a weighted combination of results from four different algorithms, where weights are determined by maximizing the likelihood of the actual magnification ratios in simulated light curves. The values in the upper right-hand corner are the 16th percentile confidence level, maximum-likelihood value, and 84th percentile confidence level, as listed in Table~\ref{tab:measurementsmucorr}.}
\label{fig:mu_combined}
\end{figure*}

\begin{figure*}[tbp]
\centering
\includegraphics[width=\textwidth]{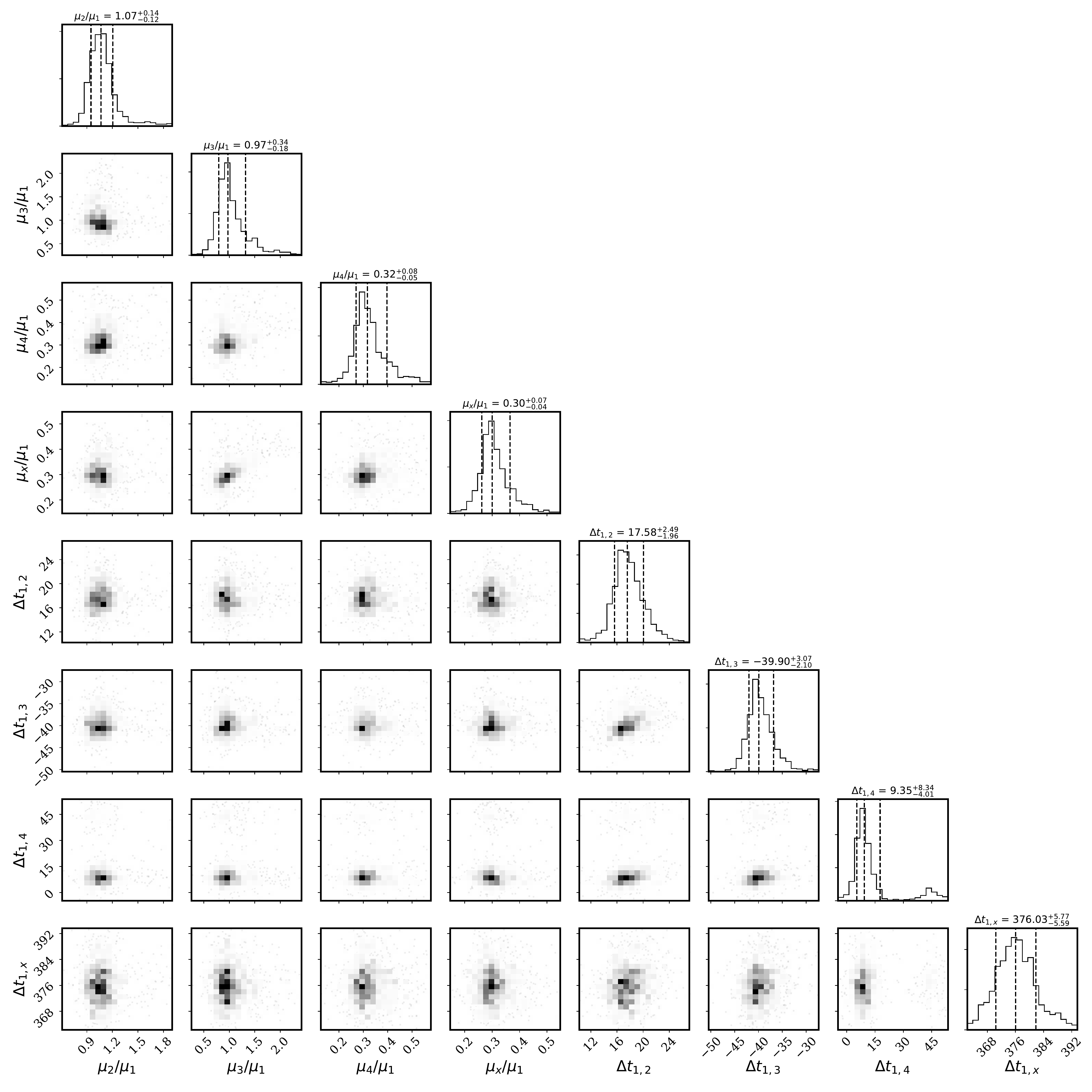}
\caption{Relative time delays and magnification ratios among images S2--SX and S1. Constraints shown are weighted combination of results from four different algorithms, where weights are determined by maximizing the likelihood of the actual time delay used to construct simulated light curves. The values shown above each histogram correspond to the 16th, 50th, and 84th percentiles. }
\label{fig:corner}
\end{figure*}

In Figure~\ref{fig:delay_combined}, we plot the combined constraints on the time delays relative to image S1.
The combined constraints on the magnification ratios between these sets of images 
are shown in Figure~\ref{fig:mu_combined}. In Tables~\ref{tab:measurementsdelaycorr} and \ref{tab:measurementsmucorr}, we list the measured maximum likelihood values and the 2.27, 16, 50, 84, and 97.73 percentiles inferred from these probability distributions. Marginal  probability distributions from the combination of four methods are shown as a ``corner plot'' in
Figure~\ref{fig:corner}.

\begin{figure*}
\centering
\gridline{\fig{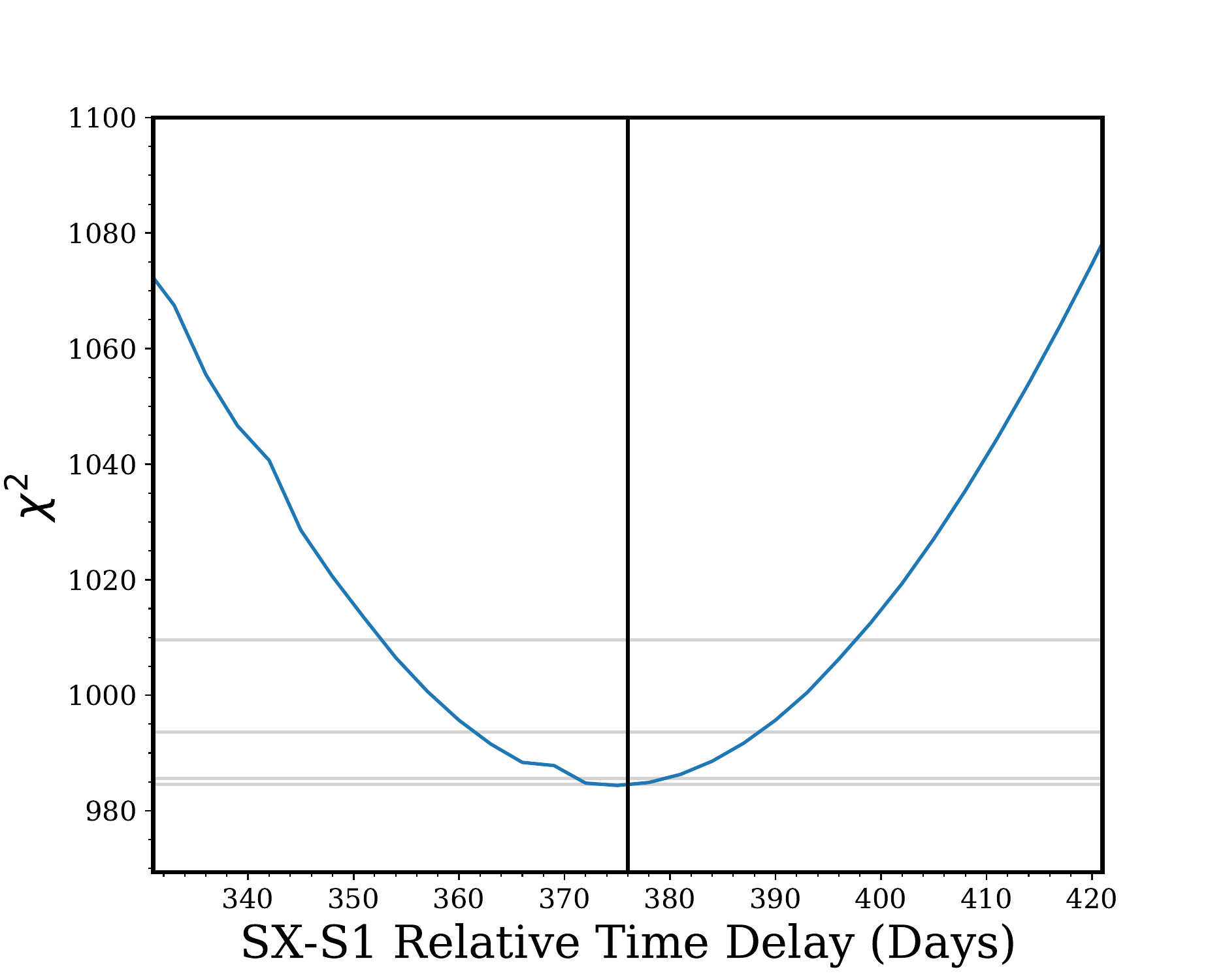}{0.65\textwidth}{}}
\caption{The $\chi^2$ value of the best-fitting piecewise polynomial model as a function of the relative time delay between SX and S1. Here we hold the value of the SX--S1 time delay fixed while allowing all other time-delay and magnification ratio parameters for all images, as well as the parameters of the light curve model to vary. The $\chi^2$ values of the best-fitting model that we obtain for each value of the SX--S1 time delay are plotted above in blue. The black vertical line shows the most probable value of the SX--S1 time delay of 376\,days. The gray horizontal lines show, from bottom to top, the minimum $\chi^2$ value of 984.6, and added increments of 1, 9, and 25. }
\label{fig:chi_sq_sx_s1}
\end{figure*}

In Figs.~\ref{fig:visual_shift_one} and \ref{fig:visual_shift_two}, we plot the light curves of SX and those of S1, S2, and S3, the images in the Einstein cross with the greatest magnification, after removing SX--S1 relative time delays of 376 days (most probable value), 360 days (3$\sigma$ offset from 376 days), and 349 days (5$\sigma$ offset from 376 days). The plotted light curves of S2 and S3 are shifted by their time delays relative to S1 as listed in Tab.~\ref{tab:measurementsdelaycorr}. 
In Fig.~\ref{fig:chi_sq_sx_s1}, we show the $\chi^2$ value of the best-fitting piecewise polynomial model against the SX--S1 relative time delay.

\subsection{Comparison of Time Delays from F125W and F160W Light Curves}
\label{sec:separatelcs}
We next apply the piecewise polynomial modeling code to the F125W and F160W light curves separately in order to evaluate whether the inferred SX--S1 time delays are consistent.  To estimate the uncertainties associated with the measurement for each filter, we use the simulated sets of light curves in the respective filters. We find that the F125W light curve yields a constraint on the SX--S1 relative time delay of $377.6^{+5.13}_{-5.04}$\,days, while the F160W light curve, whose measurements have, on average, lower signal-to-noise ratios, yields 
$370.6^{+9.78}_{-10.07}$\,days. The estimates from the F125W and F160W light curves are in 0.6$\sigma$ agreement.

\subsection{Alternative Assumptions}
We explore the effects of alternative assumptions about the lens model, the mass density of star making up the ICL, and the symmetry of the SN photosphere. In Table~\ref{tab:lensparamsmore}, we list the convergence and shear at the positions of images S1--SX according to the Grillo-g and Sharon-g models. Table~\ref{tab:sxdelayvariants} lists the constraints on the SX--S1 time delay when we instead adopt the Grillo-g and Sharon-g lensing parameters. 
We also obtain inferences on the SX--S1 time delay after constructing an elliptical photosphere with a ratio of major axes of $b/a = 0.8$, and, separately, after doubling the inferred value of $\kappa_{\star}$, the density of the stars that account for the intracluster medium. Table~\ref{tab:sxdelayvariants} shows that adopting these alternative assumptions yields almost identical constraints on the relative time delay between images SX and S1.

\subsection{Covariance Matrix}
\label{sec:covariance}

We next derive a covariance matrix for $\Delta t$ and $\mu/\mu_1$. The covariance matrix can be used to compute likelihoods for model predictions, although we develop a more precise approach with less sensitivity to outlying values in the accompanying paper. 
To compute the measurement covariance matrix, we use the 1000 simulations of the light curves of images S1--SX, which include the effects of microlensing and photometric noise.  

We can compute the covariance matrix as
\begin{equation}
\operatorname{K}_{X_i X_j} = \operatorname{cov}[X_i, X_j] = \operatorname{E}[(X_i - \operatorname{E}[X_i])(X_j - \operatorname{E}[X_j])].
\end{equation}
Here we have $X$ and $Y$ correspond to the distributions of residuals of the estimated value subtracted from the actual value
for the simulation,
\begin{equation}
\delta \Delta t^{\rm est}_{i,1} = \Delta t^{\rm est}_{i,1} - \Delta t^{\rm true}_{i,1}.
\end{equation}
As described in Section~\ref{sec:Combining}, we have identified the weights $w_{j,{\rm obs}}$ for each model $M_j$ and observable such that the linear combination of the probability distributions,
\begin{equation}
p( \mathcal{O} | D ) = \sum_{j=1}^{K} w_{j,{\rm obs}}\,p( {\rm observable} | D; M_j),
\end{equation}
maximizes the likelihood of the value of the observable used to create the simulated light curve. 
Consequently, we have a set of weights $w_{j, {\rm obs}}$ where $\sum_j w_{j, {\rm obs}} = 1$ 
for each relative delay $\Delta t_{1,j}$ and magnification ratio $\mu_{k}/\mu_{1}$ between the $j$th image and S1.

In general, considering a discrete set of possible realizations $(x_i, y_i)$ of $(X,Y)$ with probability $p_i$,
the covariance is
\begin{equation}
\operatorname{cov} (X,Y)=\sum_{i=1}^n p_i (x_i-E(X)) (y_i-E(Y)).
\end{equation}
In our case, the probability $p_i$ depends on 
the fitting method from which the $x_i$ or $y_i$ are drawn.
The probability $p_i = (1/n_{\rm sim}) w_{i,X} \times w_{j,Y}$.

In our case, to create our realizations of the stacked probability distribution, we need to draw from 
the probability distributions.
The probability of each pair of realizations corresponds to the product of the weights $w_{i,X} \times w_{j,Y}$ that 
the respective fitting methods receive for the observable.
We can write
\begin{multline}
\operatorname{cov} (X,Y) =  
\sum_{i=1}^{n_{\rm sim}} \sum_{j=1}^{n_{\rm mod}} \sum_{k=1}^{n_{\rm mod}}  \frac{1}{n_{\rm sim}} w_{j,X} \times w_{k,Y} \times \\ (x_{i,j}-E(X)) (y_{i,k}-E(Y)),
\end{multline}

\begin{multline}
\operatorname{cov} (X,Y) =  
\sum_{j=1}^{n_{\rm mod}} \sum_{k=1}^{n_{\rm mod}} w_{j,X} \times w_{k,Y} \times \\ \sum_{i=1}^{n_{\rm sim}}  \frac{1}{n_{\rm sim}}  (x_{i,j}-E(X)) (y_{i,k}-E(Y)).
\end{multline}
While there may be model covariance between measured or predicted image separations and 
the time delay and magnification, we do not include it here.

Unlike the effects of microlensing, the effects of millilensing were not included when we created simulated light curves.  Since the Einstein radii of stars in the foreground cluster can be comparable in size to the expanding SN photosphere, microlensing can cause changes in the shape of the light curve as well as the mean magnification of each image \citep[e.g.,][]{doblerkeeton06}. Given the much greater masses of subhalo deflectors expected to be responsible for millilensing, their effect should be to change the mean magnification but not affect the shape of the light curve.

\begin{figure*}
\centering
\gridline{\fig{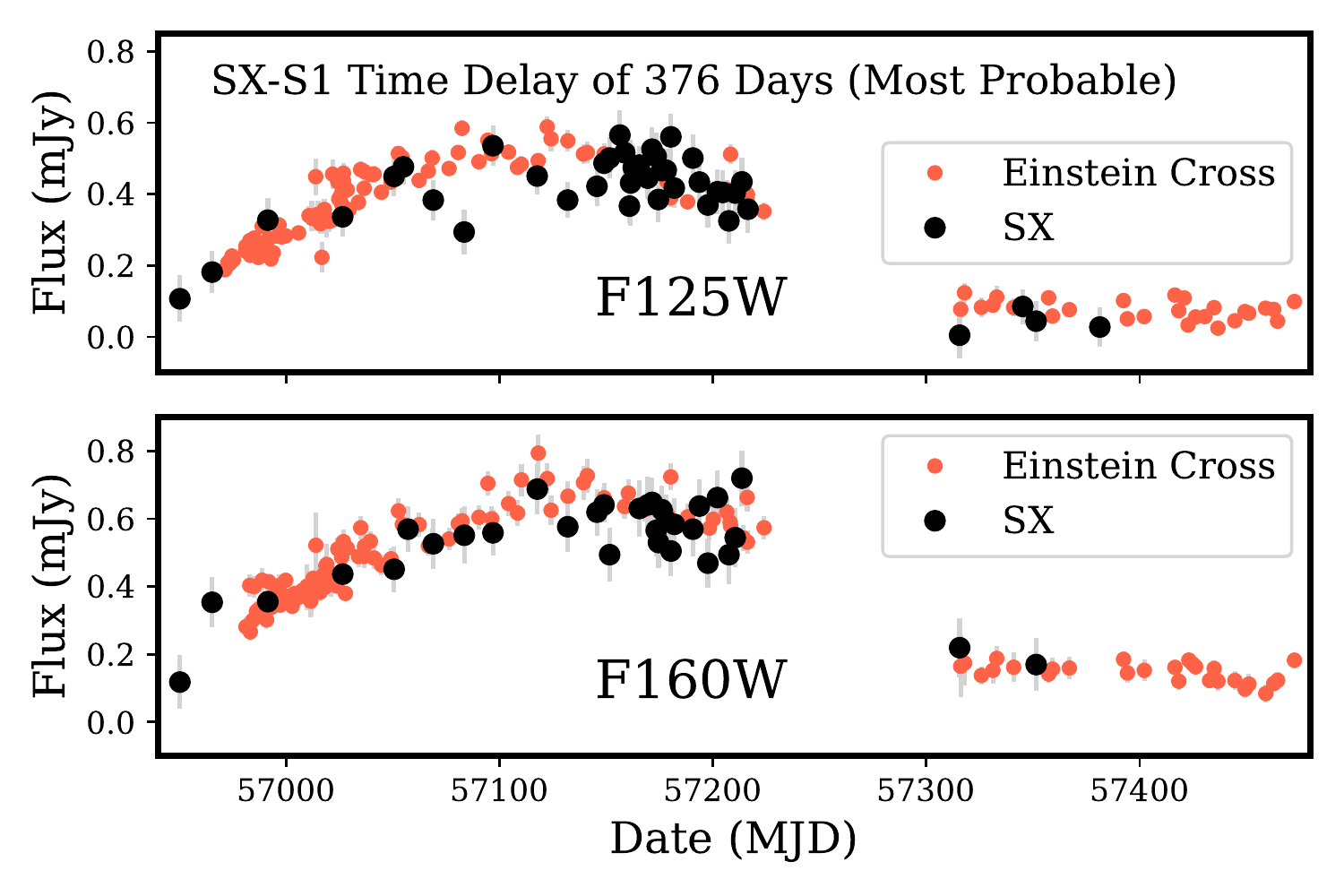}{0.75\textwidth}{}}
\gridline{\fig{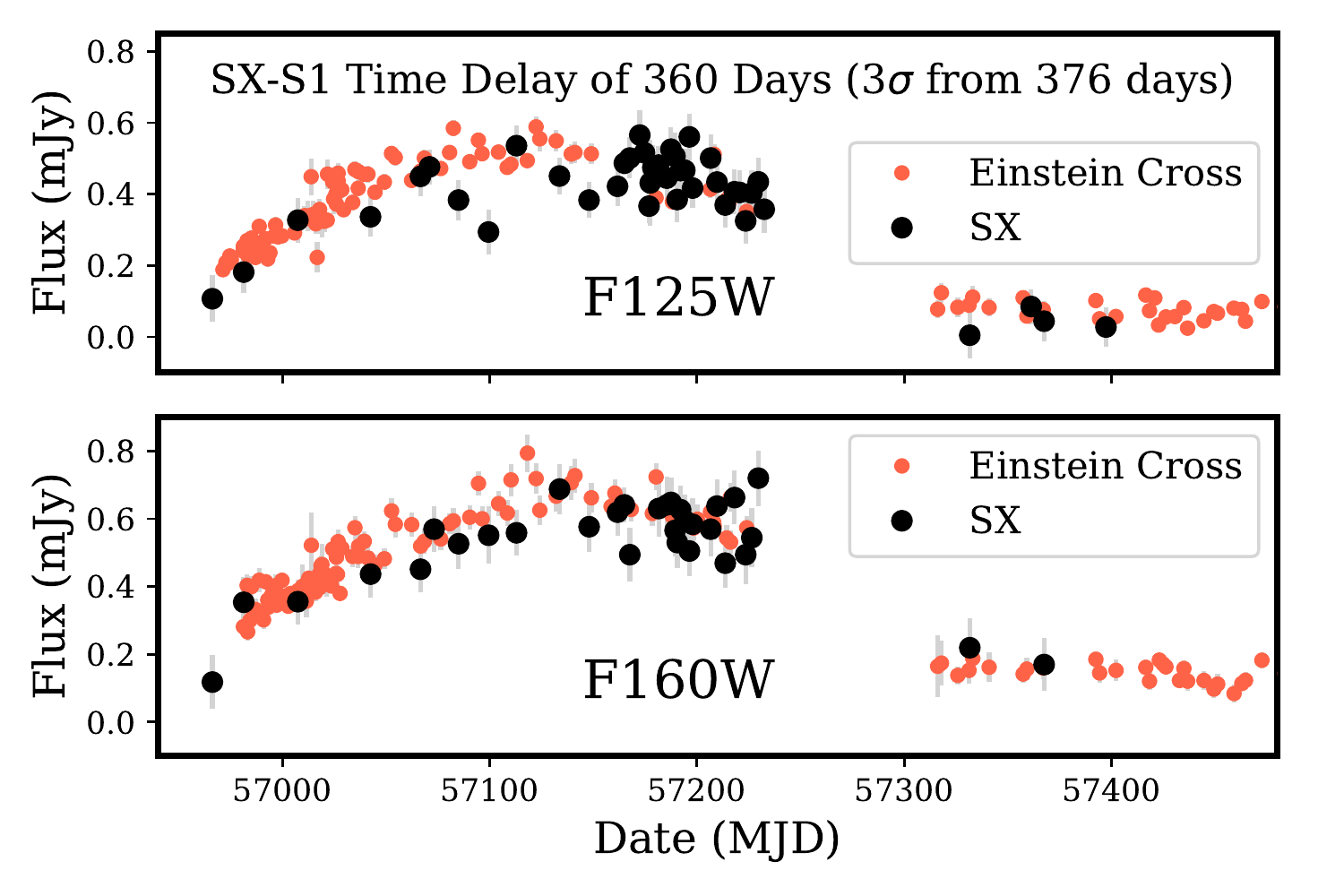}{0.75\textwidth}{}}
\caption{The light curves of the most highly magnified Einstein-cross images S1--S3 and that of image SX, after removing the best-fitting SX--S1 time delay of 376 days (upper panel), and, for comparison, an SX--S1 time delay of 360 days (lower panel). Given our constraints on the SX--S1 time delay of \corrpmdelaysxsoneCombinedchabriermedian, 360 days is offset by 3$\sigma$ from the best-fitting value.
 We have also shifted the plotted light curves of images S2 and S3 to remove their time delays relative to S1, using the values listed in Tab.~\ref{tab:measurementsdelaycorr}. We do not include the light curve of S4 in this comparison, given its smaller signal-to-noise. Fig.~\ref{fig:chi_sq_sx_s1} plots the $\chi^2$ value of the best-fitting piecewise polynomial model as a function of the SX--S1 relative time delay. }
\label{fig:visual_shift_one}
\end{figure*}

\begin{figure*}
\centering
\gridline{\fig{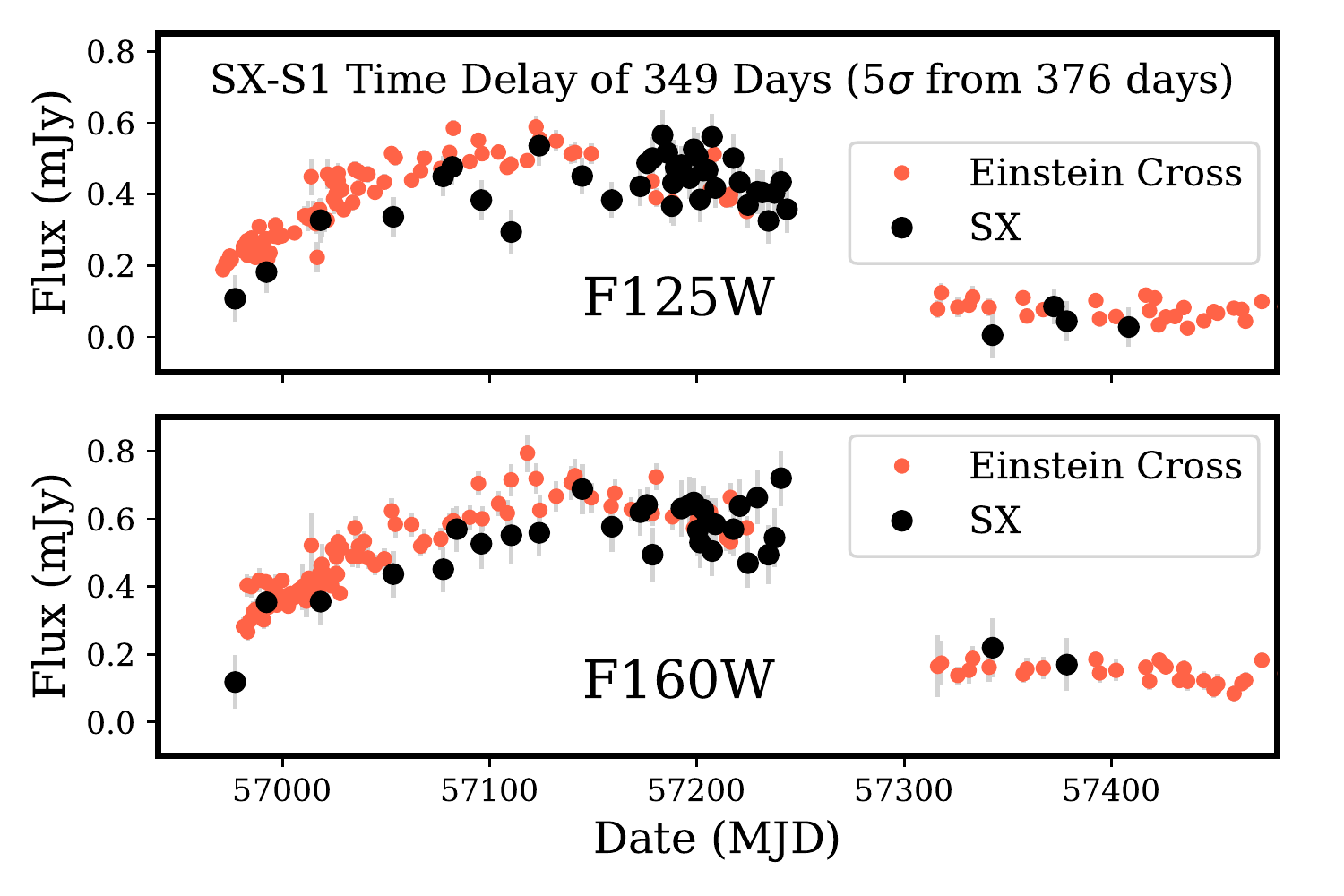}{0.8\textwidth}{}}
\caption{Continuation of Fig.~\ref{fig:visual_shift_one}. The light curves of the most highly magnified Einstein-cross images S1--S3 and that of image SX, after removing an SX--S1 time delay of 349 days (lower panel). Given our constraints on the SX--S1 time delay of \corrpmdelaysxsoneCombinedchabriermedian, 349 days is offset by 5$\sigma$ from the best-fitting value. Fig.~\ref{fig:chi_sq_sx_s1} plots the $\chi^2$ value of the best-fitting piecewise polynomial model as a function of the SX--S1 relative time delay. }
\label{fig:visual_shift_two}
\end{figure*}

 \begin{deluxetable*}{c|cc}
    \label{tab:measurementsdelaycorr}
  \tablecolumns{3}
  \tablecaption{\sc Measured Values of Time Delays}
 \tablehead{ Method & \multicolumn{2}{c}{Time Delay ($\Delta t$; days)} \\
  & S2--S1 & S3--S1}
  \startdata
Polyn.       & \corrtwosigdelaystwosoneKellychabriermedian       & \corrtwosigdelaysthreesoneKellychabriermedian   \\    
GPR & \corrtwosigdelaystwosoneThorpchabriermedian       & \corrtwosigdelaysthreesoneThorpchabriermedian       \\
{\tt SNTD}       & \corrtwosigdelaystwosoneRPthreerdchabriermedian & \corrtwosigdelaysthreesoneRPthreerdchabriermedian \\
{\tt PyCS}       & \corrtwosigdelaystwosoneMillonsplchabriermedian   & \corrtwosigdelaysthreesoneMillonsplchabriermedian \\ 
{Combined}       & \corrtwosigdelaystwosoneCombinedchabriermedian   & \corrtwosigdelaysthreesoneCombinedchabriermedian  \\
\tableline
& S4--S1 & SX--S1 \\
\tableline
Polyn. & \corrtwosigdelaysfoursoneKellychabriermedian       & \corrtwosigdelaysxsoneKellychabriermedian \\
GPR & \corrtwosigdelaysfoursoneThorpchabriermedian       & \corrtwosigdelaysxsoneThorpchabriermedian \\
{\tt SNTD} & \corrtwosigdelaysfoursoneRPthreerdchabriermedian & \corrtwosigdelaysxsoneRPthreerdchabriermedian \\
{\tt PyCS} & \corrtwosigdelaysfoursoneMillonsplchabriermedian   & \corrtwosigdelaysxsoneMillonsplchabriermedian \\
{Combined}  & \corrtwosigdelaysfoursoneCombinedchabriermedian   & \corrtwosigdelaysxsoneCombinedchabriermedian\\
\enddata
\tablecomments{Each measurement is reported as the measured (maximum likelihood) value, followed in parentheses by the 2.27, 16, 50, 84, and 97.73 percentiles of the probability distribution.}
\end{deluxetable*}

 \begin{deluxetable*}{c|cc}
    \label{tab:measurementsmucorr}
  \tablecolumns{3}
  \tablecaption{\sc Measured Values of Magnification Ratios}
 \tablehead{ Method & \multicolumn{2}{c}{Magnification Ratio ($\mu/\mu_1$)} \\
  & S2--S1 & S3--S1}
  \startdata
Polyn.       & \corrtwosigmustwosoneKellychabriermedian       & \corrtwosigmusthreesoneKellychabriermedian   \\    
GPR & \corrtwosigmustwosoneThorpchabriermedian       & \corrtwosigmusthreesoneThorpchabriermedian       \\
{\tt SNTD}       & \corrtwosigmustwosoneRPthreerdchabriermedian & \corrtwosigmusthreesoneRPthreerdchabriermedian \\
{\tt PyCS}       & \corrtwosigmustwosoneMillonsplchabriermedian   & \corrtwosigmusthreesoneMillonsplchabriermedian \\ 
{Combined}       & \corrtwosigmustwosoneCombinedchabriermedian   & \corrtwosigmusthreesoneCombinedchabriermedian  \\
\tableline
& S4--S1 & SX--S1 \\
\tableline
Polyn. & \corrtwosigmusfoursoneKellychabriermedian       & \corrtwosigmusxsoneKellychabriermedian \\
GPR & \corrtwosigmusfoursoneThorpchabriermedian       & \corrtwosigmusxsoneThorpchabriermedian \\
{\tt SNTD} & \corrtwosigmusfoursoneRPthreerdchabriermedian & \corrtwosigmusxsoneRPthreerdchabriermedian \\
{\tt PyCS} & \corrtwosigmusfoursoneMillonsplchabriermedian   & \corrtwosigmusxsoneMillonsplchabriermedian \\
{Combined}  & \corrtwosigmusfoursoneCombinedchabriermedian   & \corrtwosigmusxsoneCombinedchabriermedian\\
\enddata
\tablecomments{Each measurement is reported as the measured (maximum likelihood) value, followed in parentheses by the 2.27, 16, 50, 84, and 97.73 percentiles of the probability distribution.}
\end{deluxetable*}

\begin{deluxetable*}{c|c|cc|cc|cc|cc|cc}
\tablecaption{Lens Model Parameters Used for Alternative Microlensing Simulations}
\tablehead{Model & Reference & \multicolumn{2}{c|}{S1} &  \multicolumn{2}{c|}{S2} &  \multicolumn{2}{c|}{S3} &  \multicolumn{2}{c|}{S4} &  \multicolumn{2}{c}{SX} \\ 
 & & $\kappa$ & $\gamma$ &  $\kappa$ & $\gamma$ &  $\kappa$ & $\gamma$ &  $\kappa$ & $\gamma$ &  $\kappa$ & $\gamma$}
\startdata
Oguri-a & \dataset[HFF Lens Models (v3)]{https://archive.stsci.edu/prepds/frontier/lensmodels/} & 0.722 & 0.093 & 0.733 & 0.361 & 0.675 & 0.230 & 0.727 & 0.466 & 0.966 & 0.488\\
Grillo-g & \dataset[CLASH-VLT Clusters]{https://www.fe.infn.it/astro/lensing/} &  0.726 & 0.078 & 0.771 & 0.330 & 0.678 & 0.154 & 0.796 & 0.431 & 0.963 & 0.458\\
Sharon-g & \dataset[HFF Lens Models (v4corr)]{https://archive.stsci.edu/prepds/frontier/lensmodels/}  & 0.775 & 0.087 & 0.780 & 0.292 & 0.725 & 0.177 & 0.837 & 0.399 & 0.997 & 0.490
\enddata
\tablecomments{\label{tab:lensparamsmore} Values of $\kappa$ and $\gamma$ at the positions of the five observed images of SN Refsdal. The Sharon values correspond to those from the v4corr model. For the Grillo-g and Sharon-g models, we linearly interpolate the $\kappa$ and $\gamma$ maps.}
\end{deluxetable*}

 \begin{deluxetable*}{c|c}
    \label{tab:sxdelayvariants}
  \tablecolumns{3}
  \tablecaption{\sc SX-S1 Time Delays For Alternative Assumptions}
 \tablehead{ Method & Time Delay ($\Delta t$; days)}
  \startdata
$\kappa$ and $\gamma$ from Grillo-g  (see Table~\ref{tab:lensparamsmore})& \corrtwosigrefereereportgrillodelaysxsoneCombinedchabriermedian \\
$\kappa$ and $\gamma$ from Sharon-g (see Table~\ref{tab:lensparamsmore})& \corrtwosigrefereereportsharondelaysxsoneCombinedchabriermedian \\ 
Elliptical Photosphere ($b/a=0.8$) and $\kappa$ and $\gamma$ from Oguri-a  & \corrtwosigrefereereportogurioheightdelaysxsoneCombinedchabriermedian   \\
Double $\kappa_{\star}$ and $\kappa$ and $\gamma$  from Oguri-a & \corrtwosigrefereereportoguridoublekappasdelaysxsoneCombinedchabriermedian \\ 
\enddata
\tablecomments{Constraints on the SX-S1 time delay using microlensing simulations carried out using alternative assumptions. These include convergence and shear values for the Grillo and Sharon models, the Oguri shear and convergence with an elliptical photosphere having a major axis ratio of $b/a = 0.8$, and an assumption of twice the inferred mass density in the intracluster medium. Comparing the inferences to each other and with Table~\ref{tab:measurementsdelaycorr} shows that the SX--S1 time-delay constraints are not sensitive to these alternate assumptions. Each measurement is reported as the measured (maximum likelihood) value, followed in parentheses by the 2.27, 16, 50, 84, and 97.73 percentiles of the probability distribution.}
\end{deluxetable*}

\begin{deluxetable*}{c|cccccccc}[tpb]
\centering
\tablehead{ & $\Delta t_{2,1}$ & $\Delta t_{3,1}$ & $\Delta t_{4,1}$ & $\Delta t_{X,1}$ & $\mu_2 / \mu_1$ & $\mu_3 / \mu_1$ & $\mu_4 / \mu_1$ & $\mu_X / \mu_1$  }
\startdata
$\Delta t_{2,1}$ & 16.85806 & 4.60409 & 6.26366 & 4.74173 & -0.07571 & -0.01513 & 0.00951 & -0.00102 \\
$\Delta t_{3,1}$ & 4.60409 & 18.74664 & 4.21964 & 4.16457 & -0.05818 & -0.06709 & -0.03741 & -0.01790 \\
$\Delta t_{4,1}$ & 6.26366 & 4.21964 & 34.00929 & 3.87574 & 0.02516 & 0.09217 & -0.04550 & 0.01435 \\
$\Delta t_{X,1}$ & 4.74173 & 4.16457 & 3.87574 & 30.64843 & -0.00417 & -0.04247 & 0.01111 & 0.01576 \\
$\mu_2 / \mu_1$ & -0.07571 & -0.05818 & 0.02516 & -0.00417 & 0.05467 & -0.00385 & 0.00659 & 0.00240 \\
$\mu_3 / \mu_1$ & -0.01513 & -0.06709 & 0.09217 & -0.04247 & -0.00385 & 0.10805 & 0.00578 & 0.01033 \\
$\mu_4 / \mu_1$ & 0.00951 & -0.03741 & -0.04550 & 0.01111 & 0.00659 & 0.00578 & 0.00663 & 0.00131 \\
$\mu_X / \mu_1$ & -0.00102 & -0.01790 & 0.01435 & 0.01576 & 0.00240 & 0.01033 & 0.00131 & 0.00376 \\
\enddata
\caption{Covariance matrix for relative time delays and magnification ratio measurements calculated by estimating input parameters from simulated light curves that include microlensing, millilensing, and photometric noise. Figure~\ref{fig:covariance} shows a graphical representation of these values. The plotted covariance estimates do not include lens model uncertainties, which are added before computing likelihood values. When computing the covariance, we have excluded light curves for which a combined estimate from all fitting codes using the codes' respective weights represents a $>30$ day outlier for a time delays, or a 1.5 outlier for a magnification.}
\label{tab:covariance}
\end{deluxetable*}

\begin{figure*}[bpt]
\centering
\includegraphics[width=0.9\textwidth]{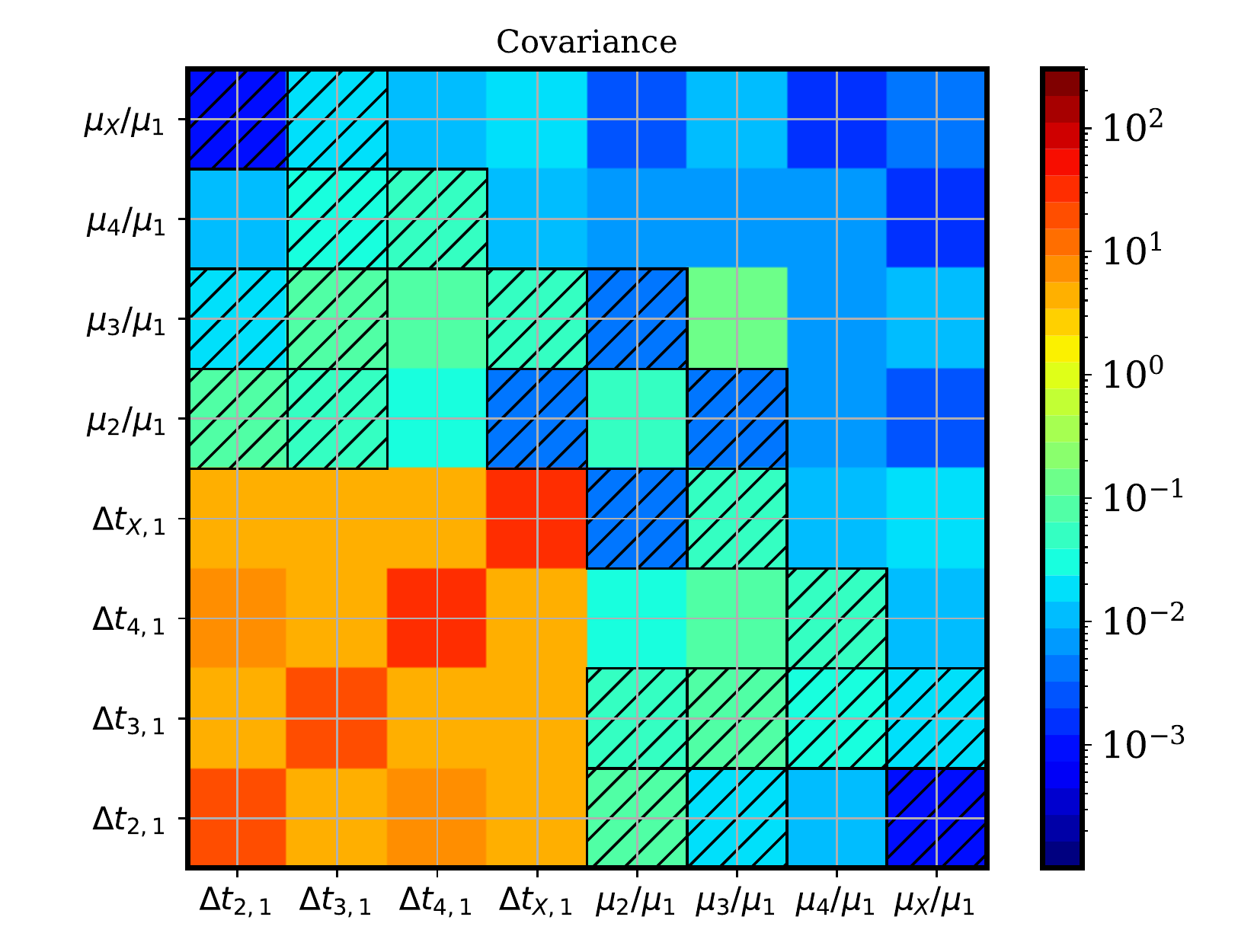}
\vspace{-2em}
\caption{Expected covariance between relative time delays and magnification ratios among the images of SN Refsdal from microlensing, millilensing, photometric uncertainty, and measurement. Hashed squares denote negative values. The plotted covariance estimates do not include lens model uncertainties, which are added before computing likelihood values. When computing the covariance, we have excluded light curves for which a combined estimate from all fitting codes using the codes' respective weights represents a $>30$ day outlier for a time delays, or a 1.5 outlier for a magnification.}
\label{fig:covariance}
\end{figure*}

We are interested in the covariance of random draws $x_i$ and $y_i$ from the linear combinations of 
distributions $X$ and $Y$ corresponding to two observables.
The probability of drawing from the distribution $X_j$ corresponding to the $j$th model is $w_{j,X}$, and
the probability of drawing from the distribution $Y_k$ corresponding to the $k$th model is $w_{k,Y}$.
Therefore, the probability of drawing from both is $p = w_{j, X} \times w_{k, Y}$.
We present this estimate for the covariance matrix in Table~\ref{tab:covariance} and Figure~\ref{fig:covariance}. When computing the covariance matrix, we have excluded light curves for which a combined estimate from all fitting codes using the codes' respective weights represents a $>30$ day outlier for a time delays, or a 1.5 outlier for a magnification.

\section{Summary and Discussion}
\label{sec:Summary}

 We have described our blind measurements of the relative time delays and magnification ratios of the first-known example of a multiply imaged supernova.
 SN Refsdal provides a powerful and unique probe of the value of H$_0$ and a novel test of galaxy-cluster models.
 Our constraints employ all available {\it HST} WFC3-IR imaging data collected during multiple follow-up campaigns.
To arrive at robust, as precise as possible, and accurate inferences about the relative time delays and magnification of the five observed images, we have used four separate light-curve fitting algorithms. Moreover, to minimize human bias, we have performed the analysis in a blind manner by adding random offsets to the light-curve dates.

To determine the uncertainties of the inferred delays and magnifications and identify the optimal combination of inferences from the light-curve fitting techniques, we constructed simulated light curves with the same cadence and S/N ratios as the observed light curves which include the effects of microlensing, millilensing, and intrinsic variation in the light curves of SN 1987A-like SNe.  We identified a powerful new statistical framework for optimally combining measurements from multiple separate light-curve fitting algorithms.

In the accompanying paper \citep{papertwo}, we use our measurements to obtain a first, blind measurement of the value of H$_0$ from a multiply imaged SN, and compare the performance of  lens model predictions for the relative magnification and time delays of the five images of SN Refsdal.

In the upcoming decade, deep imaging surveys that use newly constructed, wide-field ground-basd and space-based telescopes will discover hundreds of strongly lensed SNe that will be multiply imaged by both galaxy-scale and cluster-scale gravitational lenses. The Legacy Survey of Space and Time (LSST) with the Vera Rubin Observatory will discover hundreds of strongly lensed SNe and provide measurements of their light curves for wide-separation systems \citep[e.g.,][]{ogurimarshall10,goldsteinnugent17,hubersuyunoebauer19,wojtakhjorthgall19}. 
After launch in the mid-2020s, the Nancy Grace Roman Space Telescope will be used for a planned two-year SN Ia program, which may yield well-measured light curves of dozens of strongly lensed SNe \citep{oguri10,pierelrodneyvernardos21}. 
While a large majority of the strongly lensed SNe will have foreground galaxy lenses, a significant fraction will be multiply imaged by foreground galaxy clusters \citep{petrushevskagoobarlagattuta18}.

As the sample of strongly lensed SNe grows to hundreds of objects, we can anticipate at least two additional avenues of exploration.  First is time-delay cosmography, not only to measure H$_0$, but also to place constraints on additional cosmological parameters such as the dark energy equation-of-state parameter \citep{linder11,treumarshall16,suyuchangcourbin18,grillorosatisuyu18}. A second possibility is to anticipate the appearance of a trailing SN image and acquire observations very shortly after the time of explosion.  This method to obtain early-time spectroscopy and photometry may provide powerful constraints on the properites of SN progenitor \citep[e.g.,][]{kasen10,foxleymarrablecollettfrohmaier20}. Early spectroscopy of lensed SNe at $z>1$ could enable tests for evolution of the SN progenitor populations across cosmic time.

We anticipate that the methods we have developed to extract the time delay and magnification ratios for SN Refsdal may be helpful to the analysis of the expected set of discoveries. In particular, our adaptation of {\tt DOLPHOT} for difference imaging should be useful for the Roman Space Telescope. Moreover, the framework for the combination of estimates from multiple light-curve fitting algorithms, informed by their performance on simulated data, we hope will be useful. 

SN Refsdal was surprising given that it was a peculiar, SN 1987A-like SN explosion of a blue supergiant star, which are not common in the nearby universe \citep{kellybrammerselsing16}. Intriguingly, a second, highly magnified blue supergiant star was discovered in the same host galaxy \citep{kellydiegorodney18}.
The dataset that we assembled for SN Refsdal required a large investment of observing time with {\it HST}. 
While the light-curve evolution of SN Refsdal is simple, comparably precise delays with much more modest light-curve sampling may be possible for spectroscopic types of SNe that are very well characterized, such as the multiply-imaged SN iPTF16geu \citep{goobaramanullahkulkarni17}. 

\acknowledgements
We thank Program Coordinators Beth Periello and Tricia Royle, and Contact Scientists Norbert Pirzkal, Ivelina Momcheva, and Kailash Sahu of the Space Telescope Science Institute (STScI), for their help carrying out the {\it HST} observations. We also appreciate useful discussions with C. Grillo, P. Rosati, and S. Suyu. NASA/STScI grants 14041, 14199, 14208, 14528, 14872,  14922, 15791, and 16134 provided financial support.

P.K. is supported by NSF grant AST-1908823 and by NASA/Keck JPL RSA 1644110.
M.O. acknowledges support by the WPI Initiative (MEXT, Japan) and by JSPS KAKENHI grants JP18K03693, JP20H00181, and JP20H05856.
T.T. acknowledges the support of NSF grant AST-1906976.
S.T. was supported by the Cambridge Centre for Doctoral Training in Data-Intensive Science funded by the UK Science and Technology Facilities Council (STFC).
J.D.R.P was supported by NASA Headquarters under the NASA FINESST award 80NSSC19K1414, and by grant HST-AR-15050 from STScI, which is operated by AURA, Inc., under NASA
contract NAS 5-26555.
J.M.D. acknowledges the support of projects  PGC2018-101814-B-100 (MCIU/AEI/MINECO/FEDER, UE) Ministerio de Ciencia, Investigaci\'on y Universidades and project MDM-2017-0765 funded by the Agencia Estatal de Investigaci\'on, Unidad de Excelencia Mar\'ia de Maeztu.
M.J. is supported by the United Kingdom Research and Innovation (UKRI) Future Leaders Fellowship ``Using Cosmic Beasts to uncover the Nature of Dark Matter" (grant MR/S017216/1).
M.M. and V.B. acknowledge support from the European Research Council (ERC) under the European Union’s Horizon 2020 research and innovation programme (COSMICLENS: grant agreement 787886).
A.V.F. is grateful for assistance from the Christopher R. Redlich Fund, the TABASGO Foundation, the U.C. Berkeley Miller Institute for Basic Research in Science (where he is a Miller Senior Fellow), and many individual donors.
The UCSC team is supported in part by NASA grant NNG17PX03C, NSF grants AST-1518052 and AST-1815935, the Gordon \& Betty Moore Foundation, the Heising-Simons Foundation, and a fellowship from the David and Lucile Packard Foundation to R.J.F. 
J.H. acknowledges support from  VILLUM FONDEN Investigator Grant 16599.
S.W.J.'s SN research at Rutgers University is supported by NSF CAREER award AST-0847157, as well as by NASA/Keck JPL RSA 1508337 and 1520634.
We acknowledge the HFF Lens Model website that is hosted by the Mikulski Archive for Space Telescopes (MAST) at STScI.

\software{Astrodrizzle (Hack et al. 2012), FAST (Kriek et al. 2009), SNTD (Pierel \& Rodney 2019), DOLPHOT (Dolphin 2000), PythonPhot (Jones et al. 2015), dynesty (Speagle 2019), PyCS (Tewes et al. 2013; Millon et al. 2020a), MULE (Dobler et al. 2015), microlens (Wambsganss 1990, 1999)}

\bibliographystyle{aasjournal}
\bibliography{ms}

\clearpage
\renewcommand\thefigure{A\arabic{figure}}   
\setcounter{figure}{0} 
\renewcommand\thetable{A\arabic{table}}   
\setcounter{table}{0} 
\appendix

\begin{figure*}[hb!]
\centering
\includegraphics[angle=0,width=6.0in]{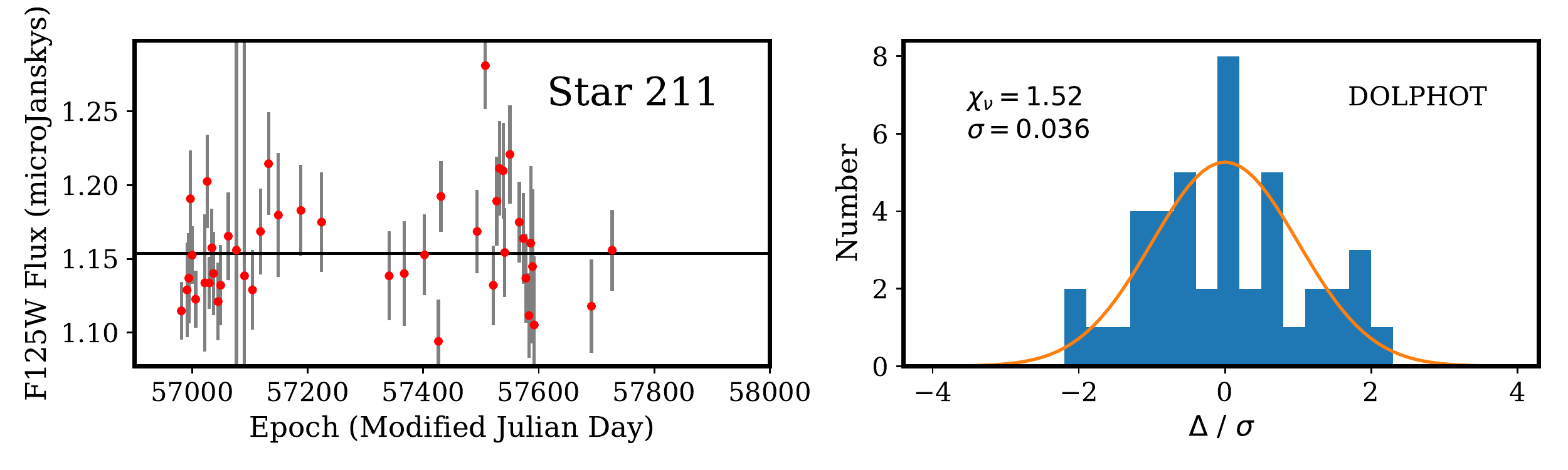}
\includegraphics[angle=0,width=6.0in]{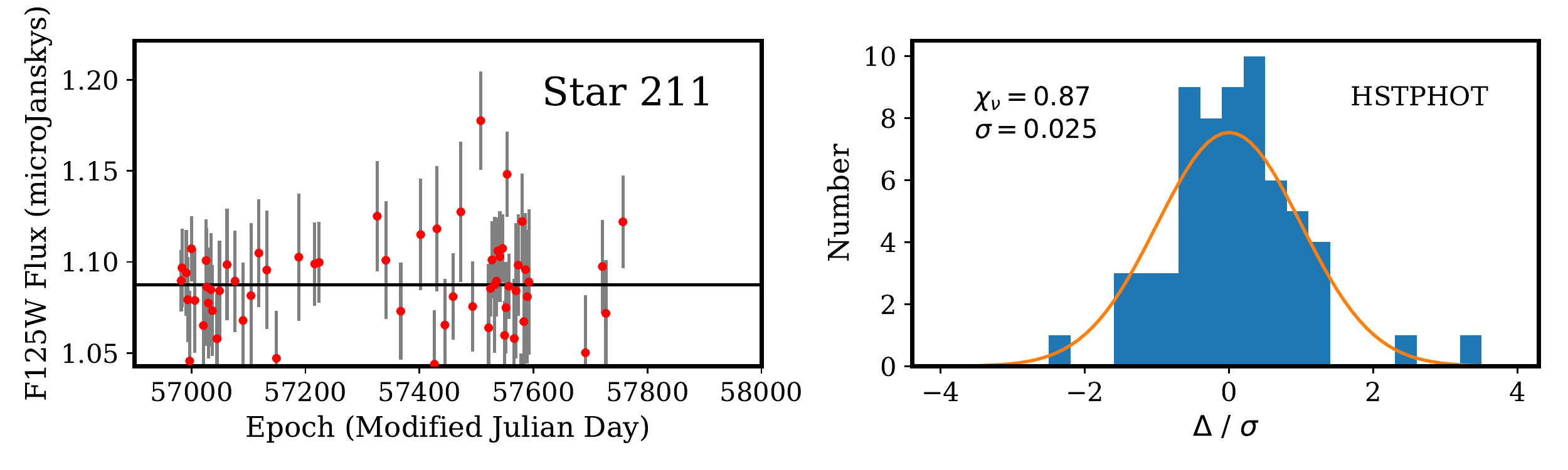}
\includegraphics[angle=0,width=6.0in]{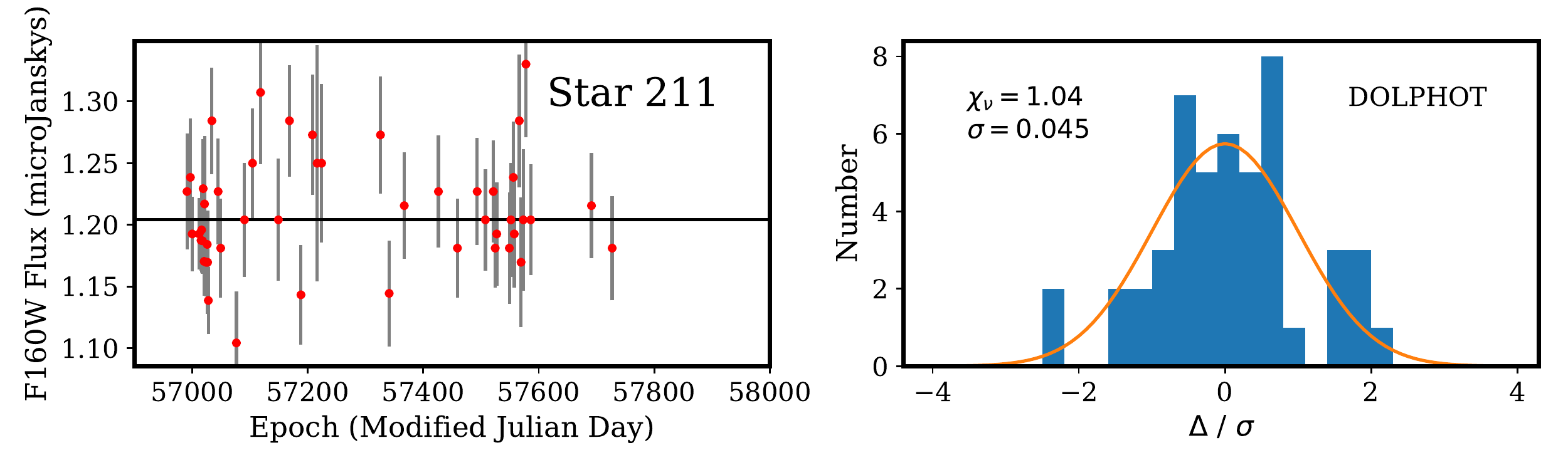}
\includegraphics[angle=0,width=6.0in]{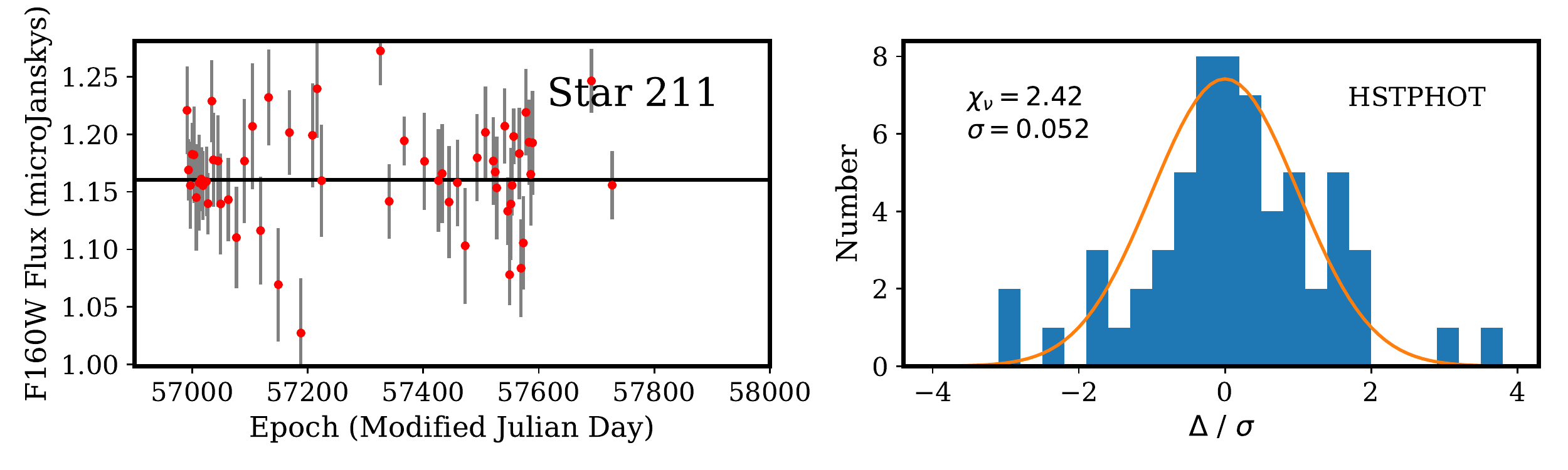}
\caption{WFC3-IR F125W and F160W photometry of Star 211 in the MACS\,J1149 galaxy-cluster field with the {\tt DOLPHOT} PSF-fitting code and aperture fluxes with {\tt HSTPHOT}. The left panel shows flux measurements and median flux values. The right panel plots residuals from the median flux value normalized by the flux uncertainty estimate. }
\label{fig:stellarphot1}
\end{figure*}

\begin{figure*}[tp]
\centering
\includegraphics[angle=0,width=6.0in]{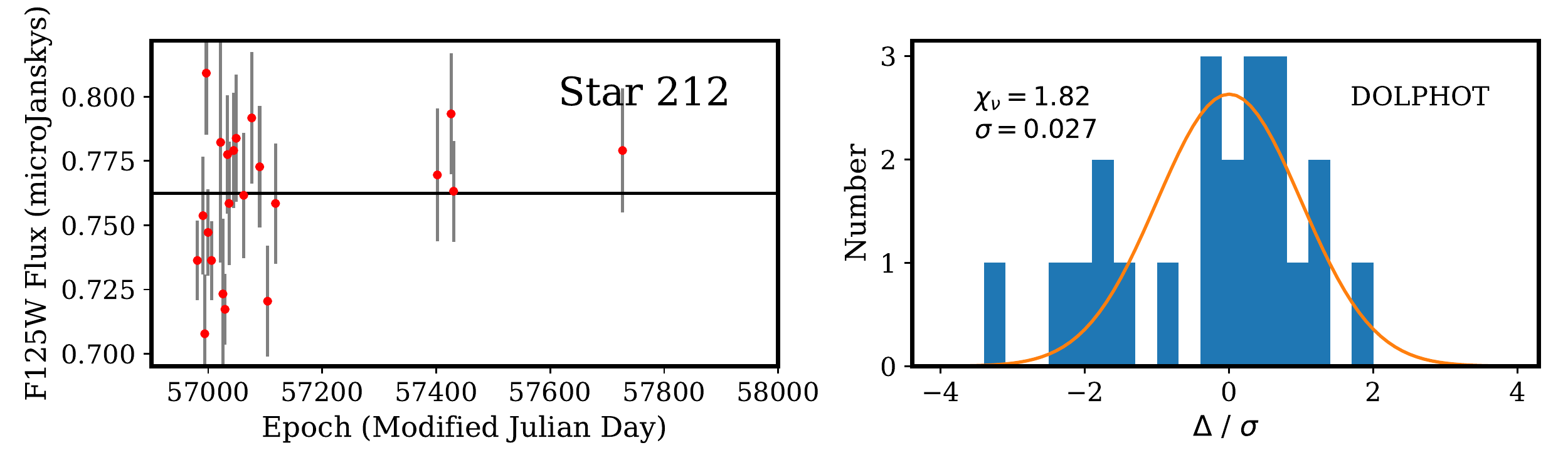}
\includegraphics[angle=0,width=6.0in]{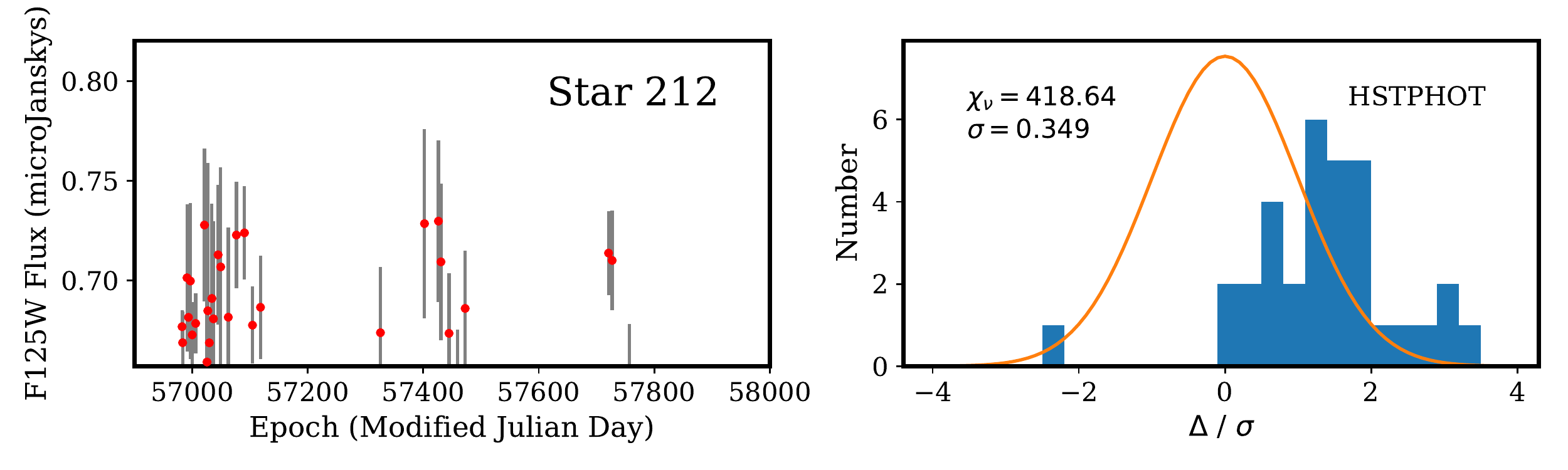}

\includegraphics[angle=0,width=6.0in]{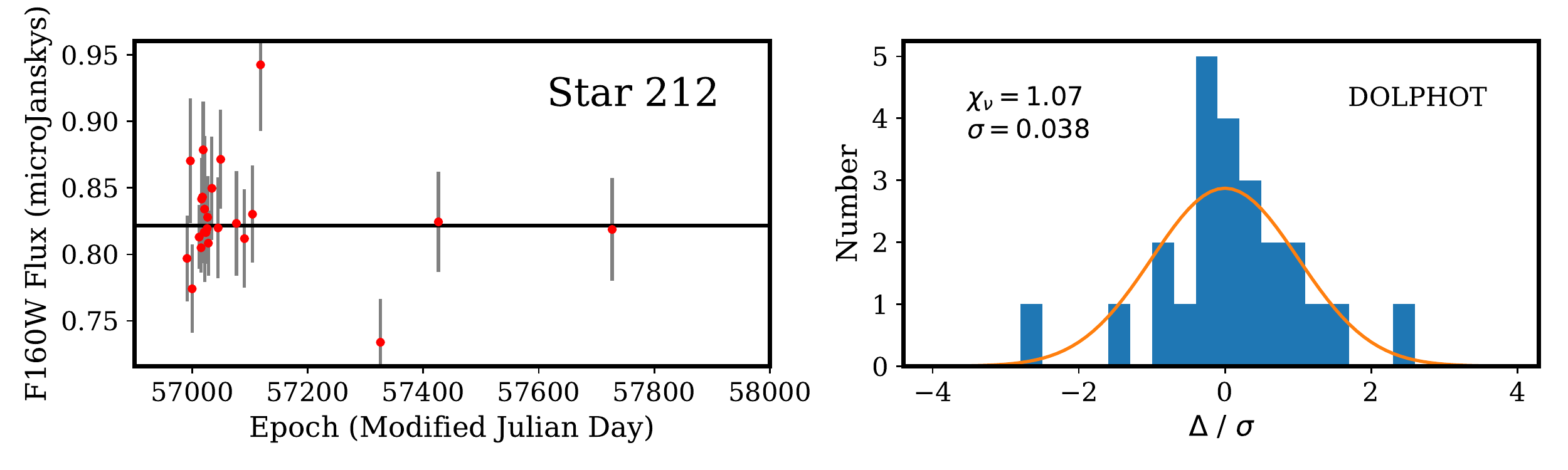}
\includegraphics[angle=0,width=6.0in]{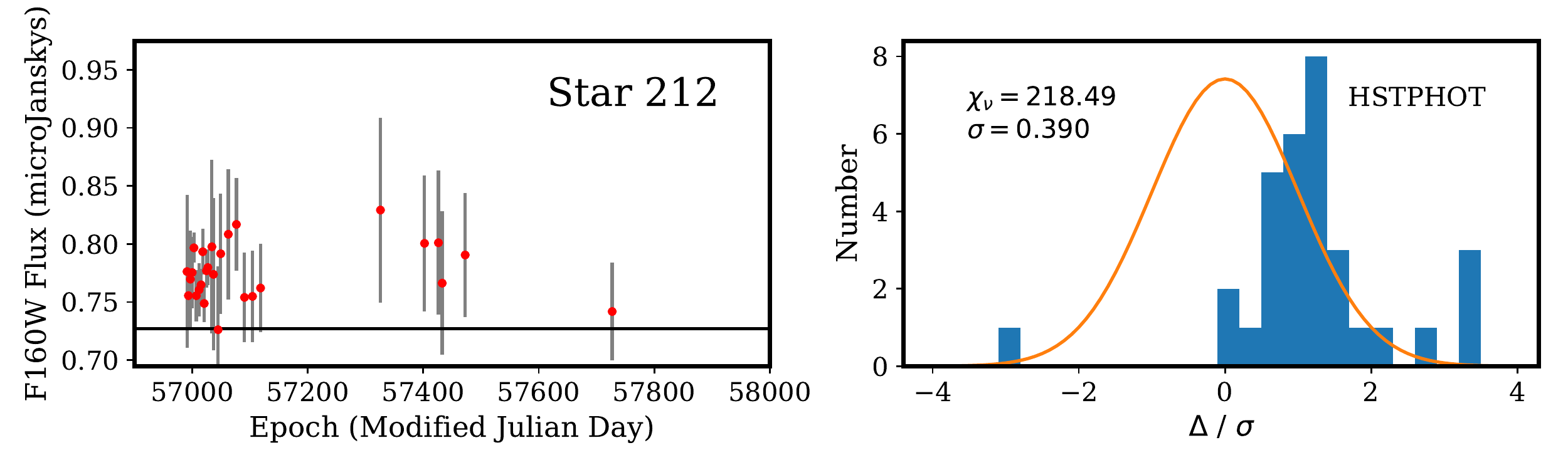}

\caption{WFC3-IR F125W and F160W photometry of Star 212 in the MACS\,J1149 galaxy-cluster field with the {\tt DOLPHOT} PSF-fitting code and aperture fluxes with {\tt HSTPHOT}. The left panel shows flux measurements and median flux values. The right panel plots residuals from the median flux value normalized by the flux uncertainty estimate. }
\label{fig:stellarphot2}
\end{figure*}

\begin{figure*}[tp]
\centering
\includegraphics[angle=0,width=6.0in]{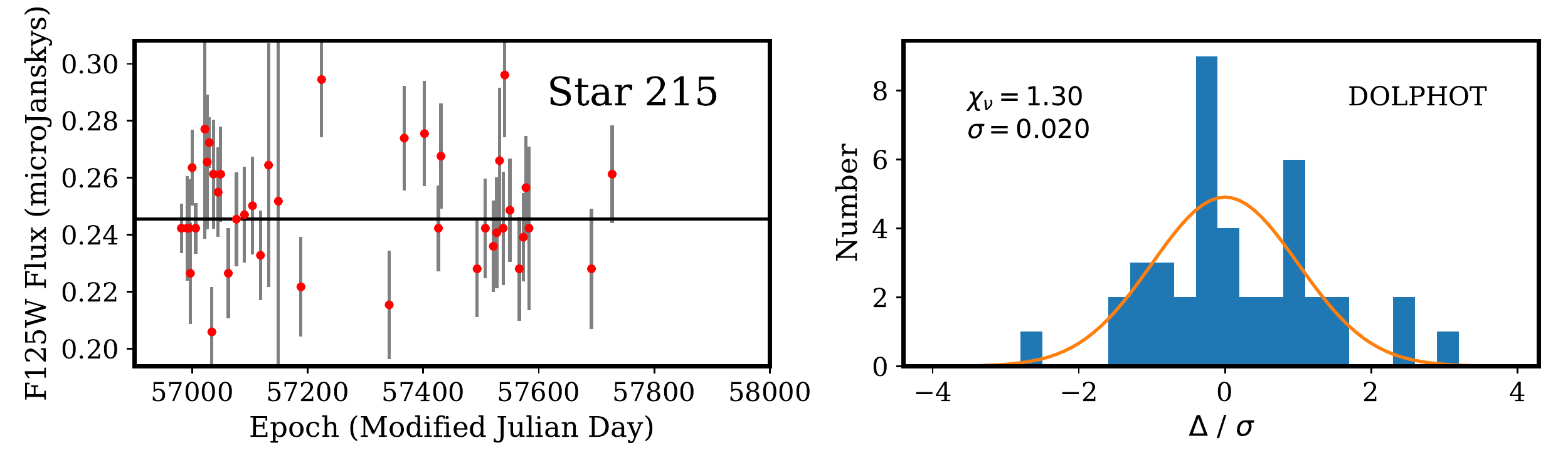}
\includegraphics[angle=0,width=6.0in]{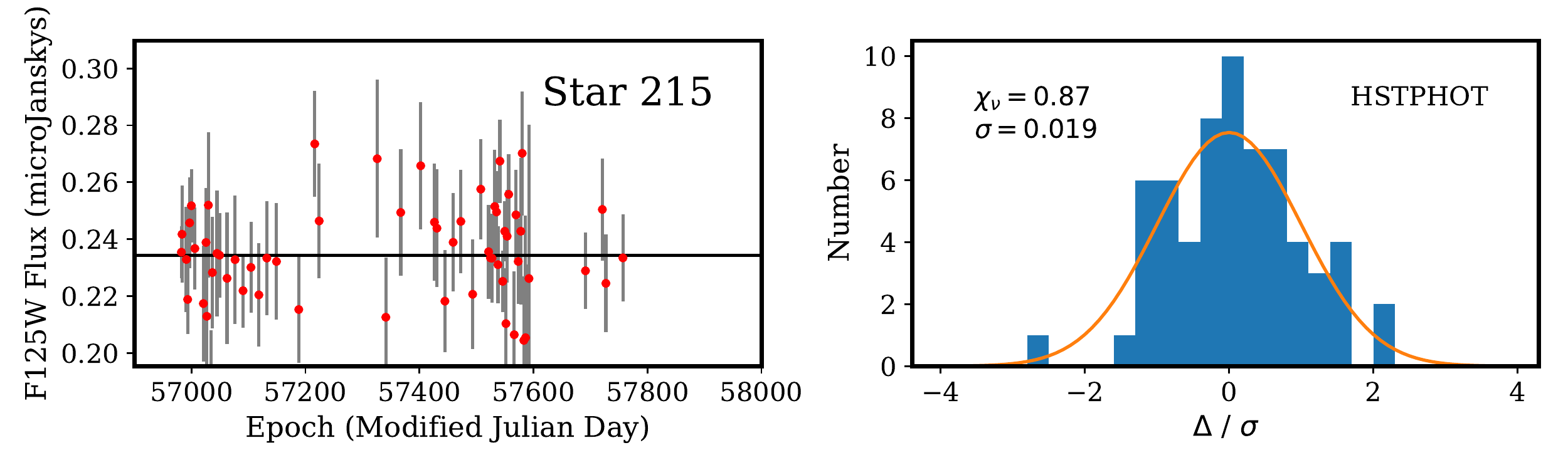}

\includegraphics[angle=0,width=6.0in]{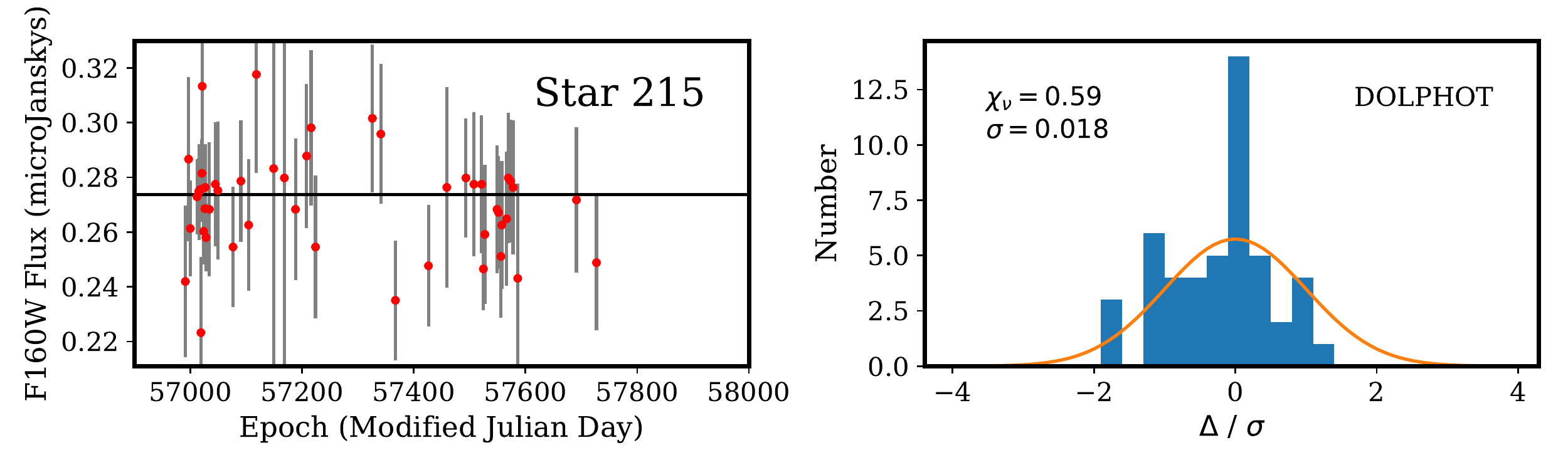}
\includegraphics[angle=0,width=6.0in]{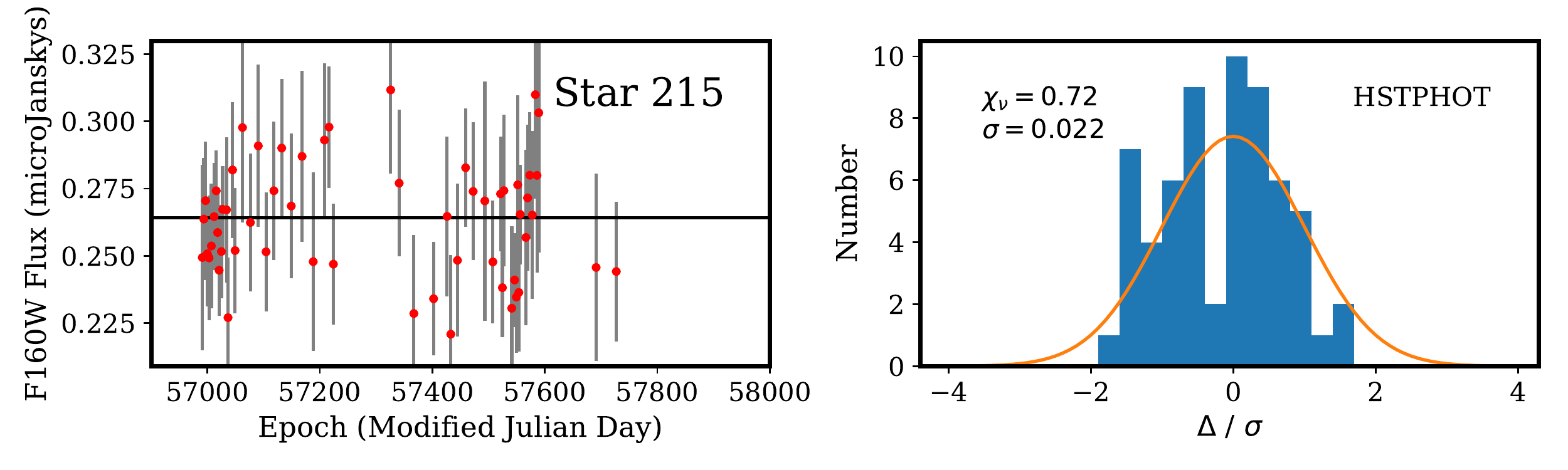}

\caption{WFC3-IR F125W and F160W photometry of Star 215 in the MACS\,J1149 galaxy-cluster field with the {\tt DOLPHOT} PSF-fitting code and aperture fluxes with {\tt HSTPHOT}. The left panel shows flux measurements and median flux values. The right panel plots residuals from the median flux value normalized by the flux uncertainty estimate. }
\label{fig:stellarphot3}
\end{figure*}

\begin{figure*}[htp!]
\centering
\includegraphics[angle=0,width=6.0in]{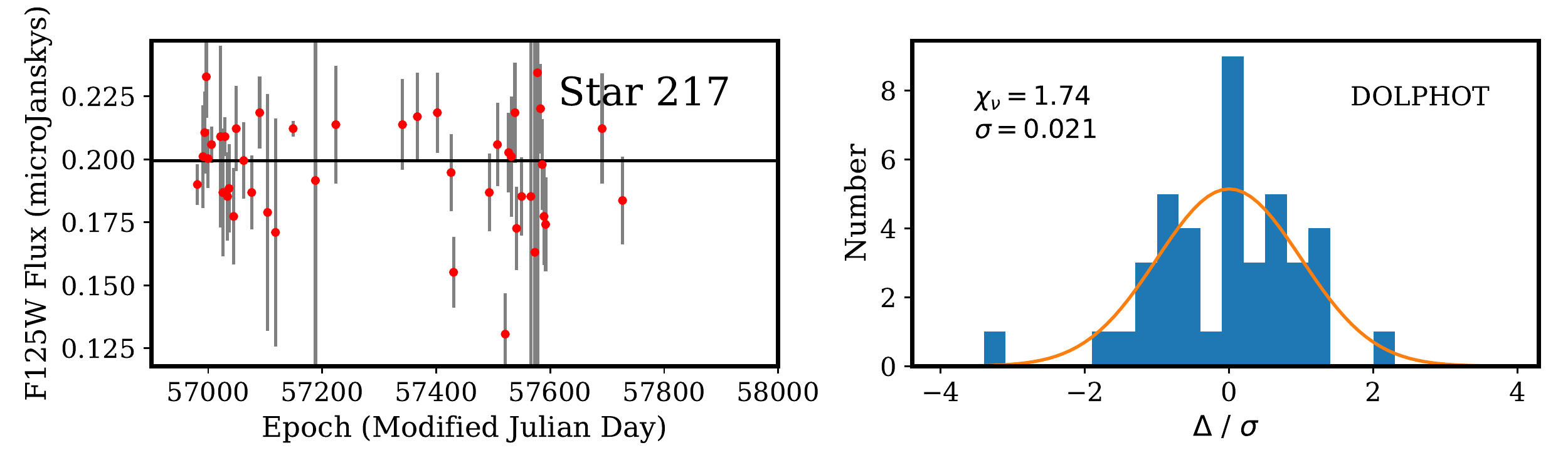}
\includegraphics[angle=0,width=6.0in]{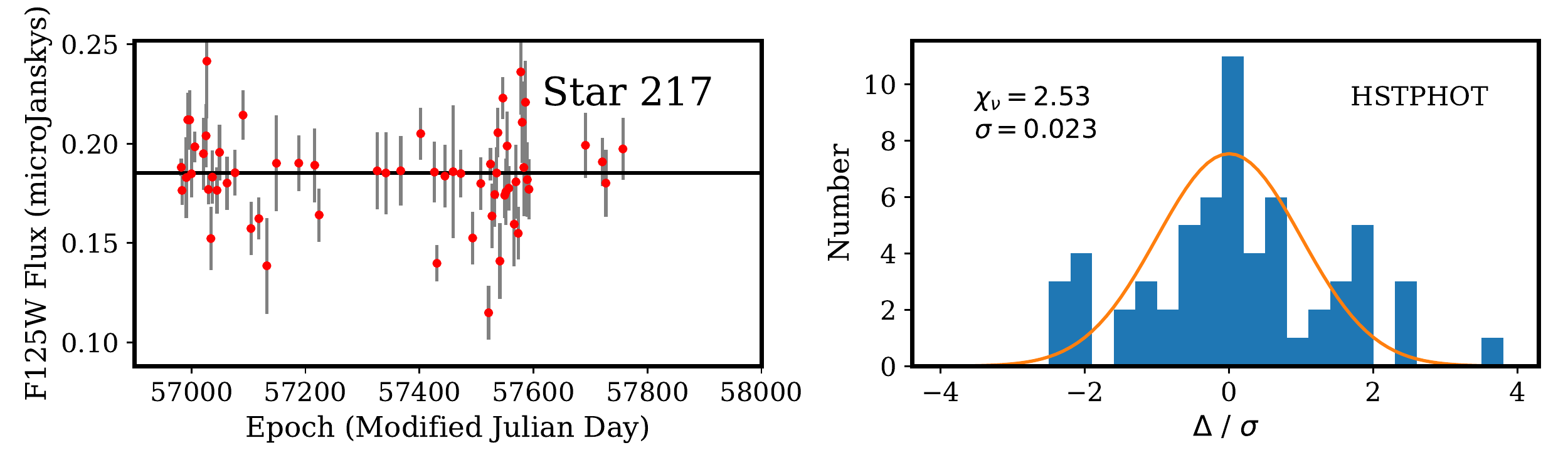}

\includegraphics[angle=0,width=6.0in]{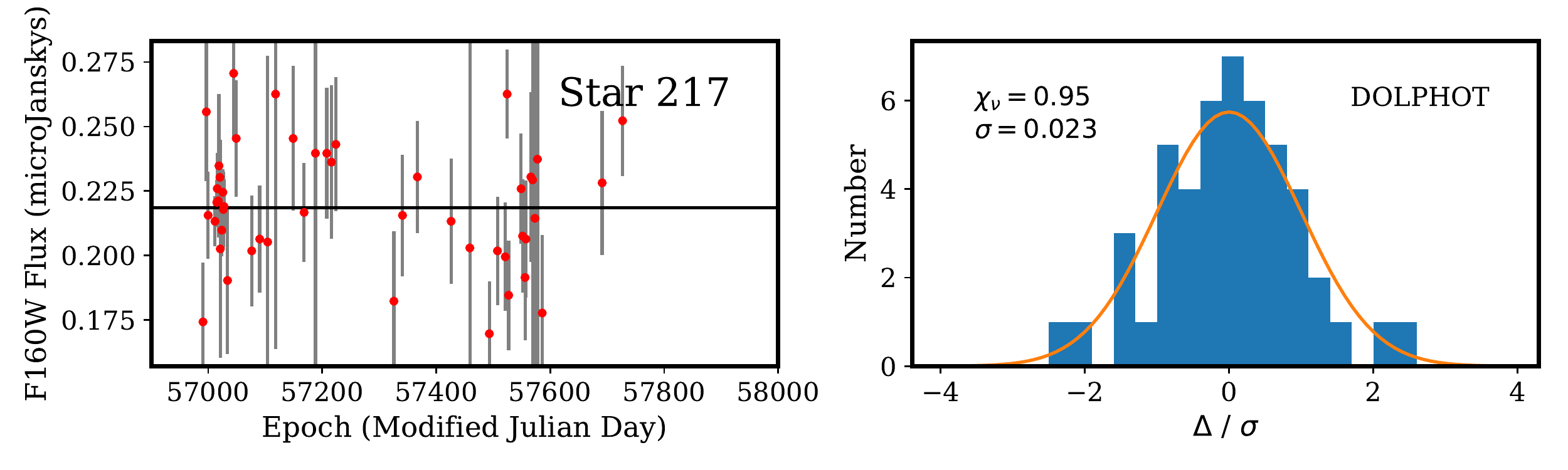}
\includegraphics[angle=0,width=6.0in]{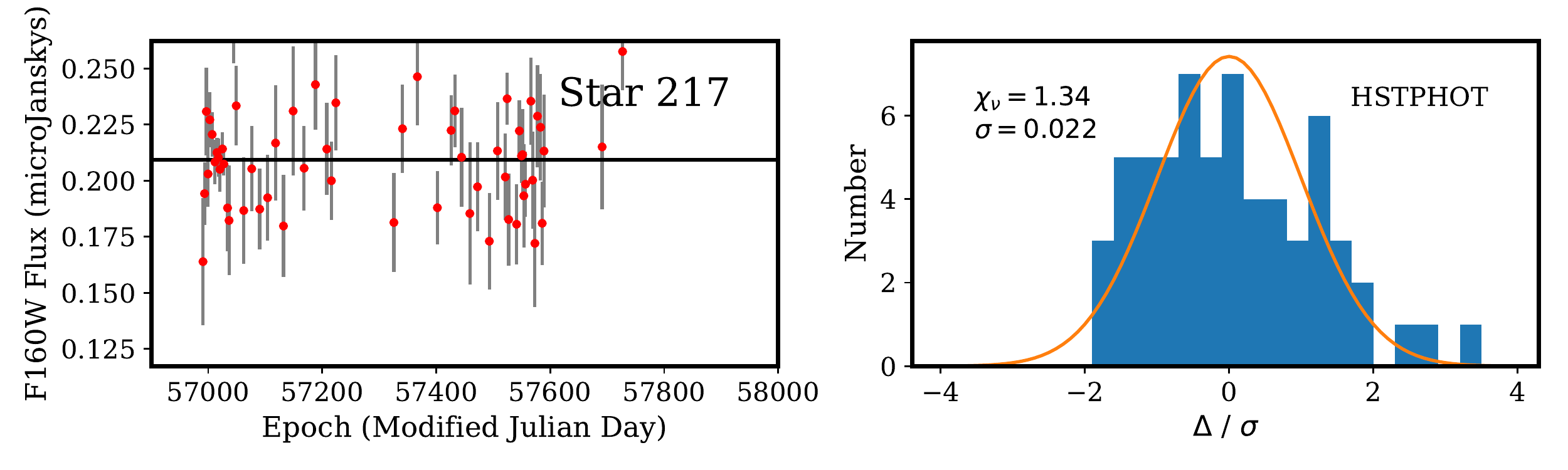}

\caption{WFC3-IR F125W and F160W photometry of Star 217 in the MACS\,J1149 galaxy-cluster field with the {\tt DOLPHOT} PSF-fitting code and aperture fluxes with {\tt HSTPHOT}. The left panel shows flux measurements and median flux values. The right panel plots residuals from the median flux value normalized by the flux uncertainty estimate. }
\label{fig:stellarphot4}
\end{figure*}

\begin{figure*}[tp]
\centering
\includegraphics[angle=0,width=6.0in]{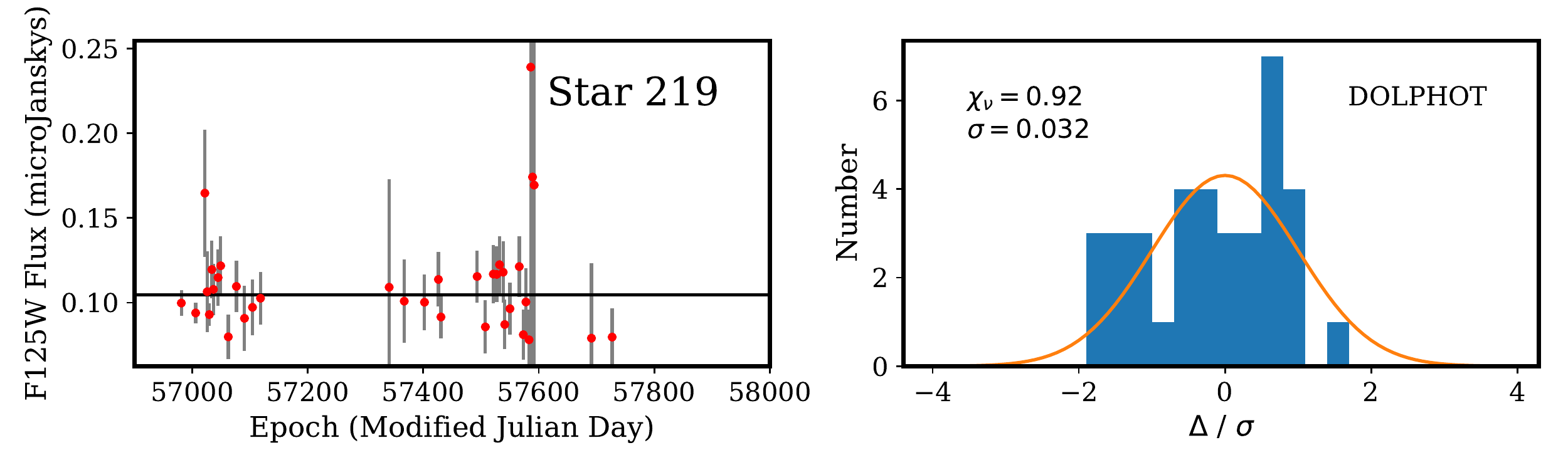}
\includegraphics[angle=0,width=6.0in]{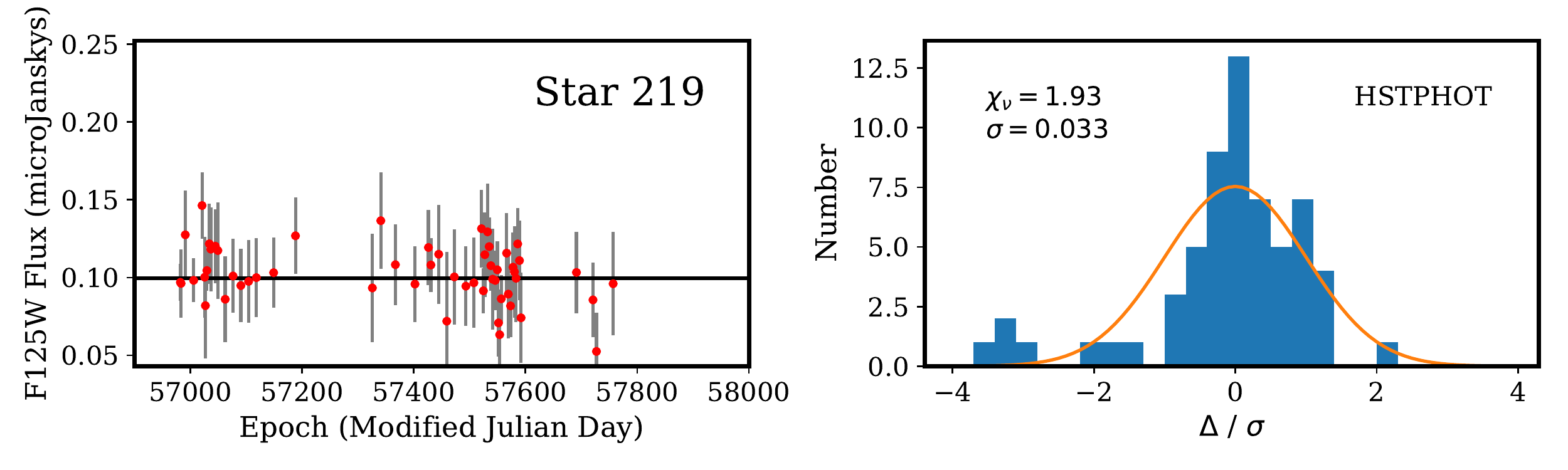}

\includegraphics[angle=0,width=6.0in]{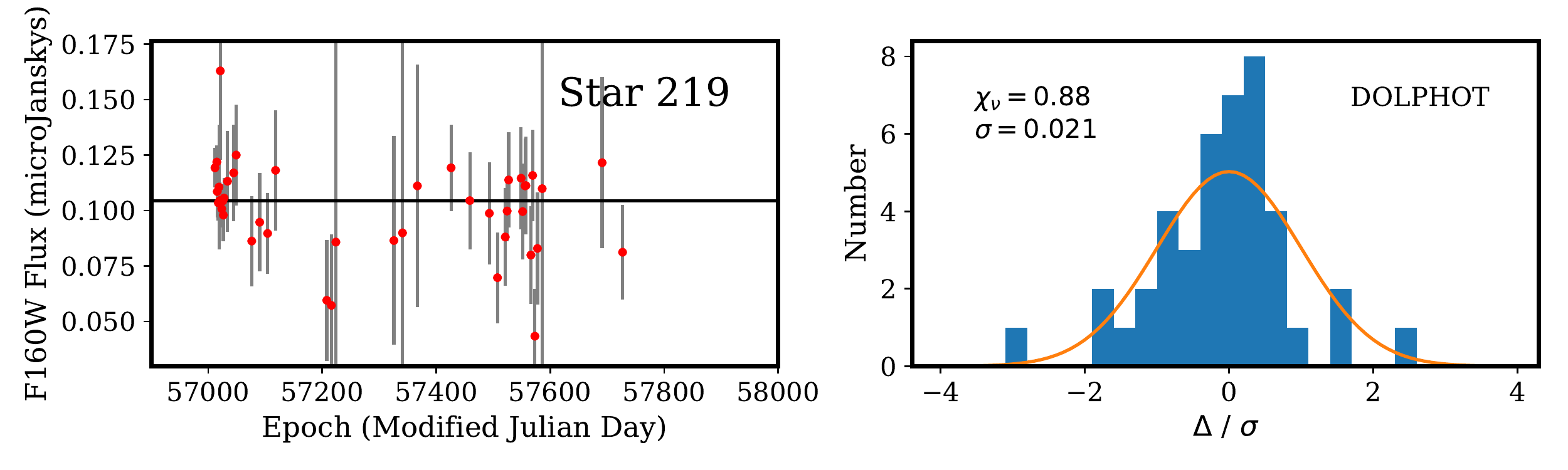}
\includegraphics[angle=0,width=6.0in]{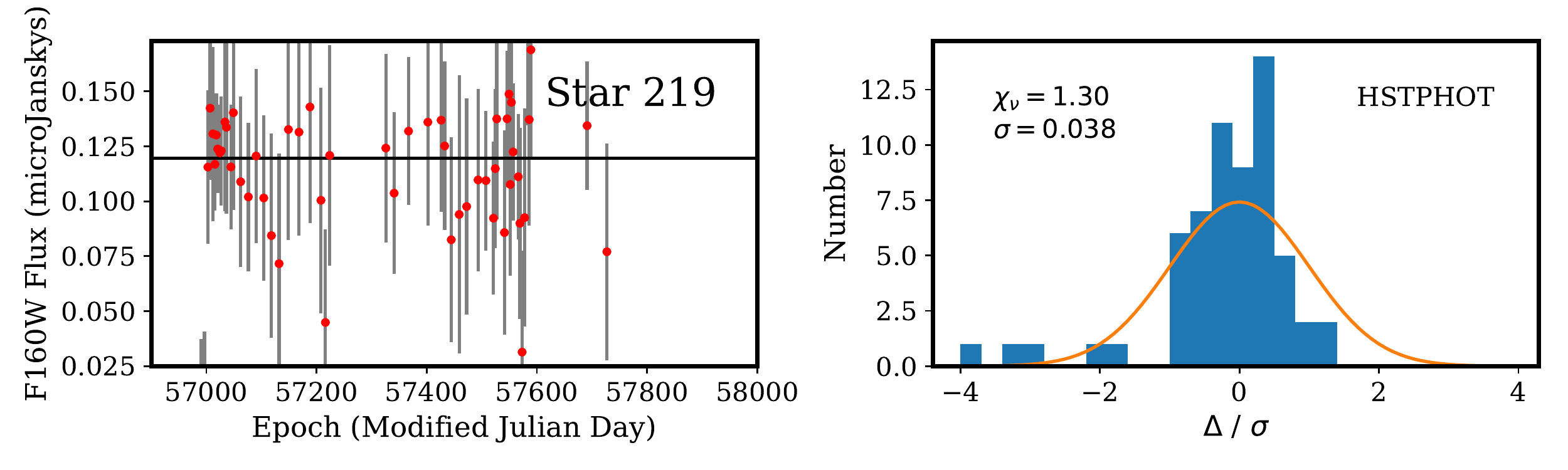}

\caption{WFC3-IR F125W and F160W photometry of Star 219 in the MACS\,J1149 galaxy-cluster field with the {\tt DOLPHOT} PSF-fitting code and aperture fluxes with {\tt HSTPHOT}. The left panel shows flux measurements and median flux values. The right panel plots residuals from the median flux value normalized by the flux uncertainty estimate. }
\label{fig:stellarphot5}
\end{figure*}

Figures~\ref{fig:stellarphot1}--\ref{fig:stellarphot5} show photometry for the five test stars, in the F125W and F160W bands and measured using the two fitting codes ({\tt DOLPHOT} and {\tt PythonPhot}).

\clearpage

\begin{deluxetable}{c|cccc|cccc}[h!]
\label{tab:salpeterRMSscatter}
  \tablecolumns{9}
  \tablecaption{\sc Salpeter RMS Scatter}
 \tablehead{ Method & \multicolumn{4}{c|}{Time Delay} & \multicolumn{4}{c}{Magnification Ratio}  \\  & S2--S1 & S3--S1 & S4--S1 & SX--S1 & 
 S2/S1 & S3/S1 & S4/S1 & SX/S1  }
  \startdata
Polynomial &  \rmsstwosonedelayKellymediansalpeter & \rmssthreesonedelayKellymediansalpeter & \rmssfoursonedelayKellymediansalpeter & \rmssxsonedelayKellymediansalpeter & \rmsstwosonemuKellymediansalpeter & \rmssthreesonemuKellymediansalpeter & \rmssfoursonemuKellymediansalpeter & \rmssxsonemuKellymediansalpeter\\
Gaussian Process & \rmsstwosonedelayThorpmediansalpeter & \rmssthreesonedelayThorpmediansalpeter & \rmssfoursonedelayThorpmediansalpeter & \rmssxsonedelayThorpmediansalpeter & \rmsstwosonemuThorpmediansalpeter & \rmssthreesonemuThorpmediansalpeter & \rmssfoursonemuThorpmediansalpeter & \rmssxsonemuThorpmediansalpeter  \\
{\tt SNTD} & \rmsstwosonedelayRPthreerdmediansalpeter & \rmssthreesonedelayRPthreerdmediansalpeter & \rmssfoursonedelayRPthreerdmediansalpeter & \rmssxsonedelayRPthreerdmediansalpeter & \rmsstwosonemuRPthreerdmediansalpeter & \rmssthreesonemuRPthreerdmediansalpeter & \rmssfoursonemuRPthreerdmediansalpeter & \rmssxsonemuRPthreerdmediansalpeter  \\
{\tt PyCS} & \rmsstwosonedelayMillonsplmediansalpeter & \rmssthreesonedelayMillonsplmediansalpeter & \rmssfoursonedelayMillonsplmediansalpeter & \rmssxsonedelayMillonsplmediansalpeter & \rmsstwosonemuMillonsplmediansalpeter & \rmssthreesonemuMillonsplmediansalpeter & \rmssfoursonemuMillonsplmediansalpeter & \rmssxsonemuMillonsplmediansalpeter  \\
\enddata
\tablecomments{The RMS scatter in the time delay columns is given in units of days.  The magnification ratio columns  are unitless, and report the scatter of flux magnification ratios (i.e., no conversion into magnitudes has been done). When computing the scatter, we have excluded $>30$ day outliers for the time delays, and 1.5 for the magnification.}
\end{deluxetable}

\begin{deluxetable}{c|cccc|cccc}[h!]
\label{tab:salpeterOutlierFraction}
  \tablecolumns{9}
  \tablecaption{\sc Salpeter $>3\sigma$ Outlier Fraction}
 \tablehead{ Method & \multicolumn{4}{c|}{Time Delay} & \multicolumn{4}{c}{Magnifation Ratio}  \\  & S2--S1 & S3--S1 & S4--S1 & SX--S1 & 
 S2/S1 & S3/S1 & S4/S1 & SX/S1  }
  \startdata
Polynomial &  \outlierfractionstwosonedelayKellymediansalpeter & \outlierfractionsthreesonedelayKellymediansalpeter & \outlierfractionsfoursonedelayKellymediansalpeter & \outlierfractionsxsonedelayKellymediansalpeter & \outlierfractionstwosonemuKellymediansalpeter & \outlierfractionsthreesonemuKellymediansalpeter & \outlierfractionsfoursonemuKellymediansalpeter & \outlierfractionsxsonemuKellymediansalpeter\\
Gaussian Process & \outlierfractionstwosonedelayThorpmediansalpeter & \outlierfractionsthreesonedelayThorpmediansalpeter & \outlierfractionsfoursonedelayThorpmediansalpeter & \outlierfractionsxsonedelayThorpmediansalpeter & \outlierfractionstwosonemuThorpmediansalpeter & \outlierfractionsthreesonemuThorpmediansalpeter & \outlierfractionsfoursonemuThorpmediansalpeter & \outlierfractionsxsonemuThorpmediansalpeter  \\
{\tt SNTD} & \outlierfractionstwosonedelayRPthreerdmediansalpeter & \outlierfractionsthreesonedelayRPthreerdmediansalpeter & \outlierfractionsfoursonedelayRPthreerdmediansalpeter & \outlierfractionsxsonedelayRPthreerdmediansalpeter & \outlierfractionstwosonemuRPthreerdmediansalpeter & \outlierfractionsthreesonemuRPthreerdmediansalpeter & \outlierfractionsfoursonemuRPthreerdmediansalpeter & \outlierfractionsxsonemuRPthreerdmediansalpeter  \\
{\tt PyCS} & \outlierfractionstwosonedelayMillonsplmediansalpeter & \outlierfractionsthreesonedelayMillonsplmediansalpeter & \outlierfractionsfoursonedelayMillonsplmediansalpeter & \outlierfractionsxsonedelayMillonsplmediansalpeter & \outlierfractionstwosonemuMillonsplmediansalpeter & \outlierfractionsthreesonemuMillonsplmediansalpeter & \outlierfractionsfoursonemuMillonsplmediansalpeter & \outlierfractionsxsonemuMillonsplmediansalpeter  \\
\enddata
\tablecaption{Outliers are defined as time delay measurements that from the mean by more than 30 days, or measured magnification
ratios that deviate by more than 1.}
\end{deluxetable}

\begin{figure*}[h!]
\centering
\gridline{\fig{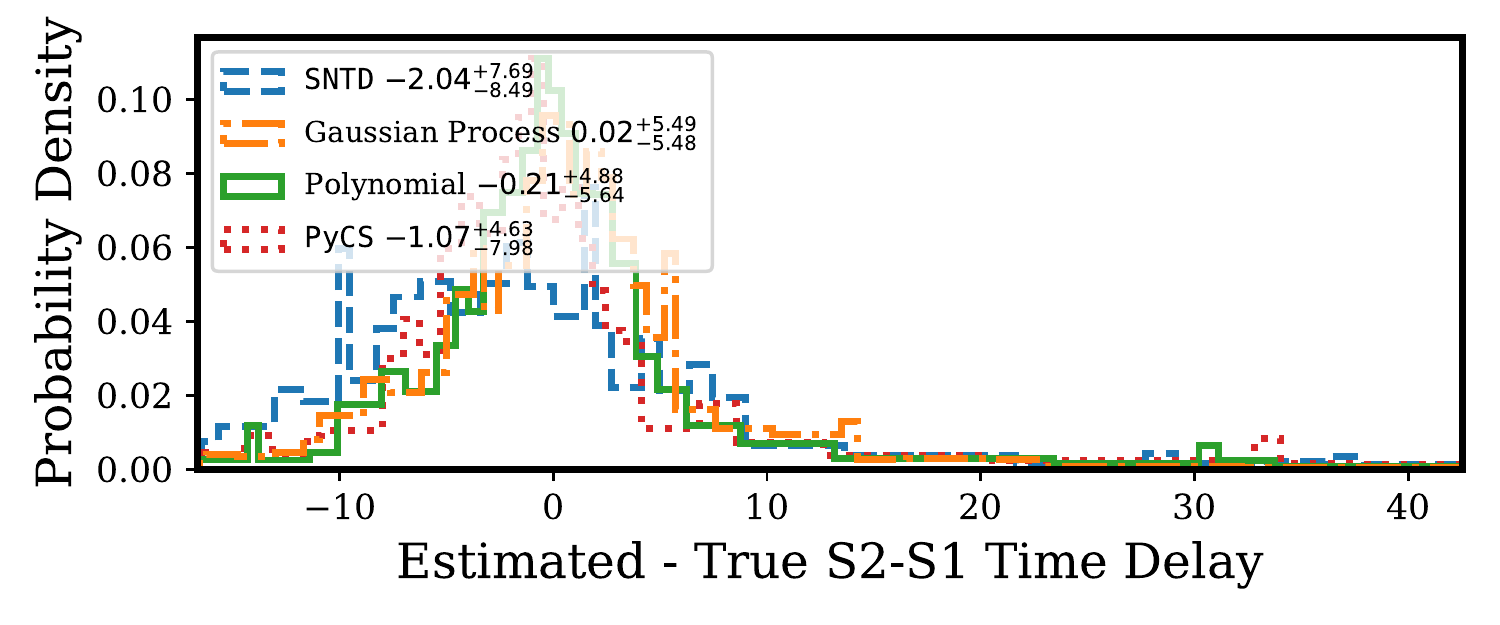}{0.49\textwidth}{}  
\fig{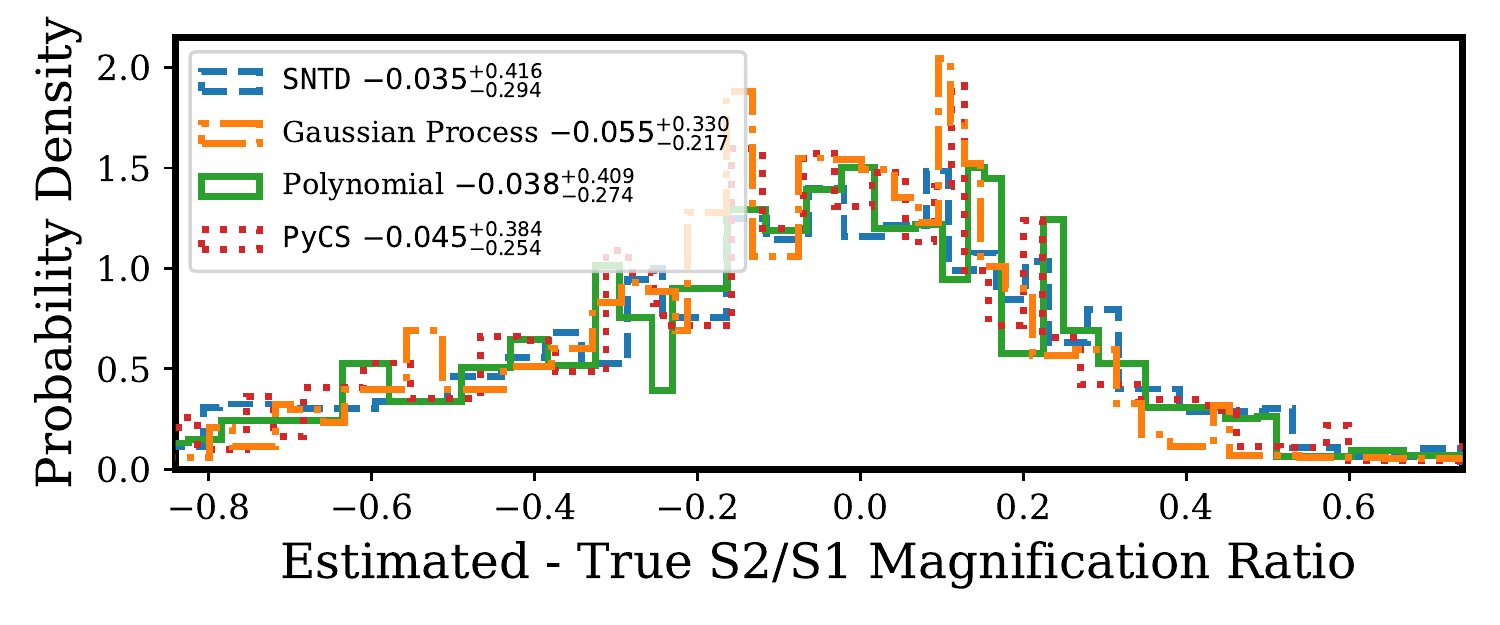}{0.49\textwidth}{}}
\vspace{-3em}
\gridline{\fig{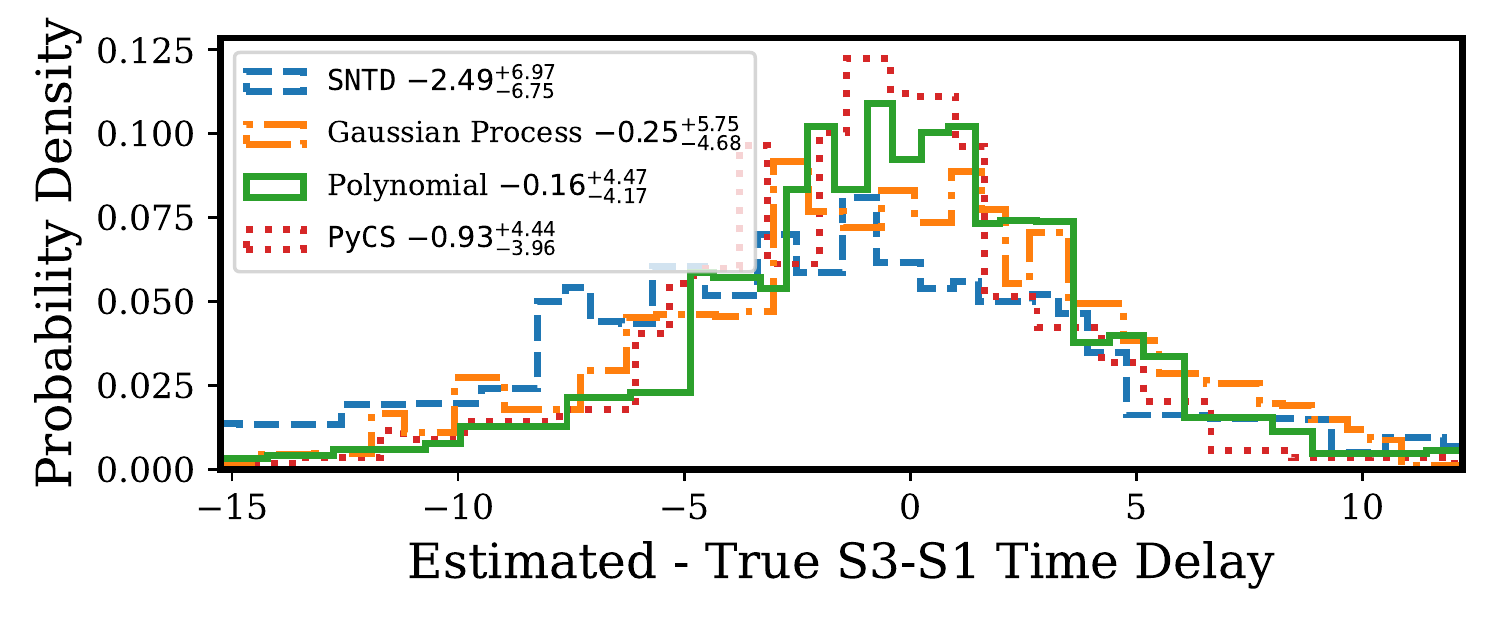}{0.49\textwidth}{}  
\fig{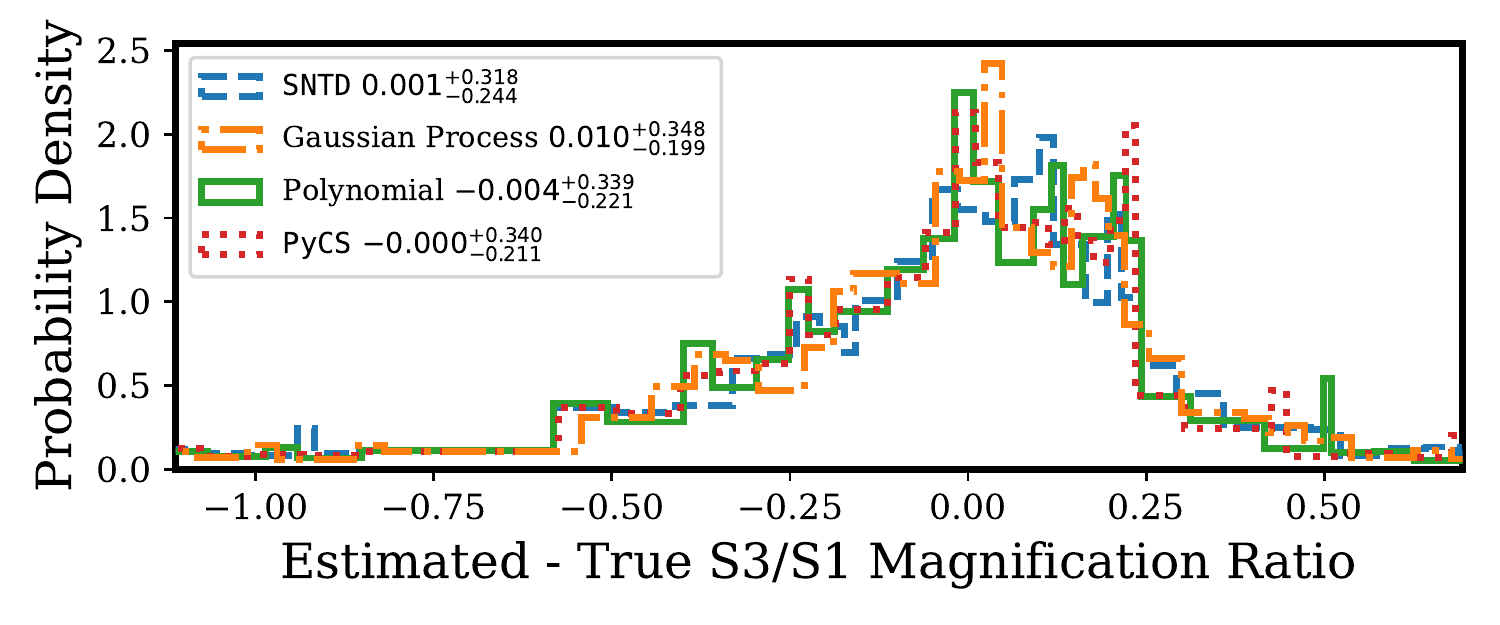}{0.49\textwidth}{}}
\vspace{-3em}
\gridline{\fig{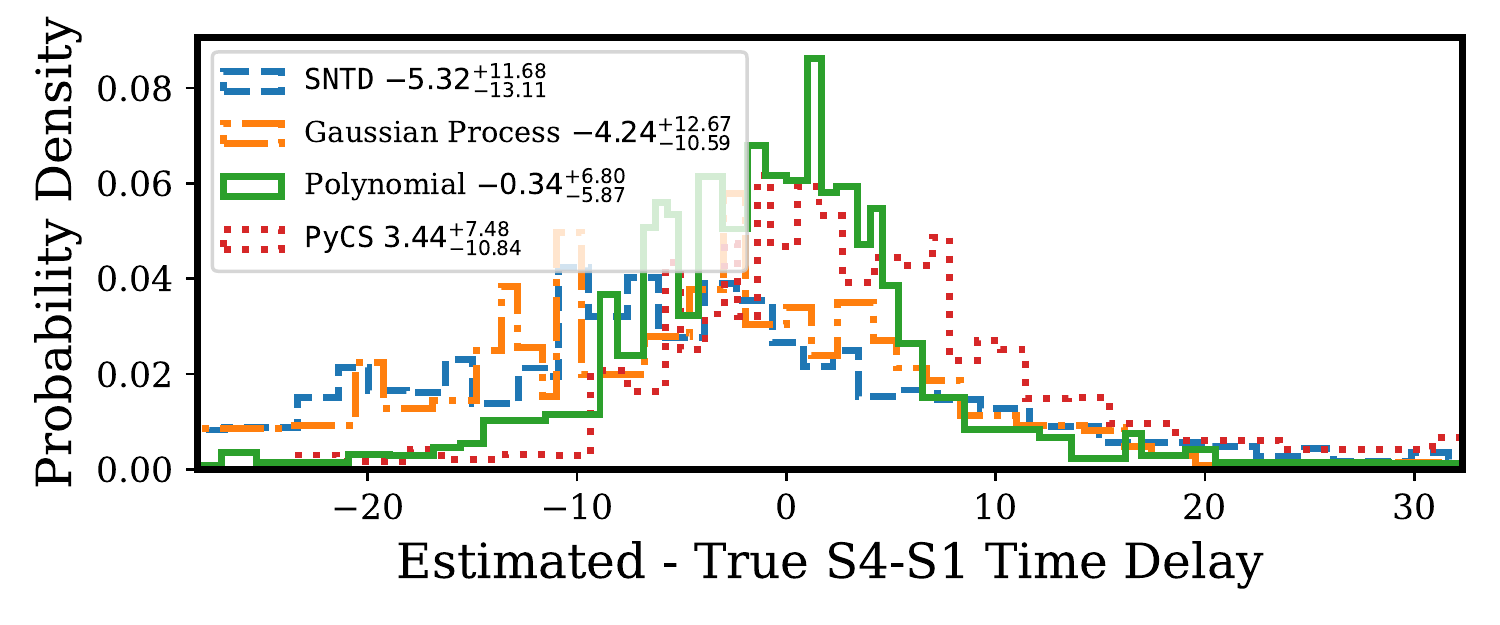}{0.49\textwidth}{}  
\fig{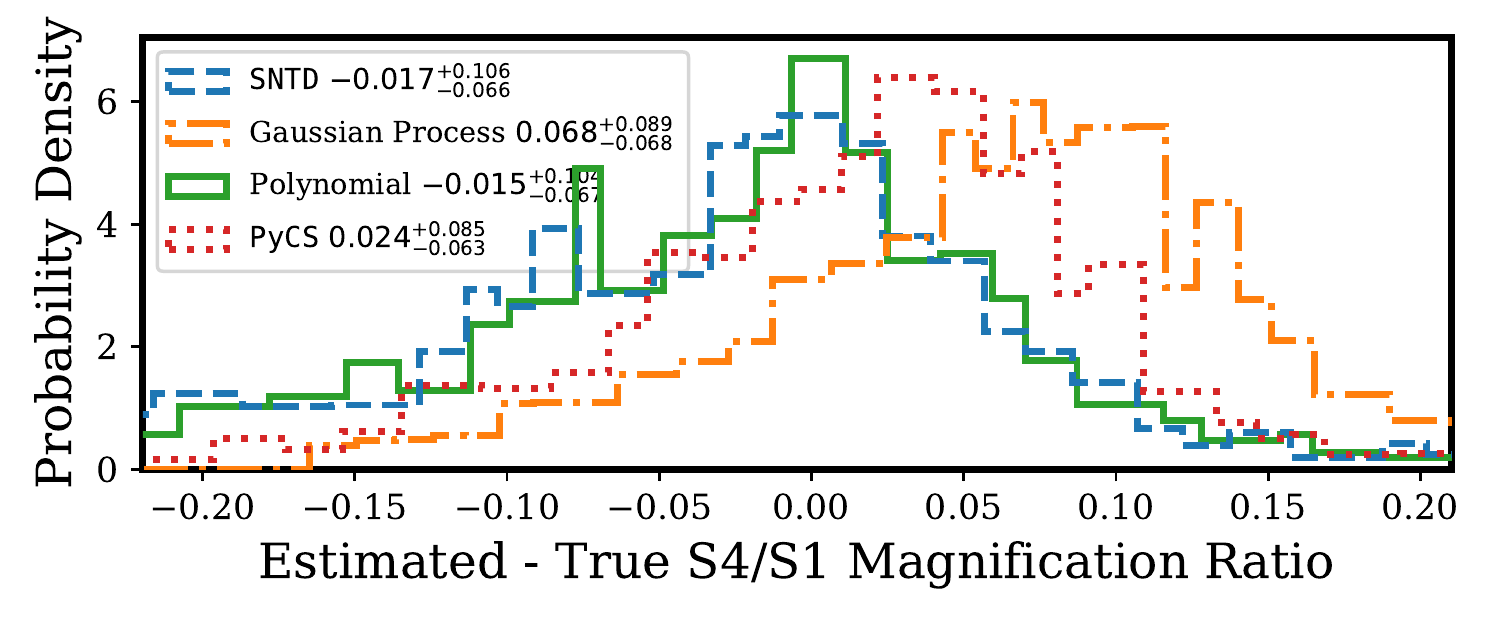}{0.49\textwidth}{}}
\vspace{-3em}
\gridline{\fig{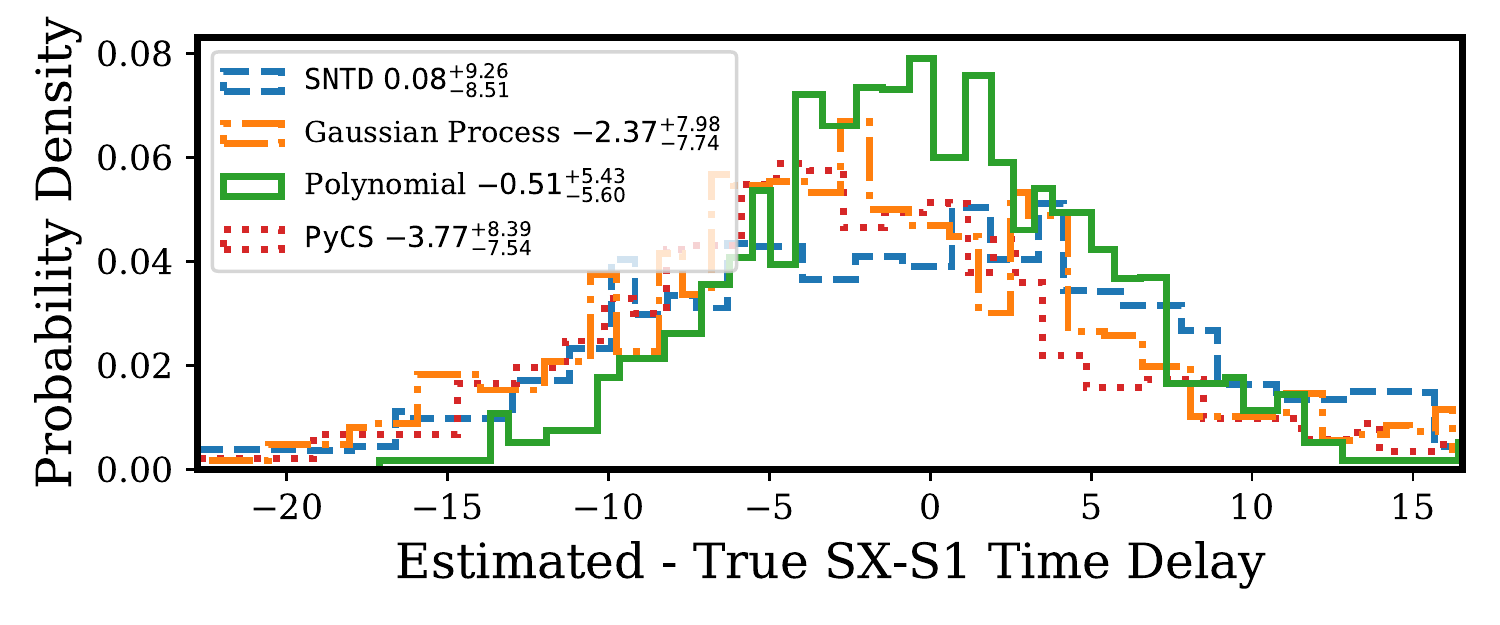}{0.49\textwidth}{}  
\fig{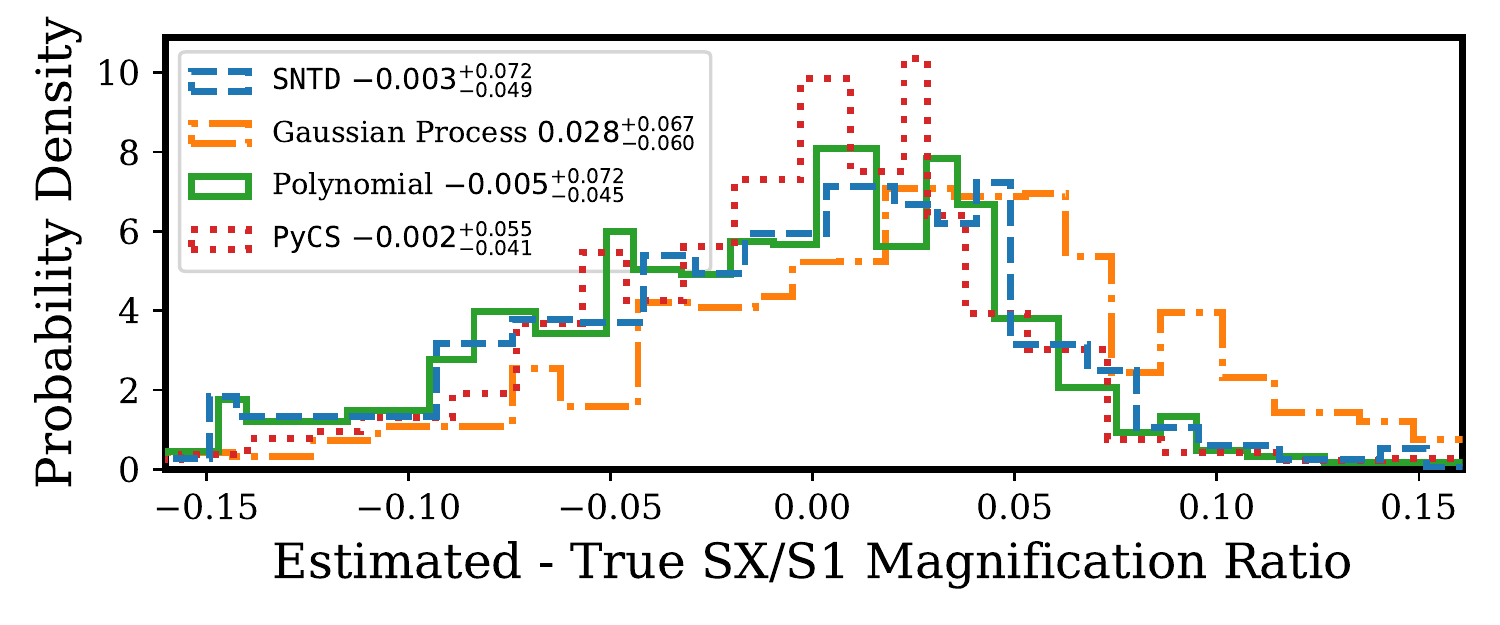}{0.49\textwidth}{}}
\vspace{-3em}
\caption{Distributions of residuals of time delay (left) and magnification ratio (right) estimated from simulated light curves by four separate algorithms given the assumption of a \citet{salpeter55} IMF. As indicated, each row shows results from the simulated data that mimics one pair of images, from S2:S1 in the top tow to SX:S1 in the bottom row.
The 16th, 50th, and 84th percentiles are listed in the legends.
}
\label{fig:residuals_salpeter}
\end{figure*}

\clearpage

\begin{figure}
\centering
\gridline{\fig{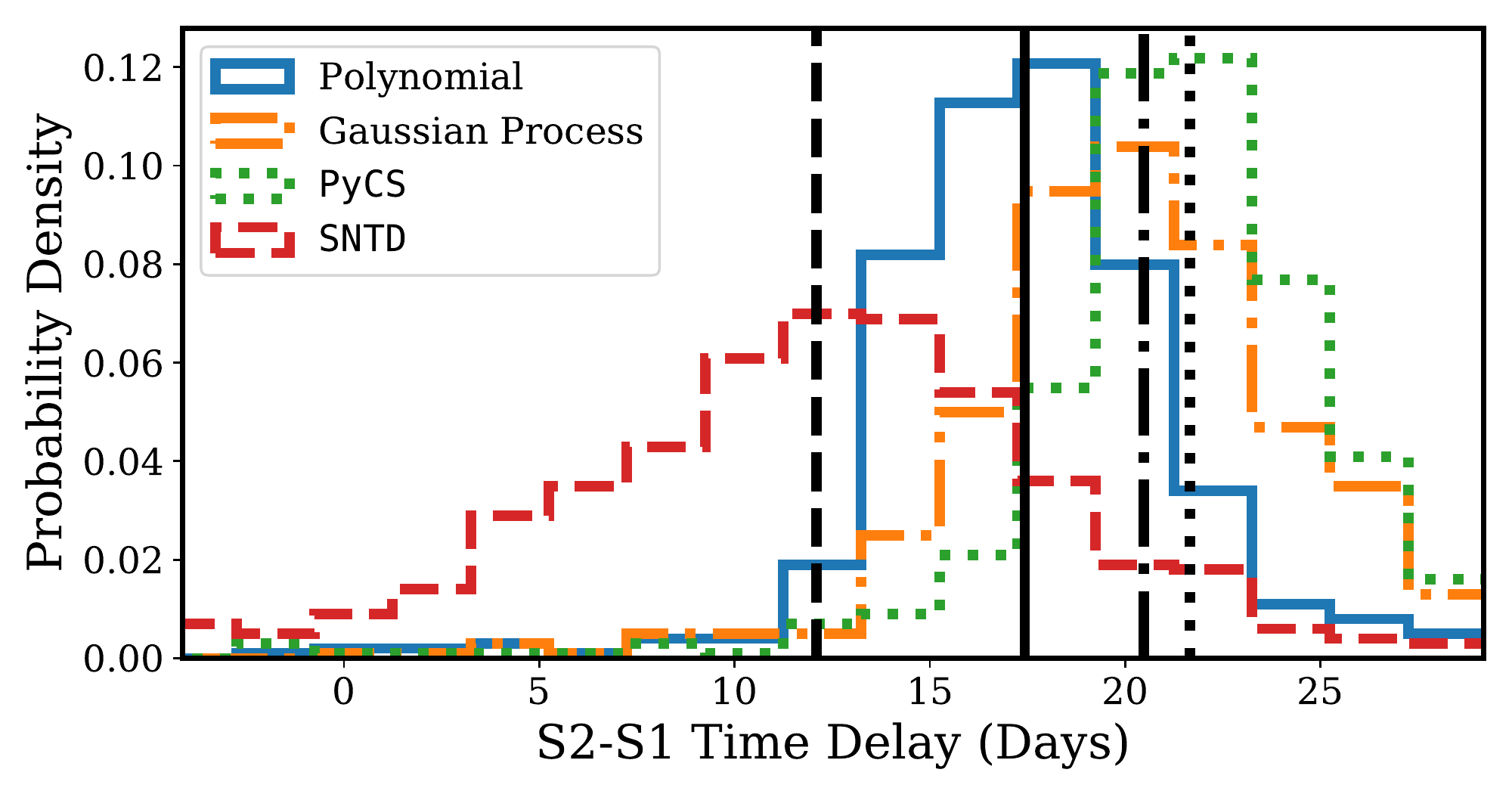}{0.49\textwidth}{}
\fig{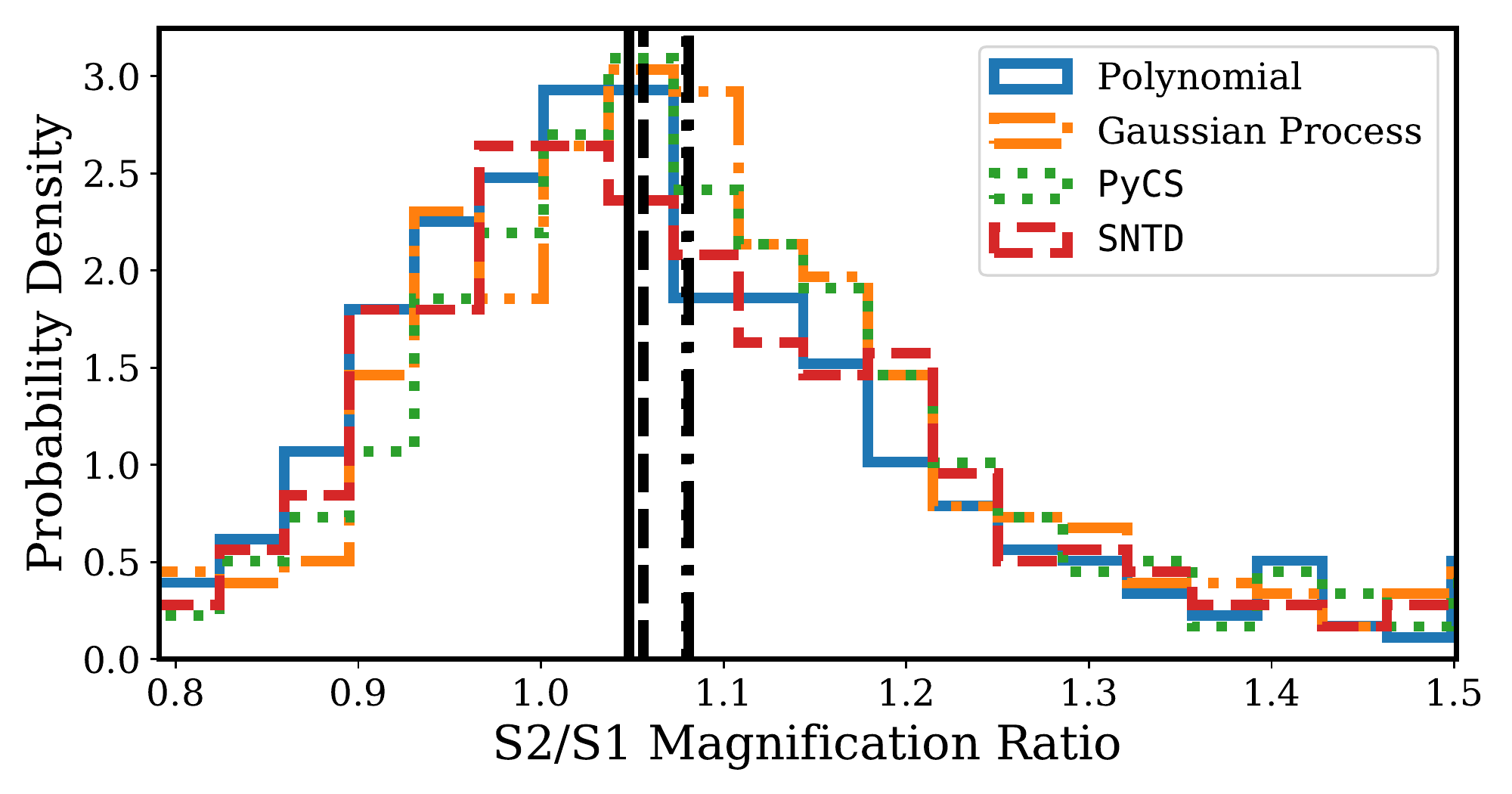}{0.49\textwidth}{}}
\vspace{-3em}
\gridline{\fig{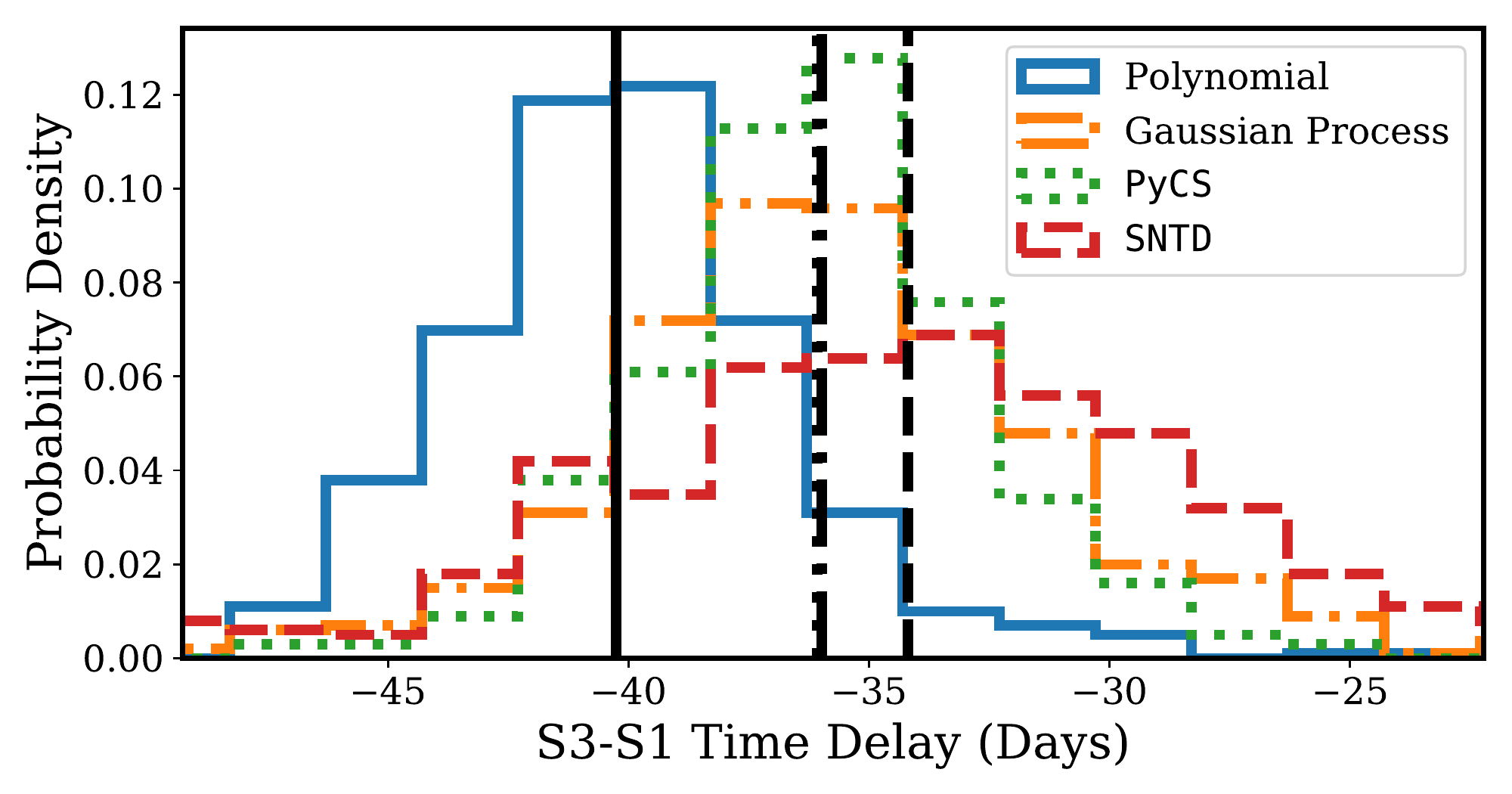}{0.49\textwidth}{}
\fig{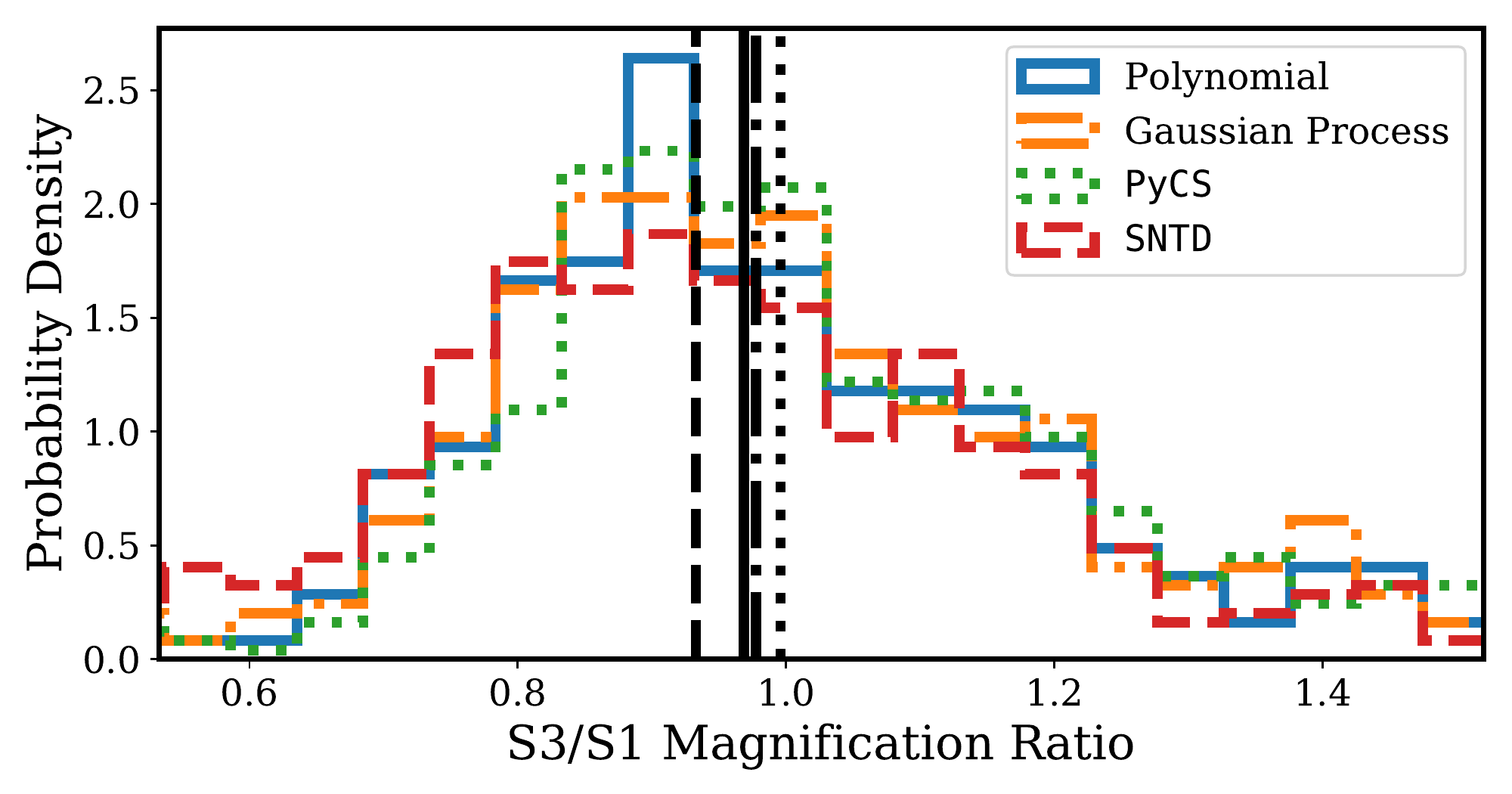}{0.49\textwidth}{}}
\vspace{-3em}
\gridline{\fig{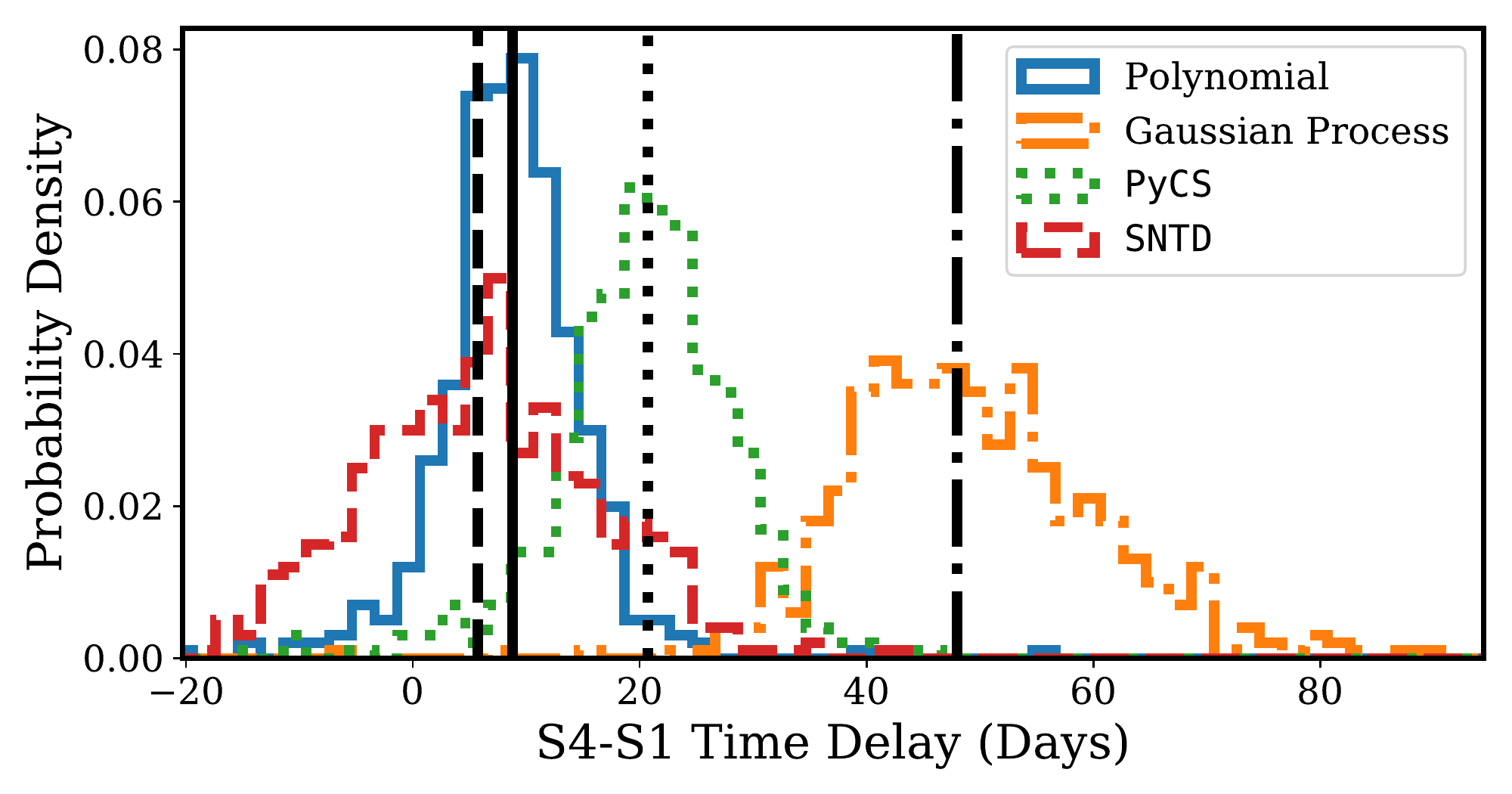}{0.49\textwidth}{}
\fig{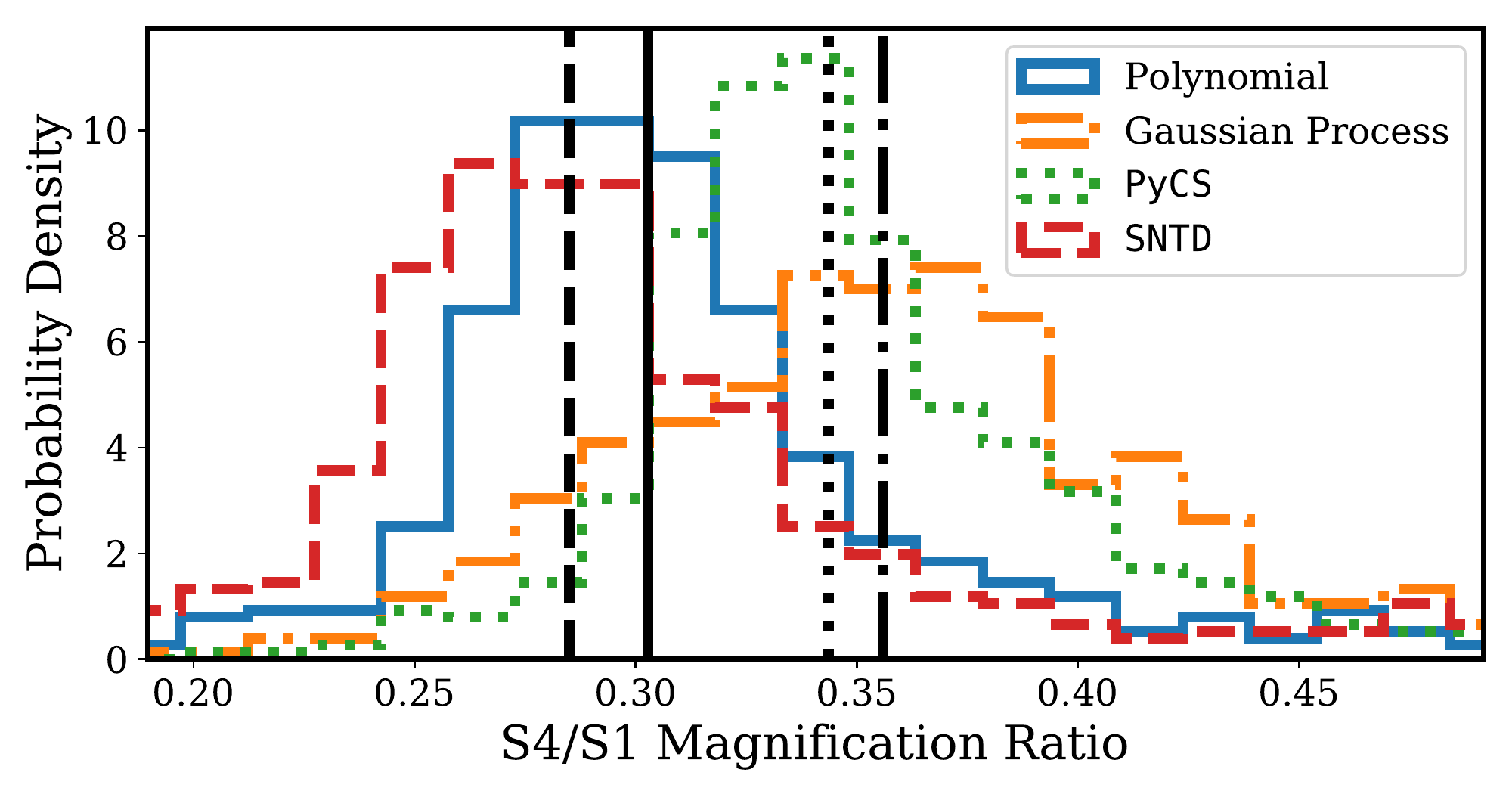}{0.49\textwidth}{}}
\vspace{-3em}
\gridline{\fig{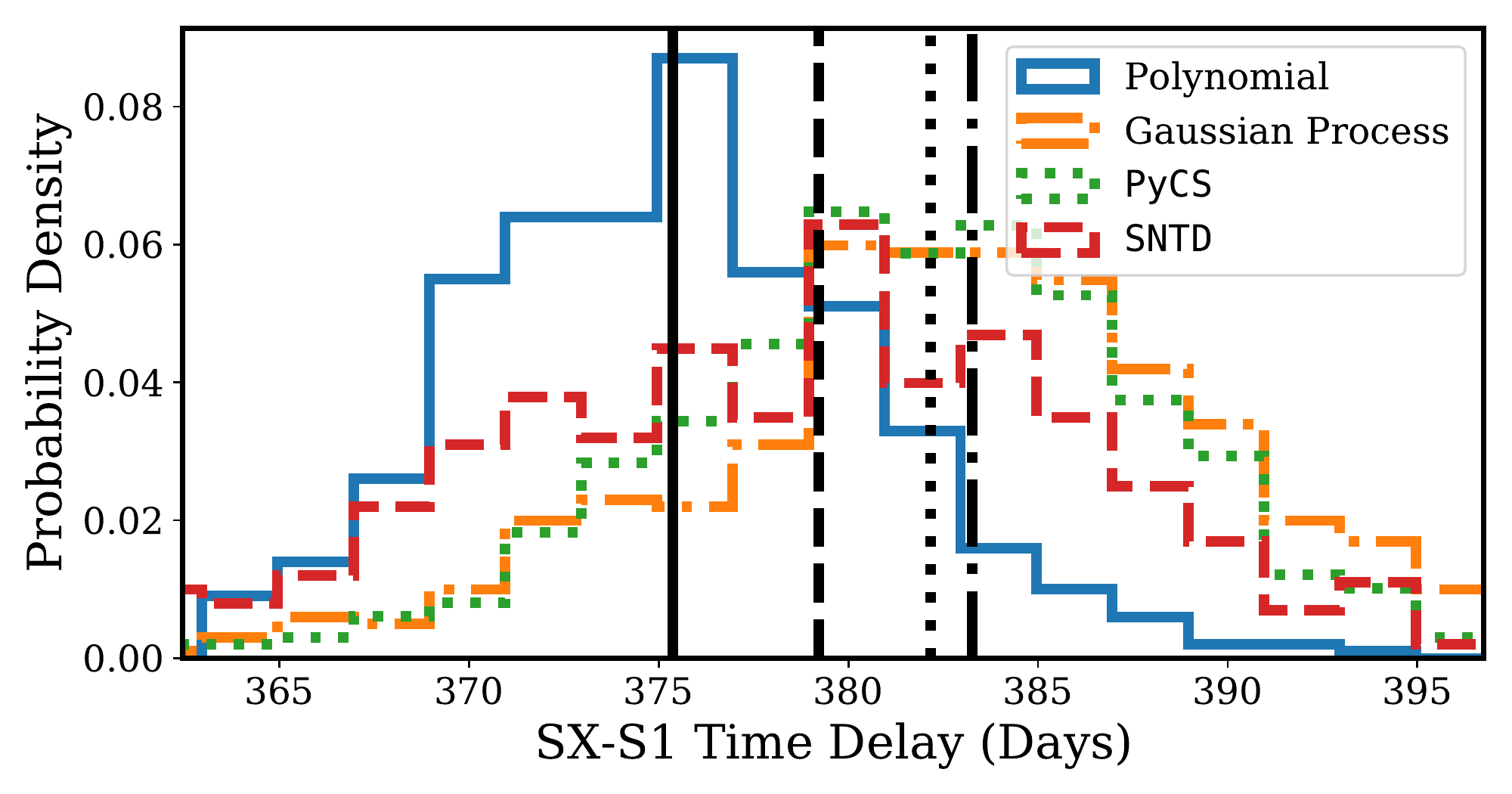}{0.49\textwidth}{}
\fig{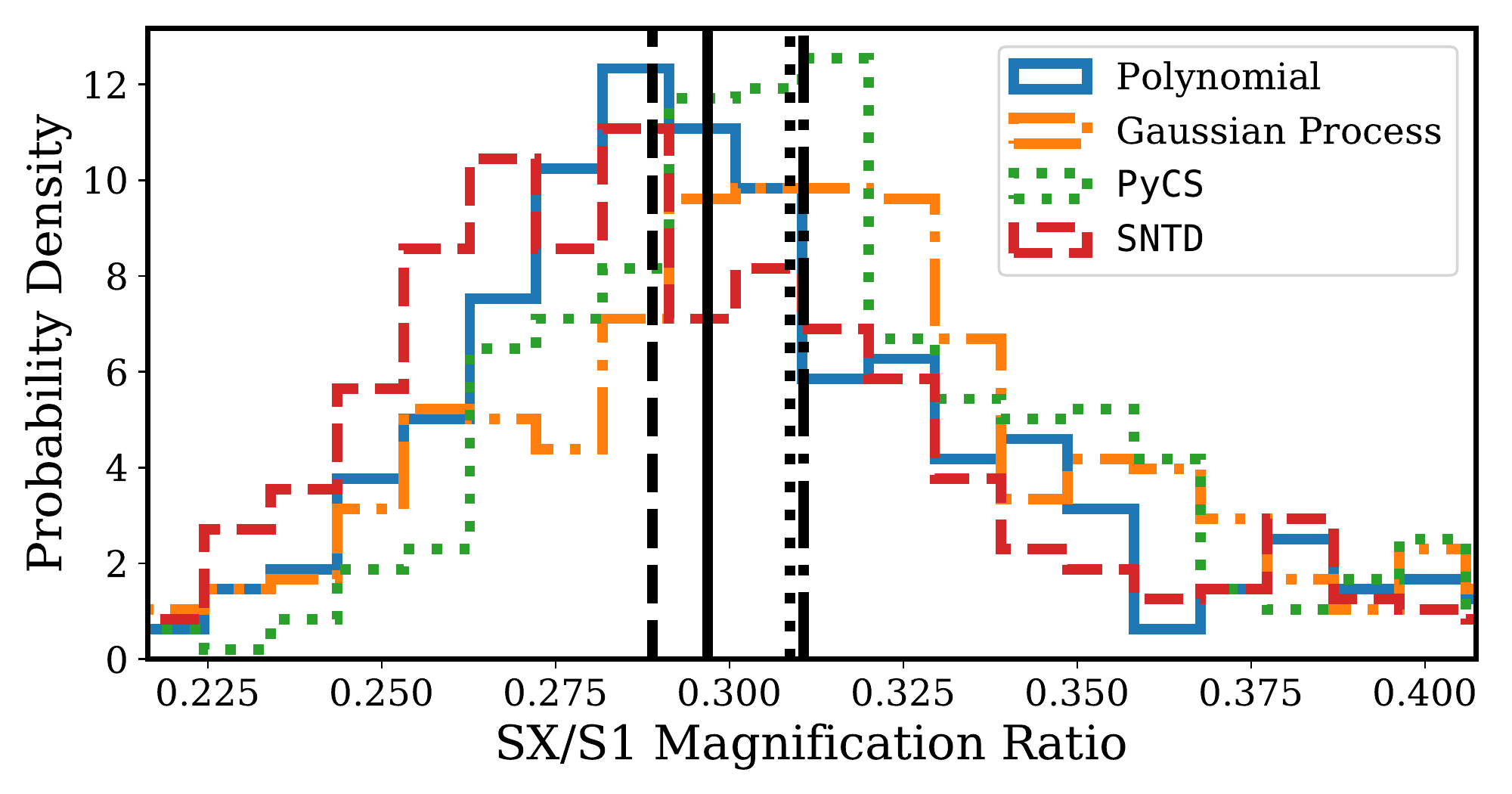}{0.49\textwidth}{}}
\vspace{-3em}
\caption{Constraints on the relative time delays and magnifications of the five images of SN Refsdal for the four fitting methods. The greatest tension is between the estimates of the S4--S1 time delay from the Gaussian process method and the piecewise polynomial method. The vertical black lines correspond to the fiftieth percentile of the respective distributions, and each vertical line has the same linestyle as that of the corresponding distribution.}
\label{fig:tension}
\end{figure}

\end{document}